\documentclass[%
 reprint,
preprintnumbers,
nofootinbib,
nobibnotes,
 amsmath,amssymb,
 aps,
]{revtex4-1}
\label{key}
\usepackage{graphicx}% Include figure files
\usepackage{dcolumn}% Align table columns on decimal point
\usepackage{bm}% bold math
\usepackage[usenames]{color}
\usepackage{hyperref}
\usepackage{cleveref}% add hypertext capabilities
\def\lsim{\mathrel{\rlap{\lower4pt\hbox{\hskip1pt$\sim$}} \raise1pt\hbox{$<$}}}
\def\gsim{\mathrel{\rlap{\lower4pt\hbox{\hskip1pt$\sim$}} \raise1pt\hbox{$>$}}}
\begin{document}

%\preprint{YHEP-COS19-005}

\title{Progress in Higgs inflation}% Force line breaks with \\
%\thanks{A footnote to the article title}%

\author{Dhong Yeon Cheong}%$^*$}
\email{dhongyeon@yonsei.ac.kr}
\author{Sung Mook Lee}
 \email{sungmook.lee@yonsei.ac.kr}
%\email{Co-first authors: DYC and SML}
 \author{Seong Chan Park}
 \email{sc.park@yonsei.ac.kr}
\affiliation{%
Department of Physics \& IPAP \& Lab for Dark Universe, Yonsei University, Seoul 03722, Korea
}%

\date{\today}% It is always \today, today,
             %  but any date may be explicitly specified

\begin{abstract}
We review the recent progress in Higgs inflation focusing on Higgs-$R^2$ inflation, primordial black hole production and the $R^3$ term. 
\end{abstract}

%\keywords{Suggested keywords}%Use showkeys class option if keyword
                              %display desired
\maketitle

%\tableofcontents

\section{Introduction}

Among many models, Higgs inflation~\cite{Bezrukov:2007ep}, equivalently Starobinsky's inflation with a $R^2$ term~\cite{Starobinsky:1980te},~\footnote{Neglecting the kinetic term during the inflation, both theories are equivalent since ${\cal L}/\sqrt{-g} \ni (M^2+\xi \phi^2) R/2 -\lambda \phi^4/4$ is mapped to $M^2 R/2+(\xi^2/4\lambda) R^2$ by solving the field equation for $\delta \phi$.} 
attracts special attention as it provides the best fit to the astrophysical and cosmological observations~\cite{Aghanim:2018eyx, Akrami:2018odb}. The success of Higgs inflation can be generalized to a broader perspective~\cite{Park:2008hz}.
However, Higgs inflation is not free from theoretical issues:
most notably, its original setup requires a large nonminimal coupling $\xi \sim 10^4$ that leads to a low cutoff $\Lambda \lsim M_P/\xi \ll M_P$~\cite{Burgess:2009ea,Barbon:2009ya,Burgess:2010zq,Lerner:2009na, Park:2018kst}. Several proposed solutions include considering a field-dependent vacuum expectation value \cite{Bezrukov:2010jz}, introducing the Higgs near-criticality \cite{Hamada:2014iga, Bezrukov:2014bra, Hamada:2014wna} or adding new degrees of freedom~\cite{Giudice:2010ka,Barbon:2015fla,Giudice:2014toa,Ema:2017rqn,Gorbunov:2018llf}.

The addition of a $R^2$ term to the gravity sector proved to be a novel setup that resolves these issues during inflation and reheating. The $R^2$ term, which may dynamically arise from radiative corrections of the nonminimal interactions~\cite{Salvio:2015kka,Calmet:2016fsr,Wang:2017fuy,Ema:2017rqn,Ghilencea:2018rqg,Ema:2019fdd,Canko:2019mud}, then pushes the theory's cutoff scale beyond the Planck scale: the new scalar field, $s$, called the scalaron emerges in association with the $R^2$ term and unitarizes the theory~\cite{Gorbunov:2018llf,He:2018mgb,He:2018gyf,Gundhi:2018wyz} just like the Higgs field does for the electroweak theory. The violent preheating in pure Higgs inflation~\cite{Ema:2016dny} is also resolved by the $R^2$ term~\cite{, He:2018gyf, He:2018mgb}.  Therefore, the most realistic approach is to consider both scalars in our setup. We refer this setup as `Higgs-$R^2$' inflation.  More theoretical discussions include Palatini formulation of gravity~\cite{Jinno:2019und} and swampland conjectures~\cite{Cheong:2018udx, Park:2018fuj}.

\section{Pure Higgs inflation} 
\subsection{Model}
As inflation has been regarded as a standard paradigm describing the early universe, it also has become important to understanding how inflation is actually realized in particle physics models. In the Standard Model(SM), there is an \textit{unique} candidate, which is the Higgs boson.

Unfortunately, the chaotic inflation type of potential $ V \propto \phi^{n}  $ is known to be inconsistent with cosmological measurements because it predicts a large tensor-to-scalar ratio $ r $ . However, an additional non-minimal coupling between the Higgs scalar $ \phi $ and the Ricci scalar $ R $ in the gravity sector flattens the potential in the large field regime in the Einstein frame and suppresses $ r $ \cite{Bezrukov:2007ep}.

The Lagrangian for the relevant inflaton and gravity sectors is
\begin{align}
S_{J} = \int d^{4}x \sqrt{-g_{J}}\left[ \frac{M_P^2}{2} \left(1 + \frac{\xi \phi^{2}}{M_{P}^{2}}\right)R_{J} - \frac{1}{2} \vert \partial_{\mu} \phi \vert^{2}  - V_{J}(\phi) \right]
\end{align}
where $V_{J}(\phi) = \frac{\lambda}{4} (\phi^{2} - v_{\rm EW}^{2})^{2}$ with $ v_{\rm EW} \simeq 246 ~\rm GeV$ and $ J $ stands for the Jordan frame. To eliminate the non-minimal coupling, we redefine the metric as
\begin{align}
	g_{\mu\nu} = \Omega(\phi)^{2} g_{J\mu\nu},
\end{align}
where
\begin{align}
	\Omega^{2} =  1 + \frac{\xi \phi^{2}}{M_{P}^{2}},
\end{align}
and we canonicalize the kinetic term with the relation
\begin{align}
	\frac{dh}{d\phi} = \left( \frac{1 + \xi(1 + 6 \xi) \phi^{2}/ M_{P}^{2}}{(1 + \xi \phi^{2} / M_{P}^{2})^{2}}	\right)^{1/2}.
\end{align}
Then, the action in the canonical Einstein frame is \footnote{In fact, the form of the action is different when the Palatini formalism is used, which regard the metrics and affine connection independently. In this review, we take the standard metric formalism.}
\begin{align} \label{Eq:dEdJ}
	S = \int d^{4}x \sqrt{-g} \left[ \frac{M_{P}^{2}}{2} R - \frac{1}{2} g^{\mu\nu} \partial_{\mu}h \partial_{\nu}h - V(h)			\right]
\end{align}
where
\begin{align}
	V(h) \equiv V_{J}(\phi(h))/\Omega^{4}(\phi(h)).
\end{align}
Approximately, the potential takes the form
\begin{align}
	V(h) \simeq
	\begin{cases}
	\frac{\lambda}{4} h^{4} &\text{for~~}\vert h \vert \ll \frac{M_{P}}{\xi} \\
	\frac{\lambda M_{P}^{2}}{4 \xi^{2}} \left(1- e^{-\sqrt{\frac{2}{3}} \frac{h}{M_{P}}}\right)^{2}
	&\text{for~~}\vert h \vert \gg \frac{M_{P}}{\xi}
	\end{cases}.
\end{align}
Note that we neglected $ v_{\rm EW} \ll M_{P}/ \xi $ and that the potential becomes asymptotically constant at large field values.
The e-folding number $ N_{e} \equiv \ln a(t_{\rm end}) / a(t_{*}) $, with `end' meaning the time at the end of the inflation and $ * $ meaning CMB pivot scale/time, is
\begin{align}
	N_{e}& = \int_{t_{\rm end}}^{\rm t_{*}} H dt 
= \int^{\phi_{*}}_{\phi_{\rm end}} \frac{1}{M_{P}^{2}}\frac{V}{dV/d\phi}\left( \frac{dh}{d\phi}\right)^{2} d\phi \\
&= \frac{3}{4} \frac{\phi_{*}^{2} - \phi_{\rm end}^{2}}{M_{P}^{2}} \simeq  \frac{3}{4} \frac{\phi_{*}^{2}}{M_{P}^{2}}.
\end{align}
Here, we use $-3 H \dot{h} \simeq V^{\prime}(h)$ during slow-roll inflation and $ \phi_{*} \gg \phi_{\rm end} $. Normally, the number of e-foldings required to solve the horizon and flatness problems is assumed to be $ N \simeq 50-60 $.

From the potential in the Einstein frame, we can calculate the slow roll parameters as
\begin{align}
\epsilon_{V} &\equiv \frac{M_{P}^{2}}{2} \left(	\frac{V^{\prime}(h)}{V(h)}	\right)^{2}  \simeq \frac{4 M_{P}^{4}}{3 \xi^{2} \phi^{4}} \simeq \frac{3}{4 N_{e}^{2}}, \\
\eta_{V} &\equiv - M_{P}^{2} \frac{V^{\prime \prime}(h)}{V(h)}  \simeq \frac{4 M_{P}^{4}}{3 \xi^{2} \phi^{4}} \left( 1 - \frac{\xi \phi^{2}}{M_{P}^{2}}\right) \simeq \frac{1}{N_{e}},
\end{align}
where $ {\prime} $ denotes the derivative with respect to $ h $.

By parameterizing the scalar and tensor power spectrum as
\begin{align}
	\mathcal{P}_{\mathcal{R}}(k) = A_{s} \left( \frac{k}{k_{*}}\right)^{1-n_{s}},&&	\mathcal{P}_{\mathcal{T}}(k) = A_{t} \left( \frac{k}{k_{*}}\right)^{n_{t}}
\end{align}
respectively, cosmological observables such as the spectral index and the tensor-to-scalar ratio are approximated with the slow-roll parameters:
\begin{align}
	n_{s} &\simeq 1- 2 \eta_{V} - 6 \epsilon_{V} \simeq 0.965 \\
	r &\equiv \frac{A_{t}}{A_{s}}= 16 \epsilon_{V} \simeq 0.003,
\end{align}
which are perfectly consistent with current Planck 2018 measurement \cite{Akrami:2018odb}. To satisfy the amplitude of the curvature power spectrum $ \log (10^{10} A_{s}) = 3.047 \pm 0.014 $, one needs other constraints for $ \xi $ and $ \lambda $. 
\begin{align}
\left.\frac{\xi^{2}}{\lambda}\right\vert_{*} \simeq 2 \times 10^{9},&& \xi \simeq 47000\sqrt{\lambda}.
\end{align}
Therefore, by assuming $ \lambda = 0.15 $, for example, we have to assume very large  non-minimal coupling $ \xi \simeq 18000 $. Such a large non-minimal coupling causes theoretical issues including naturalness, and more seriously, the unitarity problem \cite{Barbon:2009ya,Burgess:2009ea,Burgess:2010zq,Lerner:2009na}. We will come back to this issue later in this review.

Note that we assumed constant $ \lambda $ and $ \xi $ without considering the quantum corrections. In the Higgs inflation case, however, quantum corrections give non-trivial modifications not only to the inflaton dynamics, but also to cosmological observables.

\subsection{Critical Higgs Inflation}
For the currently known Higgs masses and top quark masses, the EW Higgs potential is known to be metastable as $ \lambda $ becomes negative when the renormalization scale is $ \mu \gtrsim \mathcal{O}(10^{10}\rm GeV) $ \cite{Degrassi:2012ry}. This fact may not be a big problem as long as the lifetime of the EW vacuum is longer than the age of the universe. However, if this is the case, the validity of the Higgs inflation scenario may be questioned \cite{Bezrukov:2014ipa}.~\footnote{Even in non-Higgs inflation cases, large quantum fluctuation in de Sitter bachground $ \mathcal{O }(H/2\pi) $ during inflation may cause a problem. See the Ref. \cite{Markkanen:2018pdo}.}

However, this result sensitively depends on the top quark mass measurement. In fact, the usually referred to top quark mass is the so-called `Monte-Carlo (MC) mass'. This is a mere parameter in MC simulations and the theoretical uncertainties on being identifed as the pole mass are large,  up to $ \mathcal{O}(1\rm GeV) $ \cite{Corcella:2019tgt}. Instead, by taking the latest pole mass from cross-section measurements \cite{Zyla:2020zbs}
\begin{align}
\left. m_{t}^{\rm pole} \right\vert_{\rm PDG}=172.4\pm 0.7~{\rm GeV},
\end{align}
the top quark mass which guarantees the Higgs potential stability $ m_{t}^{\rm pole} \lesssim 171.4 {\rm GeV} $ is within the $ 2 \sigma $ bound. In this review, we will assume absolute stability of the Higgs potential.

On the other hand, considering the effects of the running of the coupling to Higgs inflation cases is still important \cite{Hamada:2014iga, Bezrukov:2014bra, Hamada:2014wna}. The quartic coupling can be parameterized as
\begin{equation}
\left. \lambda\left(\mu \right) \right|_{\mu = \phi}= \lambda_\text{min} + \frac{\beta_2^\text{SM}}{\left(16 \pi^2\right)^2} \ln^2\left(\frac{\phi}{\phi_\text{min}}\right)
\label{Eq:lambdarunning}
\end{equation}
with $\beta_2^{\text{SM}} \approx 0.5$, $\mu_\text{min} = \phi_\text{min} \sim 10^{17} - 10^{18}\,\, \text{GeV}$ as denoted in \cite{Degrassi:2012ry, Buttazzo:2013uya}.\footnote{In fact, due to the non-renormalizability of the theory, there exists a dependence on the way to choose the renormalization scale, which is also called `prescription'. In this review, we choose $ \mu = \phi $, where $ \phi $ is the Jordan frame Higgs field value. For the meaning of the prescription dependence in detail, see Ref. \cite{Hamada:2016onh}.}

One of the major consequences is that small non-minimal couplings, $ \xi \gtrsim \mathcal{O}(10)$, are allowed, assuming $ \lambda_{*} \sim \mathcal{O}(10^{-3}) $. Another possible result is that the form of the potential could have an inflection point when the values of $ \lambda $ are tuned to be nearly zero, and this type of inflation is referred to as `critical Higgs inflation (CHI)'. This fact has motivated efforts to look into the possibilities of generating primordial black holes (PBH) on the model, as the inflection shape potential is a well-known class of models to induce large curvature power spectrum on small scales producing a significant amount of PBHs.

\subsection{Unitarity Problem}
One should be careful when dealing with the cut-off scale of the Higgs inflation due to the existence of large non-minimal coupling \cite{Burgess:2009ea,Burgess:2010zq,Barbon:2009ya}.

For the small field region $ \phi < M_{P} / \xi $, from Eq.~(\ref{Eq:dEdJ}), fields in each frames are related by
\begin{align}
	h \simeq \phi + \sqrt{\frac{3}{2}} \frac{\xi \phi^{2}}{M_{P}}.
\end{align}
This means that the kinetic term of the field $ h $ in the Einstein frame contains derivative couplings
\begin{align}
	 (\partial_{\mu} h)^{2}= (\partial_{\mu} \phi)^{2} +  \frac{3}{2} \frac{\xi^{2}}{M_{P}^{2}} \phi^{2} (\partial_{\mu} \phi)^{2} + \cdots
\end{align}
From the second term, one concludes that the theory becomes strongly coupled at $ E \gtrsim  \Lambda\equiv M/\xi $. 

During inflation, the fluctuation is defined with respect to the classical background field value (denoted with `bar') as
\begin{align}
	g_{\mu\nu} = \bar{g}_{\mu\nu} + \delta g_{\mu\nu},&&\phi = \bar{\phi} + \delta \phi.
\end{align}
Therefore, in the region where $ \bar{\phi} > M_{P}/\sqrt{\xi} $ corresponding to the inflationary region,\begin{align}
\frac{\xi^{2}}{M_{P}^{2}} \bar{\phi}^{2} (\partial_{\mu} \delta \phi)^{2},
\end{align}
implying
\begin{align}
	\Lambda_{\rm inf}^{J} \simeq \sqrt{\xi} \bar{\phi}.
\end{align}
Therefore, the cut-off scale during inflation is safely higher than the energy scale of inflation \cite{Bezrukov:2010jz}.

However, after inflation, the inflaton rolls down to the minimum of the potential and starts to oscillate coherently. Lastly, the Higgs field decays to SM particles and loses its energy. These procedures are called `reheating'. During the reheating phase, the decay of the longitudinal gauge boson is violent and the momentum of the produced particles is $ k \simeq \mathcal{O}(\sqrt{\lambda} M_{P}) $, which is larger than the cut-off scale during the reheating, $ M_{P} / \xi $ \footnote{In fact, the Higgs boson decay to the longitudinal mode of the gauge boson may depend sensitively on higher order operators. See the Ref. \cite{Hamada:2020kuy}.} \cite{Ema:2016dny}. Above the cut-off scale, decay processes violate unitarity, becoming strongly coupled, and lose its predictivity.

To unitarize the Higgs inflation during reheating, there has been a lot of attempts to raise the cut-off scale of the Higgs inflation by introducing additional degrees of freedom \cite{Giudice:2010ka,Barbon:2015fla,Giudice:2014toa}. One of the simplest and minimal ways is to consider the $ R^{2} $ corrections \cite{Ema:2017rqn, Gorbunov:2018llf,He:2018mgb,He:2018gyf,Gundhi:2018wyz}, as described in the next section.

\section{Higgs-$R^2$ inflation} 
\subsection{Model} 
The Higgs-$R^2$ inflation is a simple UV extension that cures the theoretical/phenomenological issues of single-field Higgs inflation~\cite{Ema:2017rqn,Gorbunov:2018llf}. The action takes the following form.

\begin{equation}
S_J  =  \int d^4 x \sqrt{-g_{J}} \left[ F(h, R_J) - \frac{1}{2} g^{\mu\nu} \nabla_\mu h \nabla_\nu h - \frac{\lambda}{4} h^4 \right] 
\end{equation}
where $ M $ is the scalaron mass, $ h $ is a scalar that stands for the Standard Model Higgs in the unitary gauge, and a conveniently defined function $F(h, R_J)$ and its derivative with respect to $R_J$
\begin{align}
F(h,R_{J})&=\frac{M_P^2}{2} \left(R_{J} + \frac{\xi h^2}{M_P^2}R_J+ \frac{R_J^2}{6M^2} \right),\\
\frac{\partial F}{\partial R_J} &= \frac{M_P^2}{2} \left(1+ \frac{\xi h^2}{M_P^2}+ \frac{R_J}{3M^2} \right).
\end{align}
The scalaron field $ s $ is defined as
\begin{align}
\sqrt{\frac{2}{3}} \frac{s}{M_{P}} 
&\equiv \ln \left(\frac{2}{M_{P}^{2}} \left\vert\frac{\partial F}{\partial R_{J}} \right\vert \right) \\
&= \ln \left(1 + \frac{\xi h^2}{M_P^2} + \frac{R_J}{3 M^2}\right) =\omega(s).
\end{align}

Through a Weyl transformation, $g_{\mu\nu} =e^{\omega(s)} g^J_{\mu\nu}$, we can get the action in the Einstein frame as
\begin{widetext}
\begin{align}
S = \int d^4 x \sqrt{-g}\left[\frac{M_P^2}{2}R - \frac{1}{2} g^{\mu\nu} \nabla_\mu s \nabla_\nu s  - \frac{1}{2} e^{-\omega(s)}g^{\mu\nu} \nabla_\mu h \nabla_\nu h - U\left(s, h\right)\right]
\end{align}
\end{widetext}
where the scalar potential is
\begin{equation}
U\left(s, h\right) \equiv e^{-2\omega(s)} \left\{ \frac{3}{4}M_P^2 M^2  \left(e^{\omega(s)} - 1 - \frac{\xi h^2}{M_P^2}\right)^2  +\frac{\lambda}{4}h^4 \right\}.
\label{eqn:einpotential}
\end{equation}
As noted in the potential Eq.~(\ref{eqn:einpotential}), higher order operators are induced in the analysis; therefore, the perturbativity of the system is guaranteed for a specific cutoff scale. Expanding the potential around $s\simeq h \simeq 0$ yields a cutoff scale 
\begin{equation}
\Lambda \sim \mathcal{O}\left(\frac{M_P^2}{\xi^2 M^2}\right)M_P\gsim M_P
\end{equation}
for scalaron masses within the following range $M  \lsim M_P/\xi$. This cutoff scale guarantees the perturbative analysis of this model throughout inflation and preheating, alleviating the unitarity problem of single-field Higgs inflation~\cite{Gorbunov:2018llf,He:2018mgb,He:2018gyf,Gundhi:2018wyz} . 

This additional $R^2$ term can also be induced through quantum loop corrections in the large-$\xi$ limit. Renormalization group equations of the system imply a new scalar degree at the mass scale $ M\sim M_P/\xi$, which in turn corresponds to the strong coupling scale of Higgs inflation~\cite{Salvio:2015kka,Calmet:2016fsr,Wang:2017fuy,Ema:2017rqn,Ghilencea:2018rqg,Ema:2019fdd,Canko:2019mud, Ema:2020evi}. Therefore, the addition of the $R^2$ term is a natural aspect in terms of both perturbative unitarity and renormalizability. 

\begin{figure}[t]
\includegraphics[width=.4\textwidth]{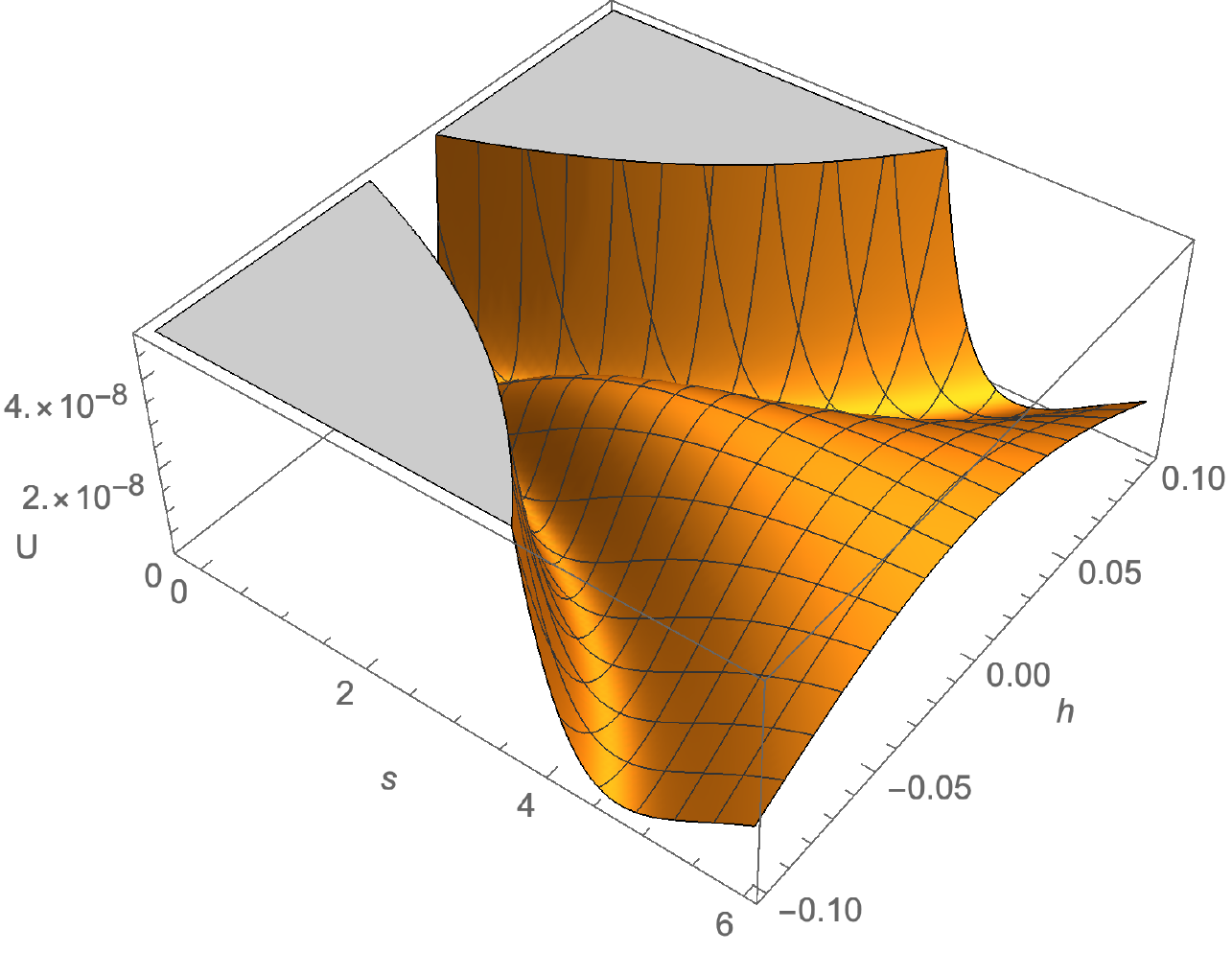}
\caption{\label{fig:potential_nonc} Shape of the potential in Eq. (\ref{eqn:einpotential}) with benchmark parameters $M = 2.06\times 10^{-4}M_P, ~ \xi = 4391.22, ~\lambda = 0.01, ~ h_\text{min} = 0.15 M_P$. The potential exhibits a valley structure in the large $s$ region. The inflaton follows this trajectory, giving successful slow-roll inflationary predictions. Inflation terminates near $s\sim M_P$, and preheating/reheating occurs. 
}
\end{figure}

\subsection{Inflation}
The additional scalar degree of freedom in the action yields a multifield inflationary potential. For a non-critical case with $\lambda = 0.01$, the potential takes a valley form as in Fig. \ref{fig:potential_nonc}. Re-formulating the action, we obtain the equation of motion 
\begin{align}
D_t \dot{\phi}^a + 3 H \dot{\phi}^a + G^{ab}D_b U=0,
\label{eq:eom}
\end{align} 
 with 
\begin{equation}
G_{ab}=\begin{pmatrix}
1&&0\\
0&&e^{-\Omega(s)}
\end{pmatrix},
\label{eq:fieldspacemetric}
\end{equation}
where $G_{ab}$ is the curved field space metric, $D_a \phi^b  = \partial_a \phi^b  + \Gamma^b_{c a} \phi^c$, $\Gamma^b_{ca} = \frac{1}{2} G^{be}\left(\partial_c G_{ae } + \partial_a G_{ec} - \partial_{e} G_{ca }\right)$, and $D_t = \dot{\phi}^{a}\nabla_a$. The inflaton will approximately roll down the valley, resulting in successful slow-roll inflation. This particular valley structure  exhibits an isocurvature mass $m_\text{iso}^2 \gg H^2$, leading to exponentially decaying isocurvature perturbations and allowing for an effective single-field description of the system. The potential along the $h$ direction is locally minimized at 
\begin{align}
h^2 = \frac{e^{\omega(s)}-1}{\frac{\xi}{M_P^2} + \frac{\lambda}{3\xi M^2}}. 
\end{align}
Inserting this into the potential and taking the large $s$ limit, we find that the inflationary potential at the plateau becomes 
\begin{align}
U_\text{inf} \approx \frac{\lambda M_P^4}{4 \xi^2 \left(1+ \frac{\lambda M_P^2}{3\xi^2 M^2}\right)}  
\end{align} 
yielding the cross-correlation between parameters $\lambda, \xi$, and $ M$. 
\begin{align} 
\frac{\xi^2}{\lambda} + \frac{M_P^2}{3M^2} \approx 2\times 10^{9}.
\end{align}

\subsection{Preheating}
After the inflationary stage, the inflaton rolls down to the minimum at $(s, h)\simeq (0, 0)$. Depending on the parameters, the inflaton oscillates in both the $s$ and the $h$ directions, making the effective single-field analysis insufficient. In this period, the quantum creation of the NG boson mode from $\mathcal{H}(x) = h(t) e^{i \theta (x)} /\sqrt{2}$ is important. This $\theta$ direction arises from the following terms in the Lagrangian~\cite{He:2018mgb}:
\begin{align}
\sqrt{-g} \mathcal{L} \supset \frac{1}{2} \dot{\theta}_c^2 - \frac{1}{2a^2} \left(\nabla \theta_c\right)^2 + \frac{1}{2} \frac{\ddot{F}}{F} \theta_c^2,
\end{align}
where $\theta_c(x) \equiv a^{3/2}(t) e^{-\omega(s)/2} h(t) \theta(x) \equiv F(t) \theta(x) $ and $a$ is the scale factor in the FRW metric. By computing the effective NG mode mass, the peak value takes the form 
\begin{align}
\left(m_{\theta_c}^\text{sp}\right)^2 \approx C_m \sqrt{3\lambda\left(M^2  - M_c^2 \right)} M_P,
\end{align}
with $C_m \approx 0.25$ and $M_c \approx 1.3\times 10^{-5} M_P$. This quantity is noticeably lower than the cutoff scale of the theory. Therefore, the violent preheating behavior is present and physical, albeit it is not as violent as it would be in the single field Higgs inflation case and does not violate the unitarity problem~\cite{Ema:2016dny}. 

For specific parameters, the inflaton may climb up the hill of the potential at $h =0$, giving a negative $m_h^2 = - 3 \omega (s) \xi M^2$ and inducing a tachyonic preheating phase~\cite{Bezrukov:2019ylq, He:2020ivk, Bezrukov:2020txg}. This in turn gives an exponential enhancement to the particle production, rapidly completing the preheating process. The criticality of the Standard Model Higgs quartic coupling may also lead to interesting and unique phenomena in the preheating procedure. We will do a detailed analysis in future works.

\section{PBH production}
As no strong signals of standard particle types of dark matter including WIMPs or axions have been found, PBHs are obtaining more attention again as a candidate for dark matter. Different from the astrophysical BH, PBHs originate from the large quantum fluctuations during the inflation.

Large efforts have been made on searching/constraning mass windows for MACHO types of dark matter and now a narrow mass range is left for PBHs to explain the whole of dark matter: $M_{\rm PBH} \in (10^{-16}, 10^{-12}) M_\odot$, as depicted in Fig.~\ref{Fig:pbhabd}. $ M_{\odot} $ denotes the solar mass. Indeed, numerous models and scenarios of inflation have been suggested to produce enough PBHs with appropriate mass ranges to explain dark matter. Future experiments, including femtolensing and gravitational waves experiments, are planned or suggested to cover these mass ranges \cite{Katz:2018zrn, Jung:2019fcs, Dasgupta:2019cae}.
Those experiments are expected to give hints to the validity of the scenarios to explain the origin of dark matter with PBHs.

One of the most realistic and minimal possibility is to consider the critical Higgs inflation model, which was motivated by the fact that the power spectrum is enhanced in ultra-slow-roll inflation and can cause large fluctuations at small scales. Unfortunately, single-field CHI turns out to generate PBHs away from the desired mass range, and its predictions itself are questioned in many studies \cite{Ezquiaga:2017fvi, Kannike:2017bxn, Germani:2017bcs, Bezrukov:2017dyv, Motohashi:2017kbs, Masina:2018ejw, Drees:2019xpp}. In this review, we summarize the recent progress on Higgs inflation providing a new possibility of PBH generation as dark matter from the Higgs-$R^2$ model \cite{Cheong:2019vzl}.

\subsection{Primordial Black Hole}
In this subsection, we briefly summarize the formulas to obtain the PBH mass spectrum given a curvature power spectrum from inflation.

When the energy density perturbation $ \delta \sim \delta\rho / \rho $~\footnote{At linear order, there is a simple relation between the energy density fluctuation and the curvature perturbation: 
\begin{align}
\delta = \frac{4}{9} \left(	\frac{k}{aH}	\right)^{2} \mathcal{R}.
\end{align}} exceeds a critical values $ \delta_{c} \simeq 0.3 - 0.5 $ \cite{Green:2004wb, Harada:2013epa, Musco:2018rwt}, the matter in the Hubble sphere with radius $ 1 / H  $ starts to collapes to a black hole when the corresponding mode re-enters to the horizon during the radiation dominated (RD) era~\cite{Carr:1975qj,  Carr:2009jm}. The energy density of the background is also determined by the Hubble parameter. Therefore, the mass of the primordial black hole is determined solely by the Hubble scale at the time of its formation \cite{Green:2004wb},
\begin{align}
	M_{\rm PBH} &= \gamma \frac{4 \pi}{3} \rho_{\rm form} H_{\rm form}^{-3} \nonumber\\
	&= 3.2\times 10^{13} \left({k}/{\text{Mpc}^{-1} }\right)^{-2}M_\odot,
\label{eqn:pbhmass}
\end{align}
where $\gamma \simeq 0.2$ represents the efficiency of the collapsing processes and $ k = a H $ is the comoving momentum scale on which the primordial black hole is generated. The variance of the density contrast $ \sigma \equiv \sqrt{\langle \delta^{2} \rangle} $ is calculated from the curvature power spectrum $ \mathcal{P}_{\mathcal{R}}(k) $ and the window function $ W(R,k) = \exp(-k^{2}R^{2}/2) $ smoothing over the comoving scale $ R\simeq 1/aH\vert_{\rm form} $:
\begin{align}
	\sigma^{2} = \int_{0}^{\infty} d \ln k~ W(R,k)^{2} \frac{16}{81}(kR)^{4} \mathcal{P}_{\mathcal{R}}(k).
\end{align}

We follow the `peaks theory' method \cite{Green:2004wb, Young:2014ana} to compute  the PBH abundance and the mass spectrum. We use the variable $ \nu_{c} \equiv \delta_{c} / \sigma $. The energy density fraction of PBHs at formation, denoted by $ 	\beta_{M_{\rm PBH}} $ can be calculated using
\begin{align}
	&\beta_{M_{\rm PBH}} \equiv \left.\frac{\rho_{\rm PBH}}{\rho_{\rm tot}}\right\vert_{\rm form} \nonumber \\
	&=\frac{R^3}{\left(2\pi\right)^{1/2}} \left(\frac{\langle k^2\rangle \left(R\right)}{3}\right)
	\left(\nu_c^2 -1\right) \exp{\left(-\frac{\nu_c^2}{2}\right)},
\end{align}
with
\begin{align}
\langle k^{2} \rangle = \frac{1}{\sigma^{2}} \int_{0}^{\infty} d \ln k ~ k^{2} W(R,k)^{2} \mathcal{P}_{\delta}(k).
\end{align}
Finally, the fraction of PBHs
against the total dark matter energy density is
\begin{align}
&f_\text{PBH}\left(M_\text{PBH}\right) 
\equiv \frac{\Omega_{\rm PBH}}{\Omega_{\rm CDM}} 
= \left(	\frac{H_{\rm form}}{H_{0}}	\right)^{2} \left( \frac{a_{\rm form}}{a_{0}}  \right)^{3} \frac{\beta_{M_{\rm PBH}}}{\Omega_{\rm CDM}} \nonumber\\
&= 2.7 \times 10^8 \left(\frac{\gamma}{0.2}\right)^{\frac{1}{2}} \left(\frac{10.75}{g_*}\right)^{\frac{1}{4}}\left(\frac{M_\odot}{M_\text{PBH}}\right)^{\frac{1}{2}} \beta_{M_\text{PBH}}
\label{eqn:pbhabundance}
\end{align}
where $g_* =106.75$ is the effective relativistic degree of freedom at the time of formation.

\subsection{PBH Production from Higgs-$ R^{2} $ Inflation}

As a minimal extension of Higgs inflation, it is highly motivated to consider the role of the $ R^{2} $ term with criticality of the self quartic coupling in generating PBHs as dark matter.

In the Higgs-$ R^{2} $ model, three relevant running parameters $(M,\xi,\lambda)$ exist. The 1-loop beta functions are \cite{Codello:2015mba, Markkanen:2018bfx, Gorbunov:2018llf, Ema:2019fdd}
\begin{align}
\beta_\alpha &= -\frac{1}{16\pi^2}\frac{\left(1+6\xi\right)^2}{18} ,\\
\beta_\xi &= - \frac{1}{16\pi^2 }\left(\xi + \frac{1}{6}\right)\left(12 \lambda + 6 y_t^2 - \frac{3}{2}g'^2 - \frac{9}{2}g^2\right) ,\\
\beta_\lambda &=
\beta_\text{SM} + \frac{1}{16\pi^2} \frac{2\xi^2 \left(1+6\xi\right)^2 M^4}{M_P^4},
\end{align}
where $\alpha = {M_P^2}/{12 M^2}$. $\beta_{\rm SM}$ stands for the other terms from the SM~\cite{DeSimone:2008ei}. Among those, as in the original critical Higgs inflation case, the running of $ \lambda $ is most important in our analysis. 

During inflation, including the inflection point, the Higgs field value is comparable to the Planck scale $h\gtrsim \mathcal{O}\left(0.1 M_P \right) $. On the other hand, the effects of the Ricci scalar $ R = 12 H^{2} $ in the de Sitter background on determining the renormalization scale is negligible due to the small Hubble parameter $H \sim 10^{-5} M_P$. Therefore, we choose our prescription $\mu = h$ and express $\lambda\left(\mu\right)$ as Eq.~(\ref{Eq:lambdarunning}). Note that the Higgs field values are independent of the frame used for Higgs-$ R^{2} $ inflation.

The field values $(s,h)=(s^*,h^*)$ and the corresponding value $\lambda_\text{min}^\text{inf}$ for the potential to have an inflection point can be determined by using the conditions
\begin{eqnarray}
&\left.{\partial U}/{\partial s} \right|_{s = s^*}= \left. {\partial U}/{\partial h} \right|_{h = h^*}=0, \label{eqn:inflection1} \nonumber\\
&\begin{vmatrix}
D_s \left(\partial_s U\right) &D_s \left(\partial_h U\right)\\
D_h \left(\partial_s U\right)& D_h \left(\partial_h U\right)\\
\end{vmatrix}
_{s = s^*, \,\,h = h^*} = 0.
\label{eqn:inflection2}
\end{eqnarray}
We then compute the $\lambda_\text{min}$ value\footnote{In fact, to generate a large enough power spectrum, $\lambda_\text{min} = \lambda_\text{min}^\text{inf} - \delta c$, with $\delta c\sim \mathcal{O}\left(10^{-7}\right)$ at the corresponding scale, the $ \lambda_{\rm min} $ must be smaller than the pure inflection value $ \lambda_{\rm min} $ by as much as $ \mathcal{O}(10^{-7}) $ so that the potential should deviate from a true inflection point.}, which we assumed to include all the information from the SM parameters.

\begin{figure}[t!]
	\includegraphics[width=8cm]{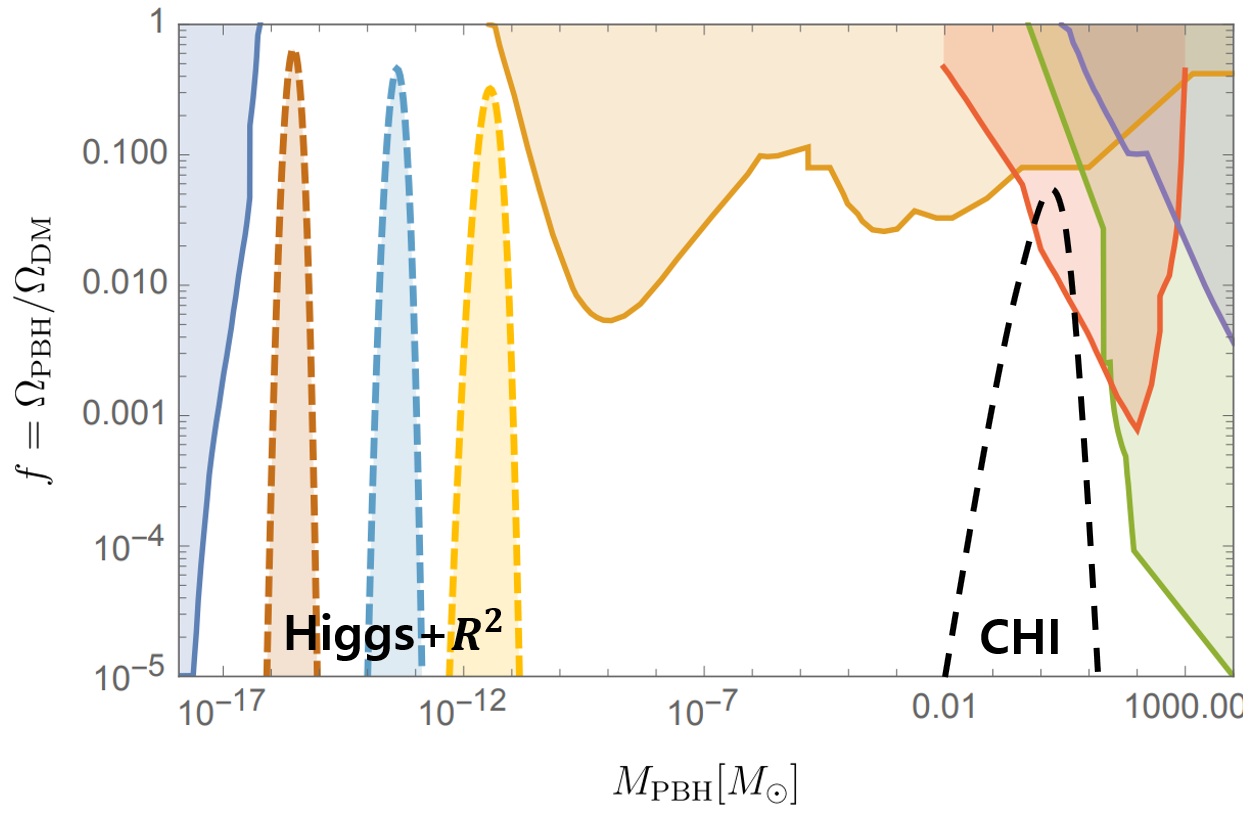}
	\caption{Current constraints on $ f_{\rm PBH} $ for various mass ranges. $M_{\rm PBH} \in (10^{-16}, 10^{-12}) M_\odot$ is still allowed for PBHs to be a total dark matter candidate. Dashed colored regions are possible mass spectra of PBHs from the Higgs - $ R^{2} $ inflation \cite{Cheong:2019vzl}. The dashed black line is from the critical Higgs inflation \cite{Ezquiaga:2017fvi}, which cannot explain total dark matter.} \label{Fig:pbhabd}
\end{figure}

Fig.~\ref{Fig:pbhabd} shows the new possibility of the Higgs-$ R^{2} $ inflation model generating a sufficient number of PBHs with $ f_{\rm PBH} \sim \mathcal{O}(1) $, in appropriate mass ranges without any strong astronomical bound. From the multi-field nature of the model with an additional scalaron direction, the inflection point is located on a relatively low scale along the inflationary trajectory.

A correlation also exists between the spectral index $ n_{s} $ and the PBH mass $ M_{\rm PBH} $. Without additional higher order corrections such as $ R^{3} $, the constraint on $ n_{s} $ narrows the possible mass ranges of PBHs from Higgs-$ R^{2} $ inflation as $ \mathcal{O}\left(10^{-16} - 10^{-15}\right)M_{\odot} $ \cite{Cheong:2019vzl}, with a $2 \sigma$ compatibility with Planck and LHC data. 

\section{$R^3$ term} 

In this section, we discuss the recent progress with additional higher order gravity terms in Higgs inflation focusing on the $R^3$ term, and analyzing the inflationary predictions and their implications on PBH production. These additional Ricci scalar terms $R^n$ are characterized in the $f(R)$ gravity class, which in turn is formulated in a scalar tensor theory as~\cite{Huang:2013hsb, Sebastiani:2013eqa, Kamada:2014gma, Artymowski:2014nva, Cheong:2020rao}
\begin{align}
S&=\frac{1}{2}\int d^4 x \sqrt{-g}\, f(R) \\
\to S&=\frac{1}{2}\int d^4 x \sqrt{-g}\, \left[ f(\phi) + f'(\phi) (R-\phi) \right]\\
&\equiv \int d^4 x \sqrt{-g} \left[ \frac{1}{2}\Omega^2 R -V(\phi)\right],
\end{align} 
with $M_P=1$ being taken for simplicity. The factors $\Omega$ and $V(\phi)$ can be expressed as  
\begin{align}
\Omega^2(\phi) &= f'(\phi),  \label{frame} \\
V(\phi) &= \frac{1}{2}\left[ \phi f'(\phi) -f(\phi) \right].
\end{align}
When transformed to the Einstein frame, the $D=k+\ell >4$ dimension cutoff scale for $f(R)=a R+ b R^{n+1}$ becomes 
\begin{align}
\Lambda_{D} = \left[\frac{k!\ell!}{\alpha_n (-2)^\ell (2/3)^{k+\ell}}\right]^{\tfrac{1}{k+\ell-4}},
\end{align}
where $\alpha_n=\frac{n \beta^{-1/n}}{2(n+1)}$. This value is generically larger than $M_P$, which guarantees that the perturbative analysis holds for generic polynomial $f(R)$ gravity theories. 

The $R^3$ extension $f(R)=a R+b R^2+c R^3$ with parameters $a=1, b=\beta/2$, and $c=\gamma/3$  yields a dual scalar theory with 
\begin{equation}
\sigma(s) \equiv e^{\sqrt{\frac{2}{3}}s}=1+\beta \phi + \gamma \phi^2,
\end{equation}
for which the solution takes the form 
\begin{eqnarray}
\phi(s) &=&\frac{\beta}{2\gamma}\left(\sqrt{1+4\frac{\gamma}{\beta^2}\left(\sigma(s)-1\right)}-1 \right) \\
&=& \frac{\sigma(s)-1}{\beta}\left[1-\frac{\gamma}{\beta}\left(\frac{\sigma(s)-1}{\beta}\right) +{\cal O}\left(\frac{\gamma}{\beta}\right)^2 \right]
\end{eqnarray}
where on the last line the conditions $\gamma \ll \beta$ and $\phi\sim 1$ are implied. This perturbative expansion gives additional terms in the Einstein potential, when expanded in powers of $(\gamma/\beta)^n$:
\begin{align}
V_E(s) \approx V_0(s) \left[1 -\frac{2}{3}\frac{\gamma}{\beta}\left(\frac{\sigma(s)-1}{\beta}\right) + \cdots \right]
\end{align}
with $V_0(s) = \frac{1}{4\beta}(1-\frac{1}{\sigma})^2 = \frac{1}{4\beta}(1-e^{-\sqrt{\frac{2}{3}s}})^2$ being the potential for the pure Starobinsky inflation scenario $(\gamma = 0)$. These additional terms alter the predictions of the slow-roll parameters and CMB observables in powers of $(\gamma/\beta^2)$. In particular, the spectral index $n_s$ and the tensor-to-scalar ratio $r$ can be expressed as 
\begin{align}
n_s
&= 1-6\epsilon(s_*) + 2\eta(s_*)\\
%&\approx 1-\frac{8(\sigma(s_*)+1)}{3(\sigma(s_*)-1)^2}-\frac{8}{9}\delta \frac{\sigma(s_*)(\sigma(s_*)-3)}{\sigma(s_*)-1} \\
&\approx 1-\frac{2}{N_e}-\frac{9}{2N_e^2} -\delta \frac{128}{81}N_e, \label{eq:nsapprox}
 %{\color{blue} \left(1+\frac{81}{64 N_e^2}\right)},
\end{align} 
and 
\begin{align}
r= 16\epsilon(s_*)
%&\approx \frac{64}{3 \sigma^2(s_*)} -\frac{128\delta}{9\sigma(s_*)}\\
\approx \frac{12}{N_e^2} - \delta \frac{256}{27}.
\label{eqn:rR3}
 %{\color{blue} \left(1-\frac{27}{32}\frac{1}{N_e^2}\right)}.
\end{align}
with $\delta \equiv \gamma/\beta^2$ and $N_e$ being the inflation duration's e-fold number.  Current Planck CMB data~\cite{Akrami:2018odb} give a rough constraint of $|\delta| \sim \mathcal{O} (10^{-4})$, which indicates that the large-scale predictions of this model are highly sensitive to the $R^3$ term. 

%\begin{figure}[t]
%\includegraphics[width=.489\textwidth]{fig_nsr}
%\caption{$(n_s,r)$ for $N_e=60$(blue), $N_e=56.9$(red) and $N_e=55$(purple) efoldings with $\delta= [-2.0,2.0]\times 10^{-4}$ vs Planck2018 $1\sigma$ (Yellow) and $2\sigma$ (Green) constraints \cite{Akrami:2018odb}. 
%\label{fig:ns-r}}
%\end{figure}

\begin{figure}[t]
\includegraphics[width=.4\textwidth]{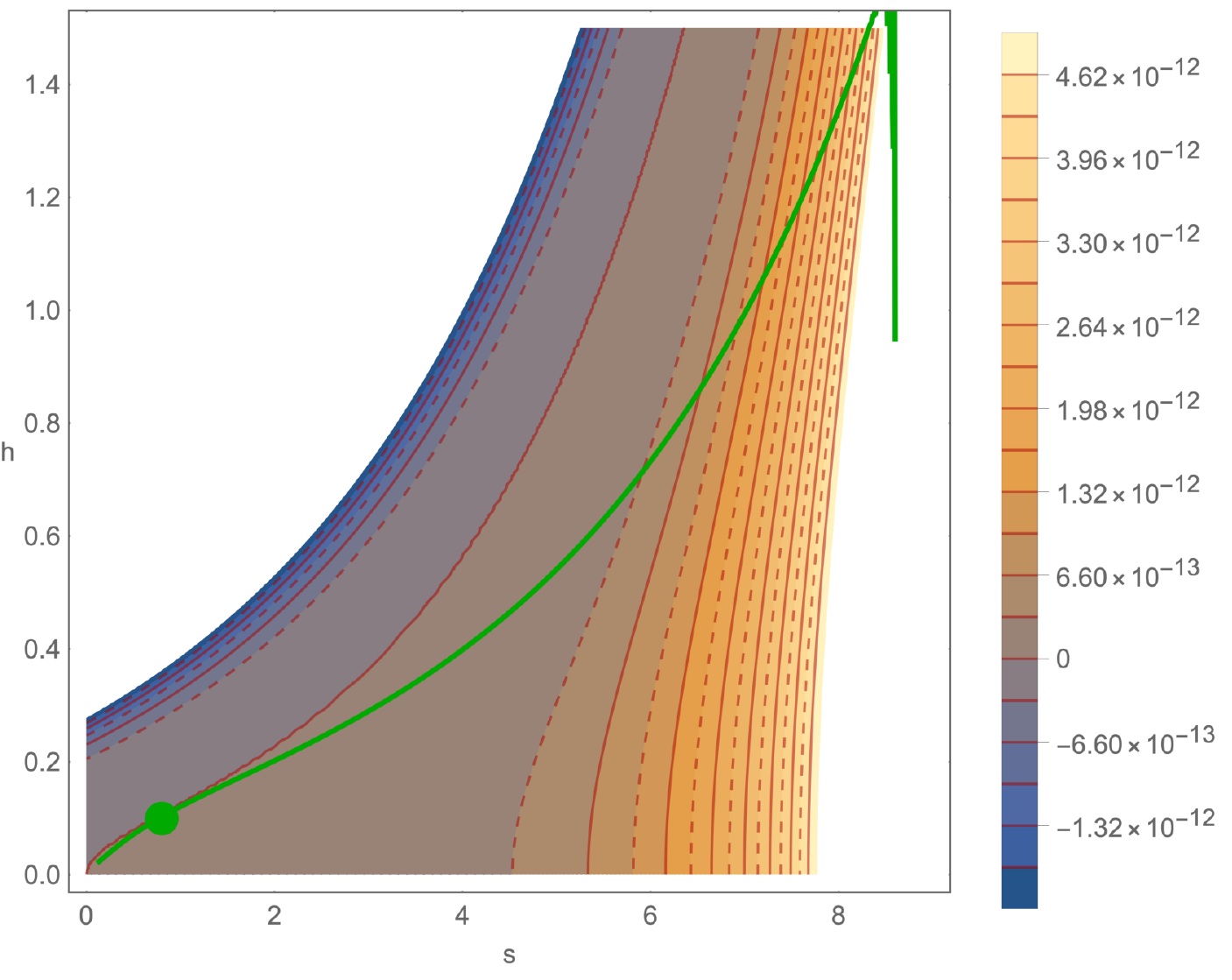}
\caption{\label{potential_variation} The potential variation in Eq.~(\ref{eqn:variation}). The marker indicates the local minimum in the inflaton's trajectory. The local mimimum of the trajectory lies on the zero variation contour while the CMB large scale plateau experiences relatively large deviations.}
\end{figure}

As this additional $R^3$ term modifies the large-scale predictions, the presence of this operator can also effect CMB predictions for PBH-compatible critical Higgs-$R^2$ inflationary scenarios. The gravitational part of the action contains the following $f(R)$ expression:
\begin{align}
f(R) = R + \xi h^2 R + \frac{1}{6M^2} R^2 + \frac{\gamma}{3} R^3,
\end{align}
for which the Einstein frame potential is 
\begin{widetext}
\begin{align}
U = e^{-2\sqrt{\frac{2}{3}s}} \left[\frac{3M^2}{4} \left(e^{\sqrt{\frac{2}{3}}s} - 1 - \xi h^2 \right)^2- \frac{9M^6}{2}\gamma\left(e^{\sqrt{\frac{2}{3}}s} - 1 - \xi h^2 \right)^3  + \frac{\lambda_\text{eff}}{4}h^4 \right].
\end{align}
\end{widetext}
By taking the running of $\lambda$ and inducing an inflection point, one can plot the contribution to the potential 
\begin{align}
\delta U = - \frac{M^2}{2}\delta e^{-2\sqrt{\frac{2}{3}s}} \left(e^{\sqrt{\frac{2}{3}}s} - 1 - \xi h^2 \right)^3
\label{eqn:variation}
\end{align}
along with the trajectory, as shown in Fig. \ref{potential_variation}. Notice that the inflection point lies on the zero contour of the potential variance, indicating that the curvature power spectrum $\mathcal{P}_\mathcal{R}$ is effectively identical to the Higgs-$R^2$ scenario while the CMB large scale predictions shift accordingly.

\begin{figure}[t]%
    {\includegraphics[width=.4\textwidth]{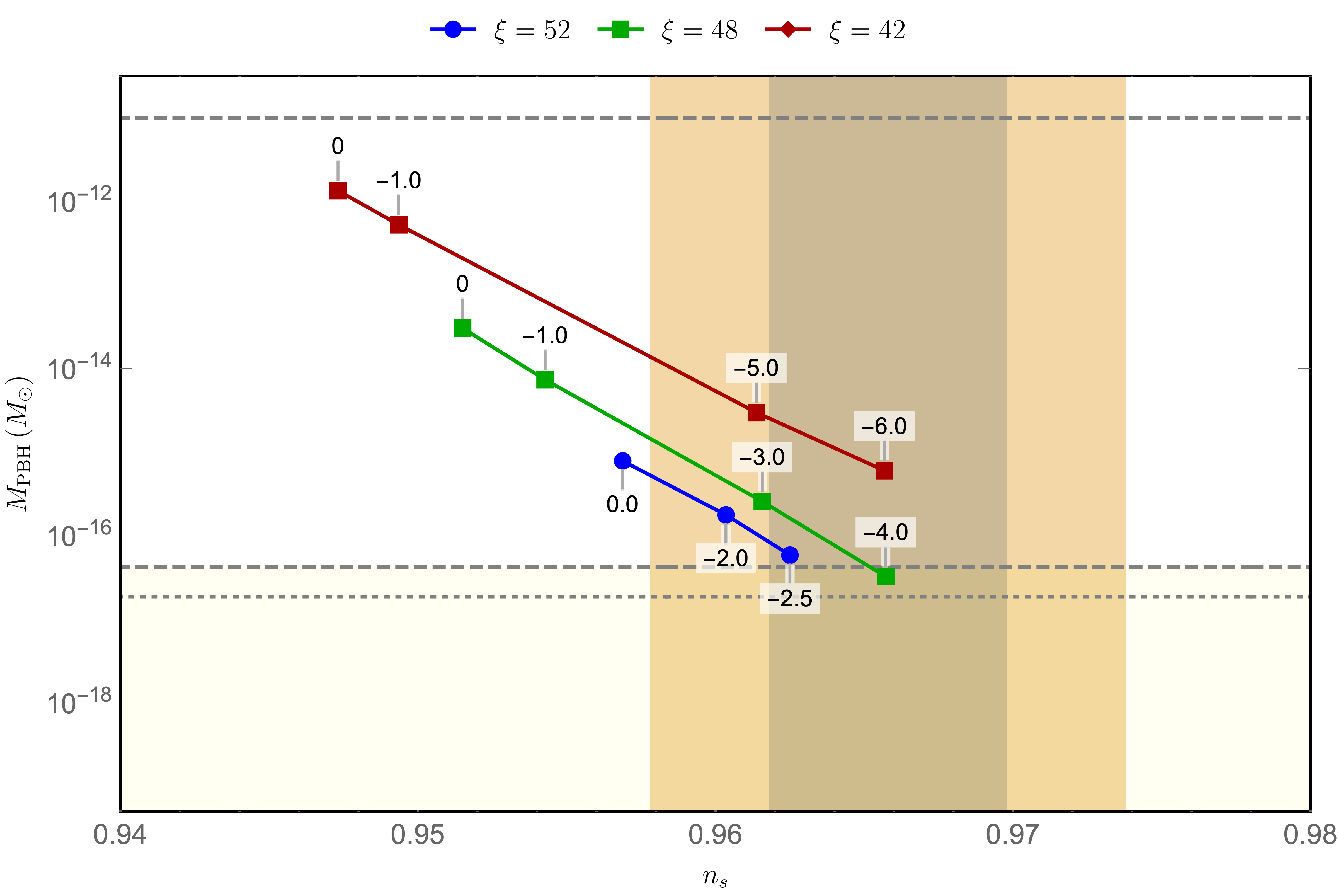} }%
    \caption{The corresponding $n_s$ - $M_\text{PBH}$ plot with an additional $\gamma R^3/ 6$ term in the action. The shifts are expressed through  $\delta = 9M^4 \gamma = [-6.0 (-4.0, -2.5), 0]\times 10^{-4}$ for $\xi=42, 48, 52 $}%
    \label{fig:nsMPBH_2}%
\end{figure}

Fig.~\ref{fig:nsMPBH_2} presents this shift of CMB/PBH observable values. As mentioned in the previous sections, the pure critical Higgs-$R^2$ scenario can give both CMB- and PBH-compatible scenarios; however its compatibility is within the $2\sigma$ range, slightly shifting its predictions from the Planck central values. The addition of higher order terms, i.e., $R^3$, shifts the CMB large-scale observables in the form of Eq.~(\ref{eq:nsapprox}) and Eq.~(\ref{eqn:rR3}), whereas the peak profile remains effectively identical to that for the corresponding Higgs-$R^2$ case. The tune of parameters therefore widens the allowed parameter region, which gives a better CMB compatibility for the Higgs-$R^2$ PBH scenario.

\section{Conclusion} 

 The only scalar field in the standard model of particle physics, the Higgs field, has been known to be responsible for electroweak symmetry breaking and the masses of elementary particles for several decades. However,  its irreplaceable roles in the early Universe have relatively recently been realized and appreciated.  In this selective review, we focus on the inflationary era starting from how the Higgs field provides the exponential expansion of the universe and produces the primordial density fluctuations. We also review the possible production of primordial black holes, which may be responsible for all dark matter  in the universe. Probably this is not the end of the story.  We strongly believe that the full power of the Higgs field in particle physics and cosmology still needs further understanding.

\begin{acknowledgments}
We thank Kazunori Kohri, Misao Sasaki, Hyun Min Lee, Shi Pi, and Fedor Bezrukov for helpful discussions and Alexei Starobinski, Minxi He, Jun'ichi Yokoyama, Ryusuke Jinno, Kohei Kamada, Kin-ya Oda,  and Mio Kubota for valuable collaborations. 
This work was supported by National Research Foundation grants funded by the Korean government (MSIT) (NRF-2018R1A4A1025334),(NRF-2019R1A2C1089334) (SCP)  and (MOE) (NRF-2020R1A6A3A13076216) (SML). The work of SML is supported by the Hyundai Motor Chung Mong-Koo Foundation.
\end{acknowledgments}

\bibliography{PBH_HIGGS}% Produces the bibliography via BibTeX.

%merlin.mbs apsrev4-1.bst 2010-07-25 4.21a (PWD, AO, DPC) hacked
%Control: key (0)
%Control: author (8) initials jnrlst
%Control: editor formatted (1) identically to author
%Control: production of article title (-1) disabled
%Control: page (0) single
%Control: year (1) truncated
%Control: production of eprint (0) enabled
\providecommand{\noopsort}[1]{}\providecommand{\singleletter}[1]{#1}%
\begin{thebibliography}{103}%
\makeatletter
\providecommand \@ifxundefined [1]{%
 \@ifx{#1\undefined}
}%
\providecommand \@ifnum [1]{%
 \ifnum #1\expandafter \@firstoftwo
 \else \expandafter \@secondoftwo
 \fi
}%
\providecommand \@ifx [1]{%
 \ifx #1\expandafter \@firstoftwo
 \else \expandafter \@secondoftwo
 \fi
}%
\providecommand \natexlab [1]{#1}%
\providecommand \enquote  [1]{``#1''}%
\providecommand \bibnamefont  [1]{#1}%
\providecommand \bibfnamefont [1]{#1}%
\providecommand \citenamefont [1]{#1}%
\providecommand \href@noop [0]{\@secondoftwo}%
\providecommand \href [0]{\begingroup \@sanitize@url \@href}%
\providecommand \@href[1]{\@@startlink{#1}\@@href}%
\providecommand \@@href[1]{\endgroup#1\@@endlink}%
\providecommand \@sanitize@url [0]{\catcode `\\12\catcode `\$12\catcode
  `\&12\catcode `\#12\catcode `\^12\catcode `\_12\catcode `\%12\relax}%
\providecommand \@@startlink[1]{}%
\providecommand \@@endlink[0]{}%
\providecommand \url  [0]{\begingroup\@sanitize@url \@url }%
\providecommand \@url [1]{\endgroup\@href {#1}{\urlprefix }}%
\providecommand \urlprefix  [0]{URL }%
\providecommand \Eprint [0]{\href }%
\providecommand \doibase [0]{http://dx.doi.org/}%
\providecommand \selectlanguage [0]{\@gobble}%
\providecommand \bibinfo  [0]{\@secondoftwo}%
\providecommand \bibfield  [0]{\@secondoftwo}%
\providecommand \translation [1]{[#1]}%
\providecommand \BibitemOpen [0]{}%
\providecommand \bibitemStop [0]{}%
\providecommand \bibitemNoStop [0]{.\EOS\space}%
\providecommand \EOS [0]{\spacefactor3000\relax}%
\providecommand \BibitemShut  [1]{\csname bibitem#1\endcsname}%
\let\auto@bib@innerbib\@empty
%</preamble>
\bibitem [{\citenamefont {Zel'dovich}\ and\ \citenamefont
  {Novikov}(1966)}]{Zel1966}%
  \BibitemOpen
  \bibfield  {author} {\bibinfo {author} {\bibfnamefont {Y.~B.}\ \bibnamefont
  {Zel'dovich}}\ and\ \bibinfo {author} {\bibfnamefont {I.~D.}\ \bibnamefont
  {Novikov}},\ }\href@noop {} {\bibfield  {journal} {\bibinfo  {journal}
  {Soviet Astronomy}\ }\textbf {\bibinfo {volume} {10}},\ \bibinfo {pages}
  {602} (\bibinfo {year} {1966})}\BibitemShut {NoStop}%
\bibitem [{\citenamefont {Hawking}(1971)}]{Hawking:1971ei}%
  \BibitemOpen
  \bibfield  {author} {\bibinfo {author} {\bibfnamefont {S.}~\bibnamefont
  {Hawking}},\ }\href@noop {} {\bibfield  {journal} {\bibinfo  {journal} {Mon.
  Not. Roy. Astron. Soc.}\ }\textbf {\bibinfo {volume} {152}},\ \bibinfo
  {pages} {75} (\bibinfo {year} {1971})}\BibitemShut {NoStop}%
%%CITATION = MNRAA,152,75;%%
\bibitem [{\citenamefont {Ivanov}\ \emph {et~al.}(1994)\citenamefont {Ivanov},
  \citenamefont {Naselsky},\ and\ \citenamefont {Novikov}}]{Ivanov:1994pa}%
  \BibitemOpen
  \bibfield  {author} {\bibinfo {author} {\bibfnamefont {P.}~\bibnamefont
  {Ivanov}}, \bibinfo {author} {\bibfnamefont {P.}~\bibnamefont {Naselsky}}, \
  and\ \bibinfo {author} {\bibfnamefont {I.}~\bibnamefont {Novikov}},\ }\href
  {\doibase 10.1103/PhysRevD.50.7173} {\bibfield  {journal} {\bibinfo
  {journal} {Phys. Rev.}\ }\textbf {\bibinfo {volume} {D50}},\ \bibinfo {pages}
  {7173} (\bibinfo {year} {1994})}\BibitemShut {NoStop}%
%%CITATION = PHRVA,D50,7173;%%
\bibitem [{\citenamefont {Blais}\ \emph {et~al.}(2002)\citenamefont {Blais},
  \citenamefont {Kiefer},\ and\ \citenamefont {Polarski}}]{Blais:2002nd}%
  \BibitemOpen
  \bibfield  {author} {\bibinfo {author} {\bibfnamefont {D.}~\bibnamefont
  {Blais}}, \bibinfo {author} {\bibfnamefont {C.}~\bibnamefont {Kiefer}}, \
  and\ \bibinfo {author} {\bibfnamefont {D.}~\bibnamefont {Polarski}},\ }\href
  {\doibase 10.1016/S0370-2693(02)01803-8} {\bibfield  {journal} {\bibinfo
  {journal} {Phys. Lett.}\ }\textbf {\bibinfo {volume} {B535}},\ \bibinfo
  {pages} {11} (\bibinfo {year} {2002})},\ \Eprint
  {http://arxiv.org/abs/astro-ph/0203520} {arXiv:astro-ph/0203520 [astro-ph]}
  \BibitemShut {NoStop}%
%%CITATION = ASTRO-PH/0203520;%%
\bibitem [{\citenamefont {Afshordi}\ \emph {et~al.}(2003)\citenamefont
  {Afshordi}, \citenamefont {McDonald},\ and\ \citenamefont
  {Spergel}}]{Afshordi:2003zb}%
  \BibitemOpen
  \bibfield  {author} {\bibinfo {author} {\bibfnamefont {N.}~\bibnamefont
  {Afshordi}}, \bibinfo {author} {\bibfnamefont {P.}~\bibnamefont {McDonald}},
  \ and\ \bibinfo {author} {\bibfnamefont {D.~N.}\ \bibnamefont {Spergel}},\
  }\href {\doibase 10.1086/378763} {\bibfield  {journal} {\bibinfo  {journal}
  {Astrophys. J.}\ }\textbf {\bibinfo {volume} {594}},\ \bibinfo {pages} {L71}
  (\bibinfo {year} {2003})},\ \Eprint {http://arxiv.org/abs/astro-ph/0302035}
  {arXiv:astro-ph/0302035 [astro-ph]} \BibitemShut {NoStop}%
%%CITATION = ASTRO-PH/0302035;%%
\bibitem [{\citenamefont {Khlopov}(2010)}]{Khlopov:2008qy}%
  \BibitemOpen
  \bibfield  {author} {\bibinfo {author} {\bibfnamefont {M.~{\relax Yu}.}\
  \bibnamefont {Khlopov}},\ }\href {\doibase 10.1088/1674-4527/10/6/001}
  {\bibfield  {journal} {\bibinfo  {journal} {Res. Astron. Astrophys.}\
  }\textbf {\bibinfo {volume} {10}},\ \bibinfo {pages} {495} (\bibinfo {year}
  {2010})},\ \Eprint {http://arxiv.org/abs/0801.0116} {arXiv:0801.0116
  [astro-ph]} \BibitemShut {NoStop}%
%%CITATION = ARXIV:0801.0116;%%
\bibitem [{\citenamefont {Frampton}\ \emph {et~al.}(2010)\citenamefont
  {Frampton}, \citenamefont {Kawasaki}, \citenamefont {Takahashi},\ and\
  \citenamefont {Yanagida}}]{Frampton:2010sw}%
  \BibitemOpen
  \bibfield  {author} {\bibinfo {author} {\bibfnamefont {P.~H.}\ \bibnamefont
  {Frampton}}, \bibinfo {author} {\bibfnamefont {M.}~\bibnamefont {Kawasaki}},
  \bibinfo {author} {\bibfnamefont {F.}~\bibnamefont {Takahashi}}, \ and\
  \bibinfo {author} {\bibfnamefont {T.~T.}\ \bibnamefont {Yanagida}},\ }\href
  {\doibase 10.1088/1475-7516/2010/04/023} {\bibfield  {journal} {\bibinfo
  {journal} {JCAP}\ }\textbf {\bibinfo {volume} {1004}},\ \bibinfo {pages}
  {023} (\bibinfo {year} {2010})},\ \Eprint {http://arxiv.org/abs/1001.2308}
  {arXiv:1001.2308 [hep-ph]} \BibitemShut {NoStop}%
%%CITATION = ARXIV:1001.2308;%%
\bibitem [{\citenamefont {Sasaki}\ \emph {et~al.}(2016)\citenamefont {Sasaki},
  \citenamefont {Suyama}, \citenamefont {Tanaka},\ and\ \citenamefont
  {Yokoyama}}]{Sasaki:2016jop}%
  \BibitemOpen
  \bibfield  {author} {\bibinfo {author} {\bibfnamefont {M.}~\bibnamefont
  {Sasaki}}, \bibinfo {author} {\bibfnamefont {T.}~\bibnamefont {Suyama}},
  \bibinfo {author} {\bibfnamefont {T.}~\bibnamefont {Tanaka}}, \ and\ \bibinfo
  {author} {\bibfnamefont {S.}~\bibnamefont {Yokoyama}},\ }\href {\doibase
  10.1103/PhysRevLett.121.059901, 10.1103/PhysRevLett.117.061101} {\bibfield
  {journal} {\bibinfo  {journal} {Phys. Rev. Lett.}\ }\textbf {\bibinfo
  {volume} {117}},\ \bibinfo {pages} {061101} (\bibinfo {year} {2016})},\
  \bibinfo {note} {[erratum: Phys. Rev. Lett.121,no.5,059901(2018)]},\ \Eprint
  {http://arxiv.org/abs/1603.08338} {arXiv:1603.08338 [astro-ph.CO]}
  \BibitemShut {NoStop}%
%%CITATION = ARXIV:1603.08338;%%
\bibitem [{\citenamefont {Carr}\ \emph
  {et~al.}(2016{\natexlab{a}})\citenamefont {Carr}, \citenamefont {Kuhnel},\
  and\ \citenamefont {Sandstad}}]{Carr:2016drx}%
  \BibitemOpen
  \bibfield  {author} {\bibinfo {author} {\bibfnamefont {B.}~\bibnamefont
  {Carr}}, \bibinfo {author} {\bibfnamefont {F.}~\bibnamefont {Kuhnel}}, \ and\
  \bibinfo {author} {\bibfnamefont {M.}~\bibnamefont {Sandstad}},\ }\href
  {\doibase 10.1103/PhysRevD.94.083504} {\bibfield  {journal} {\bibinfo
  {journal} {Phys. Rev.}\ }\textbf {\bibinfo {volume} {D94}},\ \bibinfo {pages}
  {083504} (\bibinfo {year} {2016}{\natexlab{a}})},\ \Eprint
  {http://arxiv.org/abs/1607.06077} {arXiv:1607.06077 [astro-ph.CO]}
  \BibitemShut {NoStop}%
%%CITATION = ARXIV:1607.06077;%%
\bibitem [{\citenamefont {Inomata}\ \emph {et~al.}(2017)\citenamefont
  {Inomata}, \citenamefont {Kawasaki}, \citenamefont {Mukaida}, \citenamefont
  {Tada},\ and\ \citenamefont {Yanagida}}]{Inomata:2017okj}%
  \BibitemOpen
  \bibfield  {author} {\bibinfo {author} {\bibfnamefont {K.}~\bibnamefont
  {Inomata}}, \bibinfo {author} {\bibfnamefont {M.}~\bibnamefont {Kawasaki}},
  \bibinfo {author} {\bibfnamefont {K.}~\bibnamefont {Mukaida}}, \bibinfo
  {author} {\bibfnamefont {Y.}~\bibnamefont {Tada}}, \ and\ \bibinfo {author}
  {\bibfnamefont {T.~T.}\ \bibnamefont {Yanagida}},\ }\href {\doibase
  10.1103/PhysRevD.96.043504} {\bibfield  {journal} {\bibinfo  {journal} {Phys.
  Rev.}\ }\textbf {\bibinfo {volume} {D96}},\ \bibinfo {pages} {043504}
  (\bibinfo {year} {2017})},\ \Eprint {http://arxiv.org/abs/1701.02544}
  {arXiv:1701.02544 [astro-ph.CO]} \BibitemShut {NoStop}%
%%CITATION = ARXIV:1701.02544;%%
\bibitem [{\citenamefont {Carr}\ and\ \citenamefont
  {Silk}(2018)}]{Carr:2018rid}%
  \BibitemOpen
  \bibfield  {author} {\bibinfo {author} {\bibfnamefont {B.}~\bibnamefont
  {Carr}}\ and\ \bibinfo {author} {\bibfnamefont {J.}~\bibnamefont {Silk}},\
  }\href {\doibase 10.1093/mnras/sty1204} {\bibfield  {journal} {\bibinfo
  {journal} {Mon. Not. Roy. Astron. Soc.}\ }\textbf {\bibinfo {volume} {478}},\
  \bibinfo {pages} {3756} (\bibinfo {year} {2018})},\ \Eprint
  {http://arxiv.org/abs/1801.00672} {arXiv:1801.00672 [astro-ph.CO]}
  \BibitemShut {NoStop}%
%%CITATION = ARXIV:1801.00672;%%
\bibitem [{\citenamefont {Niikura}\ \emph {et~al.}(2019)\citenamefont {Niikura}
  \emph {et~al.}}]{Niikura:2017zjd}%
  \BibitemOpen
  \bibfield  {author} {\bibinfo {author} {\bibfnamefont {H.}~\bibnamefont
  {Niikura}} \emph {et~al.},\ }\href {\doibase 10.1038/s41550-019-0723-1}
  {\bibfield  {journal} {\bibinfo  {journal} {Nat. Astron.}\ }\textbf {\bibinfo
  {volume} {3}},\ \bibinfo {pages} {524} (\bibinfo {year} {2019})},\ \Eprint
  {http://arxiv.org/abs/1701.02151} {arXiv:1701.02151 [astro-ph.CO]}
  \BibitemShut {NoStop}%
%%CITATION = ARXIV:1701.02151;%%
\bibitem [{\citenamefont {Katz}\ \emph {et~al.}(2018)\citenamefont {Katz},
  \citenamefont {Kopp}, \citenamefont {Sibiryakov},\ and\ \citenamefont
  {Xue}}]{Katz:2018zrn}%
  \BibitemOpen
  \bibfield  {author} {\bibinfo {author} {\bibfnamefont {A.}~\bibnamefont
  {Katz}}, \bibinfo {author} {\bibfnamefont {J.}~\bibnamefont {Kopp}}, \bibinfo
  {author} {\bibfnamefont {S.}~\bibnamefont {Sibiryakov}}, \ and\ \bibinfo
  {author} {\bibfnamefont {W.}~\bibnamefont {Xue}},\ }\href {\doibase
  10.1088/1475-7516/2018/12/005} {\bibfield  {journal} {\bibinfo  {journal}
  {JCAP}\ }\textbf {\bibinfo {volume} {1812}},\ \bibinfo {pages} {005}
  (\bibinfo {year} {2018})},\ \Eprint {http://arxiv.org/abs/1807.11495}
  {arXiv:1807.11495 [astro-ph.CO]} \BibitemShut {NoStop}%
%%CITATION = ARXIV:1807.11495;%%
\bibitem [{\citenamefont {Carr}\ \emph {et~al.}(2010)\citenamefont {Carr},
  \citenamefont {Kohri}, \citenamefont {Sendouda},\ and\ \citenamefont
  {Yokoyama}}]{Carr:2009jm}%
  \BibitemOpen
  \bibfield  {author} {\bibinfo {author} {\bibfnamefont {B.~J.}\ \bibnamefont
  {Carr}}, \bibinfo {author} {\bibfnamefont {K.}~\bibnamefont {Kohri}},
  \bibinfo {author} {\bibfnamefont {Y.}~\bibnamefont {Sendouda}}, \ and\
  \bibinfo {author} {\bibfnamefont {J.}~\bibnamefont {Yokoyama}},\ }\href
  {\doibase 10.1103/PhysRevD.81.104019} {\bibfield  {journal} {\bibinfo
  {journal} {Phys. Rev.}\ }\textbf {\bibinfo {volume} {D81}},\ \bibinfo {pages}
  {104019} (\bibinfo {year} {2010})},\ \Eprint {http://arxiv.org/abs/0912.5297}
  {arXiv:0912.5297 [astro-ph.CO]} \BibitemShut {NoStop}%
%%CITATION = ARXIV:0912.5297;%%
\bibitem [{\citenamefont {Carr}\ \emph
  {et~al.}(2016{\natexlab{b}})\citenamefont {Carr}, \citenamefont {Kohri},
  \citenamefont {Sendouda},\ and\ \citenamefont {Yokoyama}}]{Carr:2016hva}%
  \BibitemOpen
  \bibfield  {author} {\bibinfo {author} {\bibfnamefont {B.~J.}\ \bibnamefont
  {Carr}}, \bibinfo {author} {\bibfnamefont {K.}~\bibnamefont {Kohri}},
  \bibinfo {author} {\bibfnamefont {Y.}~\bibnamefont {Sendouda}}, \ and\
  \bibinfo {author} {\bibfnamefont {J.}~\bibnamefont {Yokoyama}},\ }\href
  {\doibase 10.1103/PhysRevD.94.044029} {\bibfield  {journal} {\bibinfo
  {journal} {Phys. Rev.}\ }\textbf {\bibinfo {volume} {D94}},\ \bibinfo {pages}
  {044029} (\bibinfo {year} {2016}{\natexlab{b}})},\ \Eprint
  {http://arxiv.org/abs/1604.05349} {arXiv:1604.05349 [astro-ph.CO]}
  \BibitemShut {NoStop}%
%%CITATION = ARXIV:1604.05349;%%
\bibitem [{\citenamefont {Laha}(2019)}]{Laha:2019ssq}%
  \BibitemOpen
  \bibfield  {author} {\bibinfo {author} {\bibfnamefont {R.}~\bibnamefont
  {Laha}},\ }\href@noop {} {\  (\bibinfo {year} {2019})},\ \Eprint
  {http://arxiv.org/abs/1906.09994} {arXiv:1906.09994 [astro-ph.HE]}
  \BibitemShut {NoStop}%
%%CITATION = ARXIV:1906.09994;%%
\bibitem [{\citenamefont {Ricotti}\ \emph {et~al.}(2008)\citenamefont
  {Ricotti}, \citenamefont {Ostriker},\ and\ \citenamefont
  {Mack}}]{Ricotti:2007au}%
  \BibitemOpen
  \bibfield  {author} {\bibinfo {author} {\bibfnamefont {M.}~\bibnamefont
  {Ricotti}}, \bibinfo {author} {\bibfnamefont {J.~P.}\ \bibnamefont
  {Ostriker}}, \ and\ \bibinfo {author} {\bibfnamefont {K.~J.}\ \bibnamefont
  {Mack}},\ }\href {\doibase 10.1086/587831} {\bibfield  {journal} {\bibinfo
  {journal} {Astrophys. J.}\ }\textbf {\bibinfo {volume} {680}},\ \bibinfo
  {pages} {829} (\bibinfo {year} {2008})},\ \Eprint
  {http://arxiv.org/abs/0709.0524} {arXiv:0709.0524 [astro-ph]} \BibitemShut
  {NoStop}%
%%CITATION = ARXIV:0709.0524;%%
\bibitem [{\citenamefont {Aloni}\ \emph {et~al.}(2017)\citenamefont {Aloni},
  \citenamefont {Blum},\ and\ \citenamefont {Flauger}}]{Blum:2016cjs}%
  \BibitemOpen
  \bibfield  {author} {\bibinfo {author} {\bibfnamefont {D.}~\bibnamefont
  {Aloni}}, \bibinfo {author} {\bibfnamefont {K.}~\bibnamefont {Blum}}, \ and\
  \bibinfo {author} {\bibfnamefont {R.}~\bibnamefont {Flauger}},\ }\href
  {\doibase 10.1088/1475-7516/2017/05/017} {\bibfield  {journal} {\bibinfo
  {journal} {JCAP}\ }\textbf {\bibinfo {volume} {1705}},\ \bibinfo {pages}
  {017} (\bibinfo {year} {2017})},\ \Eprint {http://arxiv.org/abs/1612.06811}
  {arXiv:1612.06811 [astro-ph.CO]} \BibitemShut {NoStop}%
%%CITATION = ARXIV:1612.06811;%%
\bibitem [{\citenamefont {Poulin}\ \emph {et~al.}(2017)\citenamefont {Poulin},
  \citenamefont {Serpico}, \citenamefont {Calore}, \citenamefont {Clesse},\
  and\ \citenamefont {Kohri}}]{Poulin:2017bwe}%
  \BibitemOpen
  \bibfield  {author} {\bibinfo {author} {\bibfnamefont {V.}~\bibnamefont
  {Poulin}}, \bibinfo {author} {\bibfnamefont {P.~D.}\ \bibnamefont {Serpico}},
  \bibinfo {author} {\bibfnamefont {F.}~\bibnamefont {Calore}}, \bibinfo
  {author} {\bibfnamefont {S.}~\bibnamefont {Clesse}}, \ and\ \bibinfo {author}
  {\bibfnamefont {K.}~\bibnamefont {Kohri}},\ }\href {\doibase
  10.1103/PhysRevD.96.083524} {\bibfield  {journal} {\bibinfo  {journal} {Phys.
  Rev.}\ }\textbf {\bibinfo {volume} {D96}},\ \bibinfo {pages} {083524}
  (\bibinfo {year} {2017})},\ \Eprint {http://arxiv.org/abs/1707.04206}
  {arXiv:1707.04206 [astro-ph.CO]} \BibitemShut {NoStop}%
%%CITATION = ARXIV:1707.04206;%%
\bibitem [{\citenamefont {Kawasaki}\ \emph {et~al.}(1998)\citenamefont
  {Kawasaki}, \citenamefont {Sugiyama},\ and\ \citenamefont
  {Yanagida}}]{Kawasaki:1997ju}%
  \BibitemOpen
  \bibfield  {author} {\bibinfo {author} {\bibfnamefont {M.}~\bibnamefont
  {Kawasaki}}, \bibinfo {author} {\bibfnamefont {N.}~\bibnamefont {Sugiyama}},
  \ and\ \bibinfo {author} {\bibfnamefont {T.}~\bibnamefont {Yanagida}},\
  }\href {\doibase 10.1103/PhysRevD.57.6050} {\bibfield  {journal} {\bibinfo
  {journal} {Phys. Rev.}\ }\textbf {\bibinfo {volume} {D57}},\ \bibinfo {pages}
  {6050} (\bibinfo {year} {1998})},\ \Eprint
  {http://arxiv.org/abs/hep-ph/9710259} {arXiv:hep-ph/9710259 [hep-ph]}
  \BibitemShut {NoStop}%
%%CITATION = HEP-PH/9710259;%%
\bibitem [{\citenamefont {Kohri}\ \emph {et~al.}(2008)\citenamefont {Kohri},
  \citenamefont {Lyth},\ and\ \citenamefont {Melchiorri}}]{Kohri:2007qn}%
  \BibitemOpen
  \bibfield  {author} {\bibinfo {author} {\bibfnamefont {K.}~\bibnamefont
  {Kohri}}, \bibinfo {author} {\bibfnamefont {D.~H.}\ \bibnamefont {Lyth}}, \
  and\ \bibinfo {author} {\bibfnamefont {A.}~\bibnamefont {Melchiorri}},\
  }\href {\doibase 10.1088/1475-7516/2008/04/038} {\bibfield  {journal}
  {\bibinfo  {journal} {JCAP}\ }\textbf {\bibinfo {volume} {0804}},\ \bibinfo
  {pages} {038} (\bibinfo {year} {2008})},\ \Eprint
  {http://arxiv.org/abs/0711.5006} {arXiv:0711.5006 [hep-ph]} \BibitemShut
  {NoStop}%
%%CITATION = ARXIV:0711.5006;%%
\bibitem [{\citenamefont {Drees}\ and\ \citenamefont
  {Erfani}(2011)}]{Drees:2011hb}%
  \BibitemOpen
  \bibfield  {author} {\bibinfo {author} {\bibfnamefont {M.}~\bibnamefont
  {Drees}}\ and\ \bibinfo {author} {\bibfnamefont {E.}~\bibnamefont {Erfani}},\
  }\href {\doibase 10.1088/1475-7516/2011/04/005} {\bibfield  {journal}
  {\bibinfo  {journal} {JCAP}\ }\textbf {\bibinfo {volume} {1104}},\ \bibinfo
  {pages} {005} (\bibinfo {year} {2011})},\ \Eprint
  {http://arxiv.org/abs/1102.2340} {arXiv:1102.2340 [hep-ph]} \BibitemShut
  {NoStop}%
%%CITATION = ARXIV:1102.2340;%%
\bibitem [{\citenamefont {Lyth}(2011)}]{Lyth:2011kj}%
  \BibitemOpen
  \bibfield  {author} {\bibinfo {author} {\bibfnamefont {D.~H.}\ \bibnamefont
  {Lyth}},\ }\href@noop {} {\  (\bibinfo {year} {2011})},\ \Eprint
  {http://arxiv.org/abs/1107.1681} {arXiv:1107.1681 [astro-ph.CO]} \BibitemShut
  {NoStop}%
%%CITATION = ARXIV:1107.1681;%%
\bibitem [{\citenamefont {Kawasaki}\ \emph {et~al.}(2013)\citenamefont
  {Kawasaki}, \citenamefont {Kitajima},\ and\ \citenamefont
  {Yanagida}}]{Kawasaki:2012wr}%
  \BibitemOpen
  \bibfield  {author} {\bibinfo {author} {\bibfnamefont {M.}~\bibnamefont
  {Kawasaki}}, \bibinfo {author} {\bibfnamefont {N.}~\bibnamefont {Kitajima}},
  \ and\ \bibinfo {author} {\bibfnamefont {T.~T.}\ \bibnamefont {Yanagida}},\
  }\href {\doibase 10.1103/PhysRevD.87.063519} {\bibfield  {journal} {\bibinfo
  {journal} {Phys. Rev.}\ }\textbf {\bibinfo {volume} {D87}},\ \bibinfo {pages}
  {063519} (\bibinfo {year} {2013})},\ \Eprint {http://arxiv.org/abs/1207.2550}
  {arXiv:1207.2550 [hep-ph]} \BibitemShut {NoStop}%
%%CITATION = ARXIV:1207.2550;%%
\bibitem [{\citenamefont {Kohri}\ \emph {et~al.}(2013)\citenamefont {Kohri},
  \citenamefont {Lin},\ and\ \citenamefont {Matsuda}}]{Kohri:2012yw}%
  \BibitemOpen
  \bibfield  {author} {\bibinfo {author} {\bibfnamefont {K.}~\bibnamefont
  {Kohri}}, \bibinfo {author} {\bibfnamefont {C.-M.}\ \bibnamefont {Lin}}, \
  and\ \bibinfo {author} {\bibfnamefont {T.}~\bibnamefont {Matsuda}},\ }\href
  {\doibase 10.1103/PhysRevD.87.103527} {\bibfield  {journal} {\bibinfo
  {journal} {Phys. Rev.}\ }\textbf {\bibinfo {volume} {D87}},\ \bibinfo {pages}
  {103527} (\bibinfo {year} {2013})},\ \Eprint {http://arxiv.org/abs/1211.2371}
  {arXiv:1211.2371 [hep-ph]} \BibitemShut {NoStop}%
%%CITATION = ARXIV:1211.2371;%%
\bibitem [{\citenamefont {Belotsky}\ \emph {et~al.}(2014)\citenamefont
  {Belotsky}, \citenamefont {Dmitriev}, \citenamefont {Esipova}, \citenamefont
  {Gani}, \citenamefont {Grobov}, \citenamefont {Khlopov}, \citenamefont
  {Kirillov}, \citenamefont {Rubin},\ and\ \citenamefont
  {Svadkovsky}}]{Belotsky:2014kca}%
  \BibitemOpen
  \bibfield  {author} {\bibinfo {author} {\bibfnamefont {K.~M.}\ \bibnamefont
  {Belotsky}}, \bibinfo {author} {\bibfnamefont {A.~D.}\ \bibnamefont
  {Dmitriev}}, \bibinfo {author} {\bibfnamefont {E.~A.}\ \bibnamefont
  {Esipova}}, \bibinfo {author} {\bibfnamefont {V.~A.}\ \bibnamefont {Gani}},
  \bibinfo {author} {\bibfnamefont {A.~V.}\ \bibnamefont {Grobov}}, \bibinfo
  {author} {\bibfnamefont {M.~{\relax Yu}.}\ \bibnamefont {Khlopov}}, \bibinfo
  {author} {\bibfnamefont {A.~A.}\ \bibnamefont {Kirillov}}, \bibinfo {author}
  {\bibfnamefont {S.~G.}\ \bibnamefont {Rubin}}, \ and\ \bibinfo {author}
  {\bibfnamefont {I.~V.}\ \bibnamefont {Svadkovsky}},\ }\href {\doibase
  10.1142/S0217732314400057} {\bibfield  {journal} {\bibinfo  {journal} {Mod.
  Phys. Lett.}\ }\textbf {\bibinfo {volume} {A29}},\ \bibinfo {pages} {1440005}
  (\bibinfo {year} {2014})},\ \Eprint {http://arxiv.org/abs/1410.0203}
  {arXiv:1410.0203 [astro-ph.CO]} \BibitemShut {NoStop}%
%%CITATION = ARXIV:1410.0203;%%
\bibitem [{\citenamefont {Clesse}\ and\ \citenamefont
  {García-Bellido}(2015)}]{Clesse:2015wea}%
  \BibitemOpen
  \bibfield  {author} {\bibinfo {author} {\bibfnamefont {S.}~\bibnamefont
  {Clesse}}\ and\ \bibinfo {author} {\bibfnamefont {J.}~\bibnamefont
  {García-Bellido}},\ }\href {\doibase 10.1103/PhysRevD.92.023524} {\bibfield
  {journal} {\bibinfo  {journal} {Phys. Rev.}\ }\textbf {\bibinfo {volume}
  {D92}},\ \bibinfo {pages} {023524} (\bibinfo {year} {2015})},\ \Eprint
  {http://arxiv.org/abs/1501.07565} {arXiv:1501.07565 [astro-ph.CO]}
  \BibitemShut {NoStop}%
%%CITATION = ARXIV:1501.07565;%%
\bibitem [{\citenamefont {Garcia-Bellido}\ and\ \citenamefont
  {Ruiz~Morales}(2017)}]{Garcia-Bellido:2017mdw}%
  \BibitemOpen
  \bibfield  {author} {\bibinfo {author} {\bibfnamefont {J.}~\bibnamefont
  {Garcia-Bellido}}\ and\ \bibinfo {author} {\bibfnamefont {E.}~\bibnamefont
  {Ruiz~Morales}},\ }\href {\doibase 10.1016/j.dark.2017.09.007} {\bibfield
  {journal} {\bibinfo  {journal} {Phys. Dark Univ.}\ }\textbf {\bibinfo
  {volume} {18}},\ \bibinfo {pages} {47} (\bibinfo {year} {2017})},\ \Eprint
  {http://arxiv.org/abs/1702.03901} {arXiv:1702.03901 [astro-ph.CO]}
  \BibitemShut {NoStop}%
%%CITATION = ARXIV:1702.03901;%%
\bibitem [{\citenamefont {Ballesteros}\ and\ \citenamefont
  {Taoso}(2018)}]{Ballesteros:2017fsr}%
  \BibitemOpen
  \bibfield  {author} {\bibinfo {author} {\bibfnamefont {G.}~\bibnamefont
  {Ballesteros}}\ and\ \bibinfo {author} {\bibfnamefont {M.}~\bibnamefont
  {Taoso}},\ }\href {\doibase 10.1103/PhysRevD.97.023501} {\bibfield  {journal}
  {\bibinfo  {journal} {Phys. Rev.}\ }\textbf {\bibinfo {volume} {D97}},\
  \bibinfo {pages} {023501} (\bibinfo {year} {2018})},\ \Eprint
  {http://arxiv.org/abs/1709.05565} {arXiv:1709.05565 [hep-ph]} \BibitemShut
  {NoStop}%
%%CITATION = ARXIV:1709.05565;%%
\bibitem [{\citenamefont {Inomata}\ \emph {et~al.}(2018)\citenamefont
  {Inomata}, \citenamefont {Kawasaki}, \citenamefont {Mukaida},\ and\
  \citenamefont {Yanagida}}]{Inomata:2017vxo}%
  \BibitemOpen
  \bibfield  {author} {\bibinfo {author} {\bibfnamefont {K.}~\bibnamefont
  {Inomata}}, \bibinfo {author} {\bibfnamefont {M.}~\bibnamefont {Kawasaki}},
  \bibinfo {author} {\bibfnamefont {K.}~\bibnamefont {Mukaida}}, \ and\
  \bibinfo {author} {\bibfnamefont {T.~T.}\ \bibnamefont {Yanagida}},\ }\href
  {\doibase 10.1103/PhysRevD.97.043514} {\bibfield  {journal} {\bibinfo
  {journal} {Phys. Rev.}\ }\textbf {\bibinfo {volume} {D97}},\ \bibinfo {pages}
  {043514} (\bibinfo {year} {2018})},\ \Eprint
  {http://arxiv.org/abs/1711.06129} {arXiv:1711.06129 [astro-ph.CO]}
  \BibitemShut {NoStop}%
%%CITATION = ARXIV:1711.06129;%%
\bibitem [{\citenamefont {Pi}\ \emph {et~al.}(2018)\citenamefont {Pi},
  \citenamefont {Zhang}, \citenamefont {Huang},\ and\ \citenamefont
  {Sasaki}}]{Pi:2017gih}%
  \BibitemOpen
  \bibfield  {author} {\bibinfo {author} {\bibfnamefont {S.}~\bibnamefont
  {Pi}}, \bibinfo {author} {\bibfnamefont {Y.-l.}\ \bibnamefont {Zhang}},
  \bibinfo {author} {\bibfnamefont {Q.-G.}\ \bibnamefont {Huang}}, \ and\
  \bibinfo {author} {\bibfnamefont {M.}~\bibnamefont {Sasaki}},\ }\href
  {\doibase 10.1088/1475-7516/2018/05/042} {\bibfield  {journal} {\bibinfo
  {journal} {JCAP}\ }\textbf {\bibinfo {volume} {1805}},\ \bibinfo {pages}
  {042} (\bibinfo {year} {2018})},\ \Eprint {http://arxiv.org/abs/1712.09896}
  {arXiv:1712.09896 [astro-ph.CO]} \BibitemShut {NoStop}%
%%CITATION = ARXIV:1712.09896;%%
\bibitem [{\citenamefont {Kohri}\ and\ \citenamefont
  {Terada}(2018)}]{Kohri:2018qtx}%
  \BibitemOpen
  \bibfield  {author} {\bibinfo {author} {\bibfnamefont {K.}~\bibnamefont
  {Kohri}}\ and\ \bibinfo {author} {\bibfnamefont {T.}~\bibnamefont {Terada}},\
  }\href {\doibase 10.1088/1361-6382/aaea18} {\bibfield  {journal} {\bibinfo
  {journal} {Class. Quant. Grav.}\ }\textbf {\bibinfo {volume} {35}},\ \bibinfo
  {pages} {235017} (\bibinfo {year} {2018})},\ \Eprint
  {http://arxiv.org/abs/1802.06785} {arXiv:1802.06785 [astro-ph.CO]}
  \BibitemShut {NoStop}%
%%CITATION = ARXIV:1802.06785;%%
\bibitem [{\citenamefont {Cai}\ \emph {et~al.}(2018)\citenamefont {Cai},
  \citenamefont {Tong}, \citenamefont {Wang},\ and\ \citenamefont
  {Yan}}]{Cai:2018tuh}%
  \BibitemOpen
  \bibfield  {author} {\bibinfo {author} {\bibfnamefont {Y.-F.}\ \bibnamefont
  {Cai}}, \bibinfo {author} {\bibfnamefont {X.}~\bibnamefont {Tong}}, \bibinfo
  {author} {\bibfnamefont {D.-G.}\ \bibnamefont {Wang}}, \ and\ \bibinfo
  {author} {\bibfnamefont {S.-F.}\ \bibnamefont {Yan}},\ }\href {\doibase
  10.1103/PhysRevLett.121.081306} {\bibfield  {journal} {\bibinfo  {journal}
  {Phys. Rev. Lett.}\ }\textbf {\bibinfo {volume} {121}},\ \bibinfo {pages}
  {081306} (\bibinfo {year} {2018})},\ \Eprint
  {http://arxiv.org/abs/1805.03639} {arXiv:1805.03639 [astro-ph.CO]}
  \BibitemShut {NoStop}%
%%CITATION = ARXIV:1805.03639;%%
\bibitem [{\citenamefont {Belotsky}\ \emph {et~al.}(2019)\citenamefont
  {Belotsky}, \citenamefont {Dokuchaev}, \citenamefont {Eroshenko},
  \citenamefont {Esipova}, \citenamefont {Khlopov}, \citenamefont {Khromykh},
  \citenamefont {Kirillov}, \citenamefont {Nikulin}, \citenamefont {Rubin},\
  and\ \citenamefont {Svadkovsky}}]{Belotsky:2018wph}%
  \BibitemOpen
  \bibfield  {author} {\bibinfo {author} {\bibfnamefont {K.~M.}\ \bibnamefont
  {Belotsky}}, \bibinfo {author} {\bibfnamefont {V.~I.}\ \bibnamefont
  {Dokuchaev}}, \bibinfo {author} {\bibfnamefont {Y.~N.}\ \bibnamefont
  {Eroshenko}}, \bibinfo {author} {\bibfnamefont {E.~A.}\ \bibnamefont
  {Esipova}}, \bibinfo {author} {\bibfnamefont {M.~{\relax Yu}.}\ \bibnamefont
  {Khlopov}}, \bibinfo {author} {\bibfnamefont {L.~A.}\ \bibnamefont
  {Khromykh}}, \bibinfo {author} {\bibfnamefont {A.~A.}\ \bibnamefont
  {Kirillov}}, \bibinfo {author} {\bibfnamefont {V.~V.}\ \bibnamefont
  {Nikulin}}, \bibinfo {author} {\bibfnamefont {S.~G.}\ \bibnamefont {Rubin}},
  \ and\ \bibinfo {author} {\bibfnamefont {I.~V.}\ \bibnamefont {Svadkovsky}},\
  }\href {\doibase 10.1140/epjc/s10052-019-6741-4} {\bibfield  {journal}
  {\bibinfo  {journal} {Eur. Phys. J.}\ }\textbf {\bibinfo {volume} {C79}},\
  \bibinfo {pages} {246} (\bibinfo {year} {2019})},\ \Eprint
  {http://arxiv.org/abs/1807.06590} {arXiv:1807.06590 [astro-ph.CO]}
  \BibitemShut {NoStop}%
%%CITATION = ARXIV:1807.06590;%%
\bibitem [{\citenamefont {Passaglia}\ \emph {et~al.}(2019)\citenamefont
  {Passaglia}, \citenamefont {Hu},\ and\ \citenamefont
  {Motohashi}}]{Passaglia:2018ixg}%
  \BibitemOpen
  \bibfield  {author} {\bibinfo {author} {\bibfnamefont {S.}~\bibnamefont
  {Passaglia}}, \bibinfo {author} {\bibfnamefont {W.}~\bibnamefont {Hu}}, \
  and\ \bibinfo {author} {\bibfnamefont {H.}~\bibnamefont {Motohashi}},\ }\href
  {\doibase 10.1103/PhysRevD.99.043536} {\bibfield  {journal} {\bibinfo
  {journal} {Phys. Rev.}\ }\textbf {\bibinfo {volume} {D99}},\ \bibinfo {pages}
  {043536} (\bibinfo {year} {2019})},\ \Eprint
  {http://arxiv.org/abs/1812.08243} {arXiv:1812.08243 [astro-ph.CO]}
  \BibitemShut {NoStop}%
%%CITATION = ARXIV:1812.08243;%%
\bibitem [{\citenamefont {Dimopoulos}\ \emph {et~al.}(2019)\citenamefont
  {Dimopoulos}, \citenamefont {Markkanen}, \citenamefont {Racioppi},\ and\
  \citenamefont {Vaskonen}}]{Dimopoulos:2019wew}%
  \BibitemOpen
  \bibfield  {author} {\bibinfo {author} {\bibfnamefont {K.}~\bibnamefont
  {Dimopoulos}}, \bibinfo {author} {\bibfnamefont {T.}~\bibnamefont
  {Markkanen}}, \bibinfo {author} {\bibfnamefont {A.}~\bibnamefont {Racioppi}},
  \ and\ \bibinfo {author} {\bibfnamefont {V.}~\bibnamefont {Vaskonen}},\
  }\href {\doibase 10.1088/1475-7516/2019/07/046} {\bibfield  {journal}
  {\bibinfo  {journal} {JCAP}\ }\textbf {\bibinfo {volume} {1907}},\ \bibinfo
  {pages} {046} (\bibinfo {year} {2019})},\ \Eprint
  {http://arxiv.org/abs/1903.09598} {arXiv:1903.09598 [astro-ph.CO]}
  \BibitemShut {NoStop}%
%%CITATION = ARXIV:1903.09598;%%
\bibitem [{\citenamefont {Mishra}\ and\ \citenamefont
  {Sahni}(2019)}]{Mishra:2019pzq}%
  \BibitemOpen
  \bibfield  {author} {\bibinfo {author} {\bibfnamefont {S.~S.}\ \bibnamefont
  {Mishra}}\ and\ \bibinfo {author} {\bibfnamefont {V.}~\bibnamefont {Sahni}},\
  }\href@noop {} {\  (\bibinfo {year} {2019})},\ \Eprint
  {http://arxiv.org/abs/1911.00057} {arXiv:1911.00057 [gr-qc]} \BibitemShut
  {NoStop}%
%%CITATION = ARXIV:1911.00057;%%
\bibitem [{\citenamefont {Cai}\ \emph {et~al.}(2019)\citenamefont {Cai},
  \citenamefont {Guo}, \citenamefont {Liu}, \citenamefont {Liu},\ and\
  \citenamefont {Yang}}]{Cai:2019bmk}%
  \BibitemOpen
  \bibfield  {author} {\bibinfo {author} {\bibfnamefont {R.-G.}\ \bibnamefont
  {Cai}}, \bibinfo {author} {\bibfnamefont {Z.-K.}\ \bibnamefont {Guo}},
  \bibinfo {author} {\bibfnamefont {J.}~\bibnamefont {Liu}}, \bibinfo {author}
  {\bibfnamefont {L.}~\bibnamefont {Liu}}, \ and\ \bibinfo {author}
  {\bibfnamefont {X.-Y.}\ \bibnamefont {Yang}},\ }\href@noop {} {\  (\bibinfo
  {year} {2019})},\ \Eprint {http://arxiv.org/abs/1912.10437} {arXiv:1912.10437
  [astro-ph.CO]} \BibitemShut {NoStop}%
%%CITATION = ARXIV:1912.10437;%%
\bibitem [{\citenamefont {Sasaki}\ \emph {et~al.}(2018)\citenamefont {Sasaki},
  \citenamefont {Suyama}, \citenamefont {Tanaka},\ and\ \citenamefont
  {Yokoyama}}]{Sasaki:2018dmp}%
  \BibitemOpen
  \bibfield  {author} {\bibinfo {author} {\bibfnamefont {M.}~\bibnamefont
  {Sasaki}}, \bibinfo {author} {\bibfnamefont {T.}~\bibnamefont {Suyama}},
  \bibinfo {author} {\bibfnamefont {T.}~\bibnamefont {Tanaka}}, \ and\ \bibinfo
  {author} {\bibfnamefont {S.}~\bibnamefont {Yokoyama}},\ }\href {\doibase
  10.1088/1361-6382/aaa7b4} {\bibfield  {journal} {\bibinfo  {journal} {Class.
  Quant. Grav.}\ }\textbf {\bibinfo {volume} {35}},\ \bibinfo {pages} {063001}
  (\bibinfo {year} {2018})},\ \Eprint {http://arxiv.org/abs/1801.05235}
  {arXiv:1801.05235 [astro-ph.CO]} \BibitemShut {NoStop}%
%%CITATION = ARXIV:1801.05235;%%
\bibitem [{\citenamefont {Bezrukov}\ and\ \citenamefont
  {Shaposhnikov}(2008)}]{Bezrukov:2007ep}%
  \BibitemOpen
  \bibfield  {author} {\bibinfo {author} {\bibfnamefont {F.~L.}\ \bibnamefont
  {Bezrukov}}\ and\ \bibinfo {author} {\bibfnamefont {M.}~\bibnamefont
  {Shaposhnikov}},\ }\href {\doibase 10.1016/j.physletb.2007.11.072} {\bibfield
   {journal} {\bibinfo  {journal} {Phys. Lett.}\ }\textbf {\bibinfo {volume}
  {B659}},\ \bibinfo {pages} {703} (\bibinfo {year} {2008})},\ \Eprint
  {http://arxiv.org/abs/0710.3755} {arXiv:0710.3755 [hep-th]} \BibitemShut
  {NoStop}%
%%CITATION = ARXIV:0710.3755;%%
\bibitem [{\citenamefont {Starobinsky}(1980)}]{Starobinsky:1980te}%
  \BibitemOpen
  \bibfield  {author} {\bibinfo {author} {\bibfnamefont {A.~A.}\ \bibnamefont
  {Starobinsky}},\ }\href {\doibase 10.1016/0370-2693(80)90670-X} {\bibfield
  {journal} {\bibinfo  {journal} {Phys. Lett.}\ }\textbf {\bibinfo {volume}
  {91B}},\ \bibinfo {pages} {99} (\bibinfo {year} {1980})},\ \bibinfo {note}
  {[,771(1980)]}\BibitemShut {NoStop}%
%%CITATION = PHLTA,91B,99;%%
\bibitem [{\citenamefont {Aghanim}\ \emph {et~al.}(2018)\citenamefont {Aghanim}
  \emph {et~al.}}]{Aghanim:2018eyx}%
  \BibitemOpen
  \bibfield  {author} {\bibinfo {author} {\bibfnamefont {N.}~\bibnamefont
  {Aghanim}} \emph {et~al.} (\bibinfo {collaboration} {Planck}),\ }\href@noop
  {} {\  (\bibinfo {year} {2018})},\ \Eprint {http://arxiv.org/abs/1807.06209}
  {arXiv:1807.06209 [astro-ph.CO]} \BibitemShut {NoStop}%
%%CITATION = ARXIV:1807.06209;%%
\bibitem [{\citenamefont {Akrami}\ \emph {et~al.}(2018)\citenamefont {Akrami}
  \emph {et~al.}}]{Akrami:2018odb}%
  \BibitemOpen
  \bibfield  {author} {\bibinfo {author} {\bibfnamefont {Y.}~\bibnamefont
  {Akrami}} \emph {et~al.} (\bibinfo {collaboration} {Planck}),\ }\href@noop {}
  {\  (\bibinfo {year} {2018})},\ \Eprint {http://arxiv.org/abs/1807.06211}
  {arXiv:1807.06211 [astro-ph.CO]} \BibitemShut {NoStop}%
%%CITATION = ARXIV:1807.06211;%%
\bibitem [{\citenamefont {Park}\ and\ \citenamefont
  {Yamaguchi}(2008)}]{Park:2008hz}%
  \BibitemOpen
  \bibfield  {author} {\bibinfo {author} {\bibfnamefont {S.~C.}\ \bibnamefont
  {Park}}\ and\ \bibinfo {author} {\bibfnamefont {S.}~\bibnamefont
  {Yamaguchi}},\ }\href {\doibase 10.1088/1475-7516/2008/08/009} {\bibfield
  {journal} {\bibinfo  {journal} {JCAP}\ }\textbf {\bibinfo {volume} {0808}},\
  \bibinfo {pages} {009} (\bibinfo {year} {2008})},\ \Eprint
  {http://arxiv.org/abs/0801.1722} {arXiv:0801.1722 [hep-ph]} \BibitemShut
  {NoStop}%
%%CITATION = ARXIV:0801.1722;%%
\bibitem [{\citenamefont {Burgess}\ \emph {et~al.}(2009)\citenamefont
  {Burgess}, \citenamefont {Lee},\ and\ \citenamefont
  {Trott}}]{Burgess:2009ea}%
  \BibitemOpen
  \bibfield  {author} {\bibinfo {author} {\bibfnamefont {C.~P.}\ \bibnamefont
  {Burgess}}, \bibinfo {author} {\bibfnamefont {H.~M.}\ \bibnamefont {Lee}}, \
  and\ \bibinfo {author} {\bibfnamefont {M.}~\bibnamefont {Trott}},\ }\href
  {\doibase 10.1088/1126-6708/2009/09/103} {\bibfield  {journal} {\bibinfo
  {journal} {JHEP}\ }\textbf {\bibinfo {volume} {09}},\ \bibinfo {pages} {103}
  (\bibinfo {year} {2009})},\ \Eprint {http://arxiv.org/abs/0902.4465}
  {arXiv:0902.4465 [hep-ph]} \BibitemShut {NoStop}%
%%CITATION = ARXIV:0902.4465;%%
\bibitem [{\citenamefont {Barbon}\ and\ \citenamefont
  {Espinosa}(2009)}]{Barbon:2009ya}%
  \BibitemOpen
  \bibfield  {author} {\bibinfo {author} {\bibfnamefont {J.~L.~F.}\
  \bibnamefont {Barbon}}\ and\ \bibinfo {author} {\bibfnamefont {J.~R.}\
  \bibnamefont {Espinosa}},\ }\href {\doibase 10.1103/PhysRevD.79.081302}
  {\bibfield  {journal} {\bibinfo  {journal} {Phys. Rev.}\ }\textbf {\bibinfo
  {volume} {D79}},\ \bibinfo {pages} {081302} (\bibinfo {year} {2009})},\
  \Eprint {http://arxiv.org/abs/0903.0355} {arXiv:0903.0355 [hep-ph]}
  \BibitemShut {NoStop}%
%%CITATION = ARXIV:0903.0355;%%
\bibitem [{\citenamefont {Burgess}\ \emph {et~al.}(2010)\citenamefont
  {Burgess}, \citenamefont {Lee},\ and\ \citenamefont
  {Trott}}]{Burgess:2010zq}%
  \BibitemOpen
  \bibfield  {author} {\bibinfo {author} {\bibfnamefont {C.~P.}\ \bibnamefont
  {Burgess}}, \bibinfo {author} {\bibfnamefont {H.~M.}\ \bibnamefont {Lee}}, \
  and\ \bibinfo {author} {\bibfnamefont {M.}~\bibnamefont {Trott}},\ }\href
  {\doibase 10.1007/JHEP07(2010)007} {\bibfield  {journal} {\bibinfo  {journal}
  {JHEP}\ }\textbf {\bibinfo {volume} {07}},\ \bibinfo {pages} {007} (\bibinfo
  {year} {2010})},\ \Eprint {http://arxiv.org/abs/1002.2730} {arXiv:1002.2730
  [hep-ph]} \BibitemShut {NoStop}%
%%CITATION = ARXIV:1002.2730;%%
\bibitem [{\citenamefont {Lerner}\ and\ \citenamefont
  {McDonald}(2010)}]{Lerner:2009na}%
  \BibitemOpen
  \bibfield  {author} {\bibinfo {author} {\bibfnamefont {R.~N.}\ \bibnamefont
  {Lerner}}\ and\ \bibinfo {author} {\bibfnamefont {J.}~\bibnamefont
  {McDonald}},\ }\href {\doibase 10.1088/1475-7516/2010/04/015} {\bibfield
  {journal} {\bibinfo  {journal} {JCAP}\ }\textbf {\bibinfo {volume} {1004}},\
  \bibinfo {pages} {015} (\bibinfo {year} {2010})},\ \Eprint
  {http://arxiv.org/abs/0912.5463} {arXiv:0912.5463 [hep-ph]} \BibitemShut
  {NoStop}%
%%CITATION = ARXIV:0912.5463;%%
\bibitem [{\citenamefont {Park}\ and\ \citenamefont
  {Shin}(2019)}]{Park:2018kst}%
  \BibitemOpen
  \bibfield  {author} {\bibinfo {author} {\bibfnamefont {S.~C.}\ \bibnamefont
  {Park}}\ and\ \bibinfo {author} {\bibfnamefont {C.~S.}\ \bibnamefont
  {Shin}},\ }\href {\doibase 10.1140/epjc/s10052-019-7037-4} {\bibfield
  {journal} {\bibinfo  {journal} {Eur. Phys. J.}\ }\textbf {\bibinfo {volume}
  {C79}},\ \bibinfo {pages} {529} (\bibinfo {year} {2019})},\ \Eprint
  {http://arxiv.org/abs/1807.09952} {arXiv:1807.09952 [hep-ph]} \BibitemShut
  {NoStop}%
%%CITATION = ARXIV:1807.09952;%%
\bibitem [{\citenamefont {Bezrukov}\ \emph {et~al.}(2011)\citenamefont
  {Bezrukov}, \citenamefont {Magnin}, \citenamefont {Shaposhnikov},\ and\
  \citenamefont {Sibiryakov}}]{Bezrukov:2010jz}%
  \BibitemOpen
  \bibfield  {author} {\bibinfo {author} {\bibfnamefont {F.}~\bibnamefont
  {Bezrukov}}, \bibinfo {author} {\bibfnamefont {A.}~\bibnamefont {Magnin}},
  \bibinfo {author} {\bibfnamefont {M.}~\bibnamefont {Shaposhnikov}}, \ and\
  \bibinfo {author} {\bibfnamefont {S.}~\bibnamefont {Sibiryakov}},\ }\href
  {\doibase 10.1007/JHEP01(2011)016} {\bibfield  {journal} {\bibinfo  {journal}
  {JHEP}\ }\textbf {\bibinfo {volume} {01}},\ \bibinfo {pages} {016} (\bibinfo
  {year} {2011})},\ \Eprint {http://arxiv.org/abs/1008.5157} {arXiv:1008.5157
  [hep-ph]} \BibitemShut {NoStop}%
%%CITATION = ARXIV:1008.5157;%%
\bibitem [{\citenamefont {Hamada}\ \emph {et~al.}(2014)\citenamefont {Hamada},
  \citenamefont {Kawai}, \citenamefont {Oda},\ and\ \citenamefont
  {Park}}]{Hamada:2014iga}%
  \BibitemOpen
  \bibfield  {author} {\bibinfo {author} {\bibfnamefont {Y.}~\bibnamefont
  {Hamada}}, \bibinfo {author} {\bibfnamefont {H.}~\bibnamefont {Kawai}},
  \bibinfo {author} {\bibfnamefont {K.-y.}\ \bibnamefont {Oda}}, \ and\
  \bibinfo {author} {\bibfnamefont {S.~C.}\ \bibnamefont {Park}},\ }\href
  {\doibase 10.1103/PhysRevLett.112.241301} {\bibfield  {journal} {\bibinfo
  {journal} {Phys. Rev. Lett.}\ }\textbf {\bibinfo {volume} {112}},\ \bibinfo
  {pages} {241301} (\bibinfo {year} {2014})},\ \Eprint
  {http://arxiv.org/abs/1403.5043} {arXiv:1403.5043 [hep-ph]} \BibitemShut
  {NoStop}%
%%CITATION = ARXIV:1403.5043;%%
\bibitem [{\citenamefont {Bezrukov}\ and\ \citenamefont
  {Shaposhnikov}(2014)}]{Bezrukov:2014bra}%
  \BibitemOpen
  \bibfield  {author} {\bibinfo {author} {\bibfnamefont {F.}~\bibnamefont
  {Bezrukov}}\ and\ \bibinfo {author} {\bibfnamefont {M.}~\bibnamefont
  {Shaposhnikov}},\ }\href {\doibase 10.1016/j.physletb.2014.05.074} {\bibfield
   {journal} {\bibinfo  {journal} {Phys. Lett.}\ }\textbf {\bibinfo {volume}
  {B734}},\ \bibinfo {pages} {249} (\bibinfo {year} {2014})},\ \Eprint
  {http://arxiv.org/abs/1403.6078} {arXiv:1403.6078 [hep-ph]} \BibitemShut
  {NoStop}%
%%CITATION = ARXIV:1403.6078;%%
\bibitem [{\citenamefont {Hamada}\ \emph {et~al.}(2015)\citenamefont {Hamada},
  \citenamefont {Kawai}, \citenamefont {Oda},\ and\ \citenamefont
  {Park}}]{Hamada:2014wna}%
  \BibitemOpen
  \bibfield  {author} {\bibinfo {author} {\bibfnamefont {Y.}~\bibnamefont
  {Hamada}}, \bibinfo {author} {\bibfnamefont {H.}~\bibnamefont {Kawai}},
  \bibinfo {author} {\bibfnamefont {K.-y.}\ \bibnamefont {Oda}}, \ and\
  \bibinfo {author} {\bibfnamefont {S.~C.}\ \bibnamefont {Park}},\ }\href
  {\doibase 10.1103/PhysRevD.91.053008} {\bibfield  {journal} {\bibinfo
  {journal} {Phys. Rev.}\ }\textbf {\bibinfo {volume} {D91}},\ \bibinfo {pages}
  {053008} (\bibinfo {year} {2015})},\ \Eprint {http://arxiv.org/abs/1408.4864}
  {arXiv:1408.4864 [hep-ph]} \BibitemShut {NoStop}%
%%CITATION = ARXIV:1408.4864;%%
\bibitem [{\citenamefont {Giudice}\ and\ \citenamefont
  {Lee}(2011)}]{Giudice:2010ka}%
  \BibitemOpen
  \bibfield  {author} {\bibinfo {author} {\bibfnamefont {G.~F.}\ \bibnamefont
  {Giudice}}\ and\ \bibinfo {author} {\bibfnamefont {H.~M.}\ \bibnamefont
  {Lee}},\ }\href {\doibase 10.1016/j.physletb.2010.10.035} {\bibfield
  {journal} {\bibinfo  {journal} {Phys. Lett.}\ }\textbf {\bibinfo {volume}
  {B694}},\ \bibinfo {pages} {294} (\bibinfo {year} {2011})},\ \Eprint
  {http://arxiv.org/abs/1010.1417} {arXiv:1010.1417 [hep-ph]} \BibitemShut
  {NoStop}%
%%CITATION = ARXIV:1010.1417;%%
\bibitem [{\citenamefont {Barbon}\ \emph {et~al.}(2015)\citenamefont {Barbon},
  \citenamefont {Casas}, \citenamefont {Elias-Miro},\ and\ \citenamefont
  {Espinosa}}]{Barbon:2015fla}%
  \BibitemOpen
  \bibfield  {author} {\bibinfo {author} {\bibfnamefont {J.~L.~F.}\
  \bibnamefont {Barbon}}, \bibinfo {author} {\bibfnamefont {J.~A.}\
  \bibnamefont {Casas}}, \bibinfo {author} {\bibfnamefont {J.}~\bibnamefont
  {Elias-Miro}}, \ and\ \bibinfo {author} {\bibfnamefont {J.~R.}\ \bibnamefont
  {Espinosa}},\ }\href {\doibase 10.1007/JHEP09(2015)027} {\bibfield  {journal}
  {\bibinfo  {journal} {JHEP}\ }\textbf {\bibinfo {volume} {09}},\ \bibinfo
  {pages} {027} (\bibinfo {year} {2015})},\ \Eprint
  {http://arxiv.org/abs/1501.02231} {arXiv:1501.02231 [hep-ph]} \BibitemShut
  {NoStop}%
%%CITATION = ARXIV:1501.02231;%%
\bibitem [{\citenamefont {Giudice}\ and\ \citenamefont
  {Lee}(2014)}]{Giudice:2014toa}%
  \BibitemOpen
  \bibfield  {author} {\bibinfo {author} {\bibfnamefont {G.~F.}\ \bibnamefont
  {Giudice}}\ and\ \bibinfo {author} {\bibfnamefont {H.~M.}\ \bibnamefont
  {Lee}},\ }\href {\doibase 10.1016/j.physletb.2014.04.020} {\bibfield
  {journal} {\bibinfo  {journal} {Phys. Lett.}\ }\textbf {\bibinfo {volume}
  {B733}},\ \bibinfo {pages} {58} (\bibinfo {year} {2014})},\ \Eprint
  {http://arxiv.org/abs/1402.2129} {arXiv:1402.2129 [hep-ph]} \BibitemShut
  {NoStop}%
%%CITATION = ARXIV:1402.2129;%%
\bibitem [{\citenamefont {Ema}(2017)}]{Ema:2017rqn}%
  \BibitemOpen
  \bibfield  {author} {\bibinfo {author} {\bibfnamefont {Y.}~\bibnamefont
  {Ema}},\ }\href {\doibase 10.1016/j.physletb.2017.04.060} {\bibfield
  {journal} {\bibinfo  {journal} {Phys. Lett.}\ }\textbf {\bibinfo {volume}
  {B770}},\ \bibinfo {pages} {403} (\bibinfo {year} {2017})},\ \Eprint
  {http://arxiv.org/abs/1701.07665} {arXiv:1701.07665 [hep-ph]} \BibitemShut
  {NoStop}%
%%CITATION = ARXIV:1701.07665;%%
\bibitem [{\citenamefont {Gorbunov}\ and\ \citenamefont
  {Tokareva}(2019)}]{Gorbunov:2018llf}%
  \BibitemOpen
  \bibfield  {author} {\bibinfo {author} {\bibfnamefont {D.}~\bibnamefont
  {Gorbunov}}\ and\ \bibinfo {author} {\bibfnamefont {A.}~\bibnamefont
  {Tokareva}},\ }\href {\doibase 10.1016/j.physletb.2018.11.015} {\bibfield
  {journal} {\bibinfo  {journal} {Phys. Lett.}\ }\textbf {\bibinfo {volume}
  {B788}},\ \bibinfo {pages} {37} (\bibinfo {year} {2019})},\ \Eprint
  {http://arxiv.org/abs/1807.02392} {arXiv:1807.02392 [hep-ph]} \BibitemShut
  {NoStop}%
%%CITATION = ARXIV:1807.02392;%%
\bibitem [{\citenamefont {Salvio}\ and\ \citenamefont
  {Mazumdar}(2015)}]{Salvio:2015kka}%
  \BibitemOpen
  \bibfield  {author} {\bibinfo {author} {\bibfnamefont {A.}~\bibnamefont
  {Salvio}}\ and\ \bibinfo {author} {\bibfnamefont {A.}~\bibnamefont
  {Mazumdar}},\ }\href {\doibase 10.1016/j.physletb.2015.09.020} {\bibfield
  {journal} {\bibinfo  {journal} {Phys. Lett.}\ }\textbf {\bibinfo {volume}
  {B750}},\ \bibinfo {pages} {194} (\bibinfo {year} {2015})},\ \Eprint
  {http://arxiv.org/abs/1506.07520} {arXiv:1506.07520 [hep-ph]} \BibitemShut
  {NoStop}%
%%CITATION = ARXIV:1506.07520;%%
\bibitem [{\citenamefont {Calmet}\ and\ \citenamefont
  {Kuntz}(2016)}]{Calmet:2016fsr}%
  \BibitemOpen
  \bibfield  {author} {\bibinfo {author} {\bibfnamefont {X.}~\bibnamefont
  {Calmet}}\ and\ \bibinfo {author} {\bibfnamefont {I.}~\bibnamefont {Kuntz}},\
  }\href {\doibase 10.1140/epjc/s10052-016-4136-3} {\bibfield  {journal}
  {\bibinfo  {journal} {Eur. Phys. J.}\ }\textbf {\bibinfo {volume} {C76}},\
  \bibinfo {pages} {289} (\bibinfo {year} {2016})},\ \Eprint
  {http://arxiv.org/abs/1605.02236} {arXiv:1605.02236 [hep-th]} \BibitemShut
  {NoStop}%
%%CITATION = ARXIV:1605.02236;%%
\bibitem [{\citenamefont {Wang}\ and\ \citenamefont
  {Wang}(2017)}]{Wang:2017fuy}%
  \BibitemOpen
  \bibfield  {author} {\bibinfo {author} {\bibfnamefont {Y.-C.}\ \bibnamefont
  {Wang}}\ and\ \bibinfo {author} {\bibfnamefont {T.}~\bibnamefont {Wang}},\
  }\href {\doibase 10.1103/PhysRevD.96.123506} {\bibfield  {journal} {\bibinfo
  {journal} {Phys. Rev.}\ }\textbf {\bibinfo {volume} {D96}},\ \bibinfo {pages}
  {123506} (\bibinfo {year} {2017})},\ \Eprint
  {http://arxiv.org/abs/1701.06636} {arXiv:1701.06636 [gr-qc]} \BibitemShut
  {NoStop}%
%%CITATION = ARXIV:1701.06636;%%
\bibitem [{\citenamefont {Ghilencea}(2018)}]{Ghilencea:2018rqg}%
  \BibitemOpen
  \bibfield  {author} {\bibinfo {author} {\bibfnamefont {D.~M.}\ \bibnamefont
  {Ghilencea}},\ }\href {\doibase 10.1103/PhysRevD.98.103524} {\bibfield
  {journal} {\bibinfo  {journal} {Phys. Rev.}\ }\textbf {\bibinfo {volume}
  {D98}},\ \bibinfo {pages} {103524} (\bibinfo {year} {2018})},\ \Eprint
  {http://arxiv.org/abs/1807.06900} {arXiv:1807.06900 [hep-ph]} \BibitemShut
  {NoStop}%
%%CITATION = ARXIV:1807.06900;%%
\bibitem [{\citenamefont {Ema}(2019)}]{Ema:2019fdd}%
  \BibitemOpen
  \bibfield  {author} {\bibinfo {author} {\bibfnamefont {Y.}~\bibnamefont
  {Ema}},\ }\href {\doibase 10.1088/1475-7516/2019/09/027} {\bibfield
  {journal} {\bibinfo  {journal} {JCAP}\ }\textbf {\bibinfo {volume} {1909}},\
  \bibinfo {pages} {027} (\bibinfo {year} {2019})},\ \Eprint
  {http://arxiv.org/abs/1907.00993} {arXiv:1907.00993 [hep-ph]} \BibitemShut
  {NoStop}%
%%CITATION = ARXIV:1907.00993;%%
\bibitem [{\citenamefont {Canko}\ \emph {et~al.}(2019)\citenamefont {Canko},
  \citenamefont {Gialamas},\ and\ \citenamefont {Kodaxis}}]{Canko:2019mud}%
  \BibitemOpen
  \bibfield  {author} {\bibinfo {author} {\bibfnamefont {D.~D.}\ \bibnamefont
  {Canko}}, \bibinfo {author} {\bibfnamefont {I.~D.}\ \bibnamefont {Gialamas}},
  \ and\ \bibinfo {author} {\bibfnamefont {G.~P.}\ \bibnamefont {Kodaxis}},\
  }\href@noop {} {\  (\bibinfo {year} {2019})},\ \Eprint
  {http://arxiv.org/abs/1901.06296} {arXiv:1901.06296 [hep-th]} \BibitemShut
  {NoStop}%
%%CITATION = ARXIV:1901.06296;%%
\bibitem [{\citenamefont {He}\ \emph {et~al.}(2019)\citenamefont {He},
  \citenamefont {Jinno}, \citenamefont {Kamada}, \citenamefont {Park},
  \citenamefont {Starobinsky},\ and\ \citenamefont {Yokoyama}}]{He:2018mgb}%
  \BibitemOpen
  \bibfield  {author} {\bibinfo {author} {\bibfnamefont {M.}~\bibnamefont
  {He}}, \bibinfo {author} {\bibfnamefont {R.}~\bibnamefont {Jinno}}, \bibinfo
  {author} {\bibfnamefont {K.}~\bibnamefont {Kamada}}, \bibinfo {author}
  {\bibfnamefont {S.~C.}\ \bibnamefont {Park}}, \bibinfo {author}
  {\bibfnamefont {A.~A.}\ \bibnamefont {Starobinsky}}, \ and\ \bibinfo {author}
  {\bibfnamefont {J.}~\bibnamefont {Yokoyama}},\ }\href {\doibase
  10.1016/j.physletb.2019.02.008} {\bibfield  {journal} {\bibinfo  {journal}
  {Phys. Lett.}\ }\textbf {\bibinfo {volume} {B791}},\ \bibinfo {pages} {36}
  (\bibinfo {year} {2019})},\ \Eprint {http://arxiv.org/abs/1812.10099}
  {arXiv:1812.10099 [hep-ph]} \BibitemShut {NoStop}%
%%CITATION = ARXIV:1812.10099;%%
\bibitem [{\citenamefont {He}\ \emph {et~al.}(2018)\citenamefont {He},
  \citenamefont {Starobinsky},\ and\ \citenamefont {Yokoyama}}]{He:2018gyf}%
  \BibitemOpen
  \bibfield  {author} {\bibinfo {author} {\bibfnamefont {M.}~\bibnamefont
  {He}}, \bibinfo {author} {\bibfnamefont {A.~A.}\ \bibnamefont {Starobinsky}},
  \ and\ \bibinfo {author} {\bibfnamefont {J.}~\bibnamefont {Yokoyama}},\
  }\href {\doibase 10.1088/1475-7516/2018/05/064} {\bibfield  {journal}
  {\bibinfo  {journal} {JCAP}\ }\textbf {\bibinfo {volume} {1805}},\ \bibinfo
  {pages} {064} (\bibinfo {year} {2018})},\ \Eprint
  {http://arxiv.org/abs/1804.00409} {arXiv:1804.00409 [astro-ph.CO]}
  \BibitemShut {NoStop}%
%%CITATION = ARXIV:1804.00409;%%
\bibitem [{\citenamefont {Gundhi}\ and\ \citenamefont
  {Steinwachs}(2018)}]{Gundhi:2018wyz}%
  \BibitemOpen
  \bibfield  {author} {\bibinfo {author} {\bibfnamefont {A.}~\bibnamefont
  {Gundhi}}\ and\ \bibinfo {author} {\bibfnamefont {C.~F.}\ \bibnamefont
  {Steinwachs}},\ }\href@noop {} {\  (\bibinfo {year} {2018})},\ \Eprint
  {http://arxiv.org/abs/1810.10546} {arXiv:1810.10546 [hep-th]} \BibitemShut
  {NoStop}%
%%CITATION = ARXIV:1810.10546;%%
\bibitem [{\citenamefont {Ema}\ \emph {et~al.}(2017)\citenamefont {Ema},
  \citenamefont {Jinno}, \citenamefont {Mukaida},\ and\ \citenamefont
  {Nakayama}}]{Ema:2016dny}%
  \BibitemOpen
  \bibfield  {author} {\bibinfo {author} {\bibfnamefont {Y.}~\bibnamefont
  {Ema}}, \bibinfo {author} {\bibfnamefont {R.}~\bibnamefont {Jinno}}, \bibinfo
  {author} {\bibfnamefont {K.}~\bibnamefont {Mukaida}}, \ and\ \bibinfo
  {author} {\bibfnamefont {K.}~\bibnamefont {Nakayama}},\ }\href {\doibase
  10.1088/1475-7516/2017/02/045} {\bibfield  {journal} {\bibinfo  {journal}
  {JCAP}\ }\textbf {\bibinfo {volume} {1702}},\ \bibinfo {pages} {045}
  (\bibinfo {year} {2017})},\ \Eprint {http://arxiv.org/abs/1609.05209}
  {arXiv:1609.05209 [hep-ph]} \BibitemShut {NoStop}%
%%CITATION = ARXIV:1609.05209;%%
\bibitem [{\citenamefont {Ezquiaga}\ \emph {et~al.}(2018)\citenamefont
  {Ezquiaga}, \citenamefont {Garcia-Bellido},\ and\ \citenamefont
  {Ruiz~Morales}}]{Ezquiaga:2017fvi}%
  \BibitemOpen
  \bibfield  {author} {\bibinfo {author} {\bibfnamefont {J.~M.}\ \bibnamefont
  {Ezquiaga}}, \bibinfo {author} {\bibfnamefont {J.}~\bibnamefont
  {Garcia-Bellido}}, \ and\ \bibinfo {author} {\bibfnamefont {E.}~\bibnamefont
  {Ruiz~Morales}},\ }\href {\doibase 10.1016/j.physletb.2017.11.039} {\bibfield
   {journal} {\bibinfo  {journal} {Phys. Lett.}\ }\textbf {\bibinfo {volume}
  {B776}},\ \bibinfo {pages} {345} (\bibinfo {year} {2018})},\ \Eprint
  {http://arxiv.org/abs/1705.04861} {arXiv:1705.04861 [astro-ph.CO]}
  \BibitemShut {NoStop}%
%%CITATION = ARXIV:1705.04861;%%
\bibitem [{\citenamefont {Kannike}\ \emph {et~al.}(2017)\citenamefont
  {Kannike}, \citenamefont {Marzola}, \citenamefont {Raidal},\ and\
  \citenamefont {Veermäe}}]{Kannike:2017bxn}%
  \BibitemOpen
  \bibfield  {author} {\bibinfo {author} {\bibfnamefont {K.}~\bibnamefont
  {Kannike}}, \bibinfo {author} {\bibfnamefont {L.}~\bibnamefont {Marzola}},
  \bibinfo {author} {\bibfnamefont {M.}~\bibnamefont {Raidal}}, \ and\ \bibinfo
  {author} {\bibfnamefont {H.}~\bibnamefont {Veermäe}},\ }\href {\doibase
  10.1088/1475-7516/2017/09/020} {\bibfield  {journal} {\bibinfo  {journal}
  {JCAP}\ }\textbf {\bibinfo {volume} {1709}},\ \bibinfo {pages} {020}
  (\bibinfo {year} {2017})},\ \Eprint {http://arxiv.org/abs/1705.06225}
  {arXiv:1705.06225 [astro-ph.CO]} \BibitemShut {NoStop}%
%%CITATION = ARXIV:1705.06225;%%
\bibitem [{\citenamefont {Germani}\ and\ \citenamefont
  {Prokopec}(2017)}]{Germani:2017bcs}%
  \BibitemOpen
  \bibfield  {author} {\bibinfo {author} {\bibfnamefont {C.}~\bibnamefont
  {Germani}}\ and\ \bibinfo {author} {\bibfnamefont {T.}~\bibnamefont
  {Prokopec}},\ }\href {\doibase 10.1016/j.dark.2017.09.001} {\bibfield
  {journal} {\bibinfo  {journal} {Phys. Dark Univ.}\ }\textbf {\bibinfo
  {volume} {18}},\ \bibinfo {pages} {6} (\bibinfo {year} {2017})},\ \Eprint
  {http://arxiv.org/abs/1706.04226} {arXiv:1706.04226 [astro-ph.CO]}
  \BibitemShut {NoStop}%
%%CITATION = ARXIV:1706.04226;%%
\bibitem [{\citenamefont {Bezrukov}\ \emph {et~al.}(2018)\citenamefont
  {Bezrukov}, \citenamefont {Pauly},\ and\ \citenamefont
  {Rubio}}]{Bezrukov:2017dyv}%
  \BibitemOpen
  \bibfield  {author} {\bibinfo {author} {\bibfnamefont {F.}~\bibnamefont
  {Bezrukov}}, \bibinfo {author} {\bibfnamefont {M.}~\bibnamefont {Pauly}}, \
  and\ \bibinfo {author} {\bibfnamefont {J.}~\bibnamefont {Rubio}},\ }\href
  {\doibase 10.1088/1475-7516/2018/02/040} {\bibfield  {journal} {\bibinfo
  {journal} {JCAP}\ }\textbf {\bibinfo {volume} {1802}},\ \bibinfo {pages}
  {040} (\bibinfo {year} {2018})},\ \Eprint {http://arxiv.org/abs/1706.05007}
  {arXiv:1706.05007 [hep-ph]} \BibitemShut {NoStop}%
%%CITATION = ARXIV:1706.05007;%%
\bibitem [{\citenamefont {Motohashi}\ and\ \citenamefont
  {Hu}(2017)}]{Motohashi:2017kbs}%
  \BibitemOpen
  \bibfield  {author} {\bibinfo {author} {\bibfnamefont {H.}~\bibnamefont
  {Motohashi}}\ and\ \bibinfo {author} {\bibfnamefont {W.}~\bibnamefont {Hu}},\
  }\href {\doibase 10.1103/PhysRevD.96.063503} {\bibfield  {journal} {\bibinfo
  {journal} {Phys. Rev.}\ }\textbf {\bibinfo {volume} {D96}},\ \bibinfo {pages}
  {063503} (\bibinfo {year} {2017})},\ \Eprint
  {http://arxiv.org/abs/1706.06784} {arXiv:1706.06784 [astro-ph.CO]}
  \BibitemShut {NoStop}%
%%CITATION = ARXIV:1706.06784;%%
\bibitem [{\citenamefont {Masina}(2018)}]{Masina:2018ejw}%
  \BibitemOpen
  \bibfield  {author} {\bibinfo {author} {\bibfnamefont {I.}~\bibnamefont
  {Masina}},\ }\href {\doibase 10.1103/PhysRevD.98.043536} {\bibfield
  {journal} {\bibinfo  {journal} {Phys. Rev.}\ }\textbf {\bibinfo {volume}
  {D98}},\ \bibinfo {pages} {043536} (\bibinfo {year} {2018})},\ \Eprint
  {http://arxiv.org/abs/1805.02160} {arXiv:1805.02160 [hep-ph]} \BibitemShut
  {NoStop}%
%%CITATION = ARXIV:1805.02160;%%
\bibitem [{\citenamefont {Drees}\ and\ \citenamefont
  {Xu}(2019)}]{Drees:2019xpp}%
  \BibitemOpen
  \bibfield  {author} {\bibinfo {author} {\bibfnamefont {M.}~\bibnamefont
  {Drees}}\ and\ \bibinfo {author} {\bibfnamefont {Y.}~\bibnamefont {Xu}},\
  }\href@noop {} {\  (\bibinfo {year} {2019})},\ \Eprint
  {http://arxiv.org/abs/1905.13581} {arXiv:1905.13581 [hep-ph]} \BibitemShut
  {NoStop}%
%%CITATION = ARXIV:1905.13581;%%
\bibitem [{\citenamefont {Sirunyan}\ \emph {et~al.}(2019)\citenamefont
  {Sirunyan} \emph {et~al.}}]{Sirunyan:2019jyn}%
  \BibitemOpen
  \bibfield  {author} {\bibinfo {author} {\bibfnamefont {A.~M.}\ \bibnamefont
  {Sirunyan}} \emph {et~al.} (\bibinfo {collaboration} {CMS}),\ }\href@noop {}
  {\  (\bibinfo {year} {2019})},\ \Eprint {http://arxiv.org/abs/1909.09193}
  {arXiv:1909.09193 [hep-ex]} \BibitemShut {NoStop}%
%%CITATION = ARXIV:1909.09193;%%
\bibitem [{\citenamefont {Codello}\ and\ \citenamefont
  {Jain}(2016)}]{Codello:2015mba}%
  \BibitemOpen
  \bibfield  {author} {\bibinfo {author} {\bibfnamefont {A.}~\bibnamefont
  {Codello}}\ and\ \bibinfo {author} {\bibfnamefont {R.~K.}\ \bibnamefont
  {Jain}},\ }\href {\doibase 10.1088/0264-9381/33/22/225006} {\bibfield
  {journal} {\bibinfo  {journal} {Class. Quant. Grav.}\ }\textbf {\bibinfo
  {volume} {33}},\ \bibinfo {pages} {225006} (\bibinfo {year} {2016})},\
  \Eprint {http://arxiv.org/abs/1507.06308} {arXiv:1507.06308 [gr-qc]}
  \BibitemShut {NoStop}%
%%CITATION = ARXIV:1507.06308;%%
\bibitem [{\citenamefont {Markkanen}\ \emph {et~al.}(2018)\citenamefont
  {Markkanen}, \citenamefont {Nurmi}, \citenamefont {Rajantie},\ and\
  \citenamefont {Stopyra}}]{Markkanen:2018bfx}%
  \BibitemOpen
  \bibfield  {author} {\bibinfo {author} {\bibfnamefont {T.}~\bibnamefont
  {Markkanen}}, \bibinfo {author} {\bibfnamefont {S.}~\bibnamefont {Nurmi}},
  \bibinfo {author} {\bibfnamefont {A.}~\bibnamefont {Rajantie}}, \ and\
  \bibinfo {author} {\bibfnamefont {S.}~\bibnamefont {Stopyra}},\ }\href
  {\doibase 10.1007/JHEP06(2018)040} {\bibfield  {journal} {\bibinfo  {journal}
  {JHEP}\ }\textbf {\bibinfo {volume} {06}},\ \bibinfo {pages} {040} (\bibinfo
  {year} {2018})},\ \Eprint {http://arxiv.org/abs/1804.02020} {arXiv:1804.02020
  [hep-ph]} \BibitemShut {NoStop}%
%%CITATION = ARXIV:1804.02020;%%
\bibitem [{\citenamefont {De~Simone}\ \emph {et~al.}(2009)\citenamefont
  {De~Simone}, \citenamefont {Hertzberg},\ and\ \citenamefont
  {Wilczek}}]{DeSimone:2008ei}%
  \BibitemOpen
  \bibfield  {author} {\bibinfo {author} {\bibfnamefont {A.}~\bibnamefont
  {De~Simone}}, \bibinfo {author} {\bibfnamefont {M.~P.}\ \bibnamefont
  {Hertzberg}}, \ and\ \bibinfo {author} {\bibfnamefont {F.}~\bibnamefont
  {Wilczek}},\ }\href {\doibase 10.1016/j.physletb.2009.05.054} {\bibfield
  {journal} {\bibinfo  {journal} {Phys. Lett.}\ }\textbf {\bibinfo {volume}
  {B678}},\ \bibinfo {pages} {1} (\bibinfo {year} {2009})},\ \Eprint
  {http://arxiv.org/abs/0812.4946} {arXiv:0812.4946 [hep-ph]} \BibitemShut
  {NoStop}%
%%CITATION = ARXIV:0812.4946;%%
\bibitem [{\citenamefont {Herranen}\ \emph {et~al.}(2014)\citenamefont
  {Herranen}, \citenamefont {Markkanen}, \citenamefont {Nurmi},\ and\
  \citenamefont {Rajantie}}]{Herranen:2014cua}%
  \BibitemOpen
  \bibfield  {author} {\bibinfo {author} {\bibfnamefont {M.}~\bibnamefont
  {Herranen}}, \bibinfo {author} {\bibfnamefont {T.}~\bibnamefont {Markkanen}},
  \bibinfo {author} {\bibfnamefont {S.}~\bibnamefont {Nurmi}}, \ and\ \bibinfo
  {author} {\bibfnamefont {A.}~\bibnamefont {Rajantie}},\ }\href {\doibase
  10.1103/PhysRevLett.113.211102} {\bibfield  {journal} {\bibinfo  {journal}
  {Phys. Rev. Lett.}\ }\textbf {\bibinfo {volume} {113}},\ \bibinfo {pages}
  {211102} (\bibinfo {year} {2014})},\ \Eprint {http://arxiv.org/abs/1407.3141}
  {arXiv:1407.3141 [hep-ph]} \BibitemShut {NoStop}%
%%CITATION = ARXIV:1407.3141;%%
\bibitem [{\citenamefont {Herranen}\ \emph {et~al.}(2017)\citenamefont
  {Herranen}, \citenamefont {Hohenegger}, \citenamefont {Osland},\ and\
  \citenamefont {Tranberg}}]{Herranen:2016xsy}%
  \BibitemOpen
  \bibfield  {author} {\bibinfo {author} {\bibfnamefont {M.}~\bibnamefont
  {Herranen}}, \bibinfo {author} {\bibfnamefont {A.}~\bibnamefont
  {Hohenegger}}, \bibinfo {author} {\bibfnamefont {A.}~\bibnamefont {Osland}},
  \ and\ \bibinfo {author} {\bibfnamefont {A.}~\bibnamefont {Tranberg}},\
  }\href {\doibase 10.1103/PhysRevD.95.023525} {\bibfield  {journal} {\bibinfo
  {journal} {Phys. Rev.}\ }\textbf {\bibinfo {volume} {D95}},\ \bibinfo {pages}
  {023525} (\bibinfo {year} {2017})},\ \Eprint
  {http://arxiv.org/abs/1608.08906} {arXiv:1608.08906 [hep-ph]} \BibitemShut
  {NoStop}%
%%CITATION = ARXIV:1608.08906;%%
\bibitem [{\citenamefont {Chataignier}\ \emph {et~al.}(2018)\citenamefont
  {Chataignier}, \citenamefont {Prokopec}, \citenamefont {Schmidt},\ and\
  \citenamefont {Swiezewska}}]{Chataignier:2018aud}%
  \BibitemOpen
  \bibfield  {author} {\bibinfo {author} {\bibfnamefont {L.}~\bibnamefont
  {Chataignier}}, \bibinfo {author} {\bibfnamefont {T.}~\bibnamefont
  {Prokopec}}, \bibinfo {author} {\bibfnamefont {M.~G.}\ \bibnamefont
  {Schmidt}}, \ and\ \bibinfo {author} {\bibfnamefont {B.}~\bibnamefont
  {Swiezewska}},\ }\href {\doibase 10.1007/JHEP03(2018)014} {\bibfield
  {journal} {\bibinfo  {journal} {JHEP}\ }\textbf {\bibinfo {volume} {03}},\
  \bibinfo {pages} {014} (\bibinfo {year} {2018})},\ \Eprint
  {http://arxiv.org/abs/1801.05258} {arXiv:1801.05258 [hep-ph]} \BibitemShut
  {NoStop}%
%%CITATION = ARXIV:1801.05258;%%
\bibitem [{\citenamefont {Degrassi}\ \emph {et~al.}(2012)\citenamefont
  {Degrassi}, \citenamefont {Di~Vita}, \citenamefont {Elias-Miro},
  \citenamefont {Espinosa}, \citenamefont {Giudice}, \citenamefont {Isidori},\
  and\ \citenamefont {Strumia}}]{Degrassi:2012ry}%
  \BibitemOpen
  \bibfield  {author} {\bibinfo {author} {\bibfnamefont {G.}~\bibnamefont
  {Degrassi}}, \bibinfo {author} {\bibfnamefont {S.}~\bibnamefont {Di~Vita}},
  \bibinfo {author} {\bibfnamefont {J.}~\bibnamefont {Elias-Miro}}, \bibinfo
  {author} {\bibfnamefont {J.~R.}\ \bibnamefont {Espinosa}}, \bibinfo {author}
  {\bibfnamefont {G.~F.}\ \bibnamefont {Giudice}}, \bibinfo {author}
  {\bibfnamefont {G.}~\bibnamefont {Isidori}}, \ and\ \bibinfo {author}
  {\bibfnamefont {A.}~\bibnamefont {Strumia}},\ }\href {\doibase
  10.1007/JHEP08(2012)098} {\bibfield  {journal} {\bibinfo  {journal} {JHEP}\
  }\textbf {\bibinfo {volume} {08}},\ \bibinfo {pages} {098} (\bibinfo {year}
  {2012})},\ \Eprint {http://arxiv.org/abs/1205.6497} {arXiv:1205.6497
  [hep-ph]} \BibitemShut {NoStop}%
%%CITATION = ARXIV:1205.6497;%%
\bibitem [{\citenamefont {Buttazzo}\ \emph {et~al.}(2013)\citenamefont
  {Buttazzo}, \citenamefont {Degrassi}, \citenamefont {Giardino}, \citenamefont
  {Giudice}, \citenamefont {Sala}, \citenamefont {Salvio},\ and\ \citenamefont
  {Strumia}}]{Buttazzo:2013uya}%
  \BibitemOpen
  \bibfield  {author} {\bibinfo {author} {\bibfnamefont {D.}~\bibnamefont
  {Buttazzo}}, \bibinfo {author} {\bibfnamefont {G.}~\bibnamefont {Degrassi}},
  \bibinfo {author} {\bibfnamefont {P.~P.}\ \bibnamefont {Giardino}}, \bibinfo
  {author} {\bibfnamefont {G.~F.}\ \bibnamefont {Giudice}}, \bibinfo {author}
  {\bibfnamefont {F.}~\bibnamefont {Sala}}, \bibinfo {author} {\bibfnamefont
  {A.}~\bibnamefont {Salvio}}, \ and\ \bibinfo {author} {\bibfnamefont
  {A.}~\bibnamefont {Strumia}},\ }\href {\doibase 10.1007/JHEP12(2013)089}
  {\bibfield  {journal} {\bibinfo  {journal} {JHEP}\ }\textbf {\bibinfo
  {volume} {12}},\ \bibinfo {pages} {089} (\bibinfo {year} {2013})},\ \Eprint
  {http://arxiv.org/abs/1307.3536} {arXiv:1307.3536 [hep-ph]} \BibitemShut
  {NoStop}%
%%CITATION = ARXIV:1307.3536;%%
\bibitem [{\citenamefont {Achucarro}\ \emph {et~al.}(2012)\citenamefont
  {Achucarro}, \citenamefont {Atal}, \citenamefont {Cespedes}, \citenamefont
  {Gong}, \citenamefont {Palma},\ and\ \citenamefont
  {Patil}}]{Achucarro:2012yr}%
  \BibitemOpen
  \bibfield  {author} {\bibinfo {author} {\bibfnamefont {A.}~\bibnamefont
  {Achucarro}}, \bibinfo {author} {\bibfnamefont {V.}~\bibnamefont {Atal}},
  \bibinfo {author} {\bibfnamefont {S.}~\bibnamefont {Cespedes}}, \bibinfo
  {author} {\bibfnamefont {J.-O.}\ \bibnamefont {Gong}}, \bibinfo {author}
  {\bibfnamefont {G.~A.}\ \bibnamefont {Palma}}, \ and\ \bibinfo {author}
  {\bibfnamefont {S.~P.}\ \bibnamefont {Patil}},\ }\href {\doibase
  10.1103/PhysRevD.86.121301} {\bibfield  {journal} {\bibinfo  {journal} {Phys.
  Rev.}\ }\textbf {\bibinfo {volume} {D86}},\ \bibinfo {pages} {121301}
  (\bibinfo {year} {2012})},\ \Eprint {http://arxiv.org/abs/1205.0710}
  {arXiv:1205.0710 [hep-th]} \BibitemShut {NoStop}%
%%CITATION = ARXIV:1205.0710;%%
\bibitem [{\citenamefont {Mulryne}\ and\ \citenamefont
  {Ronayne}(2016)}]{Mulryne:2016mzv}%
  \BibitemOpen
  \bibfield  {author} {\bibinfo {author} {\bibfnamefont {D.~J.}\ \bibnamefont
  {Mulryne}}\ and\ \bibinfo {author} {\bibfnamefont {J.~W.}\ \bibnamefont
  {Ronayne}},\ }\href@noop {} {\  (\bibinfo {year} {2016})},\ \Eprint
  {http://arxiv.org/abs/1609.00381} {arXiv:1609.00381 [astro-ph.CO]}
  \BibitemShut {NoStop}%
%%CITATION = ARXIV:1609.00381;%%
\bibitem [{\citenamefont {Mulryne}\ \emph {et~al.}(2011)\citenamefont
  {Mulryne}, \citenamefont {Seery},\ and\ \citenamefont
  {Wesley}}]{Mulryne:2010rp}%
  \BibitemOpen
  \bibfield  {author} {\bibinfo {author} {\bibfnamefont {D.~J.}\ \bibnamefont
  {Mulryne}}, \bibinfo {author} {\bibfnamefont {D.}~\bibnamefont {Seery}}, \
  and\ \bibinfo {author} {\bibfnamefont {D.}~\bibnamefont {Wesley}},\ }\href
  {\doibase 10.1088/1475-7516/2011/04/030} {\bibfield  {journal} {\bibinfo
  {journal} {JCAP}\ }\textbf {\bibinfo {volume} {1104}},\ \bibinfo {pages}
  {030} (\bibinfo {year} {2011})},\ \Eprint {http://arxiv.org/abs/1008.3159}
  {arXiv:1008.3159 [astro-ph.CO]} \BibitemShut {NoStop}%
%%CITATION = ARXIV:1008.3159;%%
\bibitem [{\citenamefont {Dias}\ \emph {et~al.}(2015)\citenamefont {Dias},
  \citenamefont {Elliston}, \citenamefont {Frazer}, \citenamefont {Mulryne},\
  and\ \citenamefont {Seery}}]{Dias:2014msa}%
  \BibitemOpen
  \bibfield  {author} {\bibinfo {author} {\bibfnamefont {M.}~\bibnamefont
  {Dias}}, \bibinfo {author} {\bibfnamefont {J.}~\bibnamefont {Elliston}},
  \bibinfo {author} {\bibfnamefont {J.}~\bibnamefont {Frazer}}, \bibinfo
  {author} {\bibfnamefont {D.}~\bibnamefont {Mulryne}}, \ and\ \bibinfo
  {author} {\bibfnamefont {D.}~\bibnamefont {Seery}},\ }\href {\doibase
  10.1088/1475-7516/2015/02/040} {\bibfield  {journal} {\bibinfo  {journal}
  {JCAP}\ }\textbf {\bibinfo {volume} {1502}},\ \bibinfo {pages} {040}
  (\bibinfo {year} {2015})},\ \Eprint {http://arxiv.org/abs/1410.3491}
  {arXiv:1410.3491 [gr-qc]} \BibitemShut {NoStop}%
%%CITATION = ARXIV:1410.3491;%%
\bibitem [{\citenamefont {Leach}\ \emph {et~al.}(2001)\citenamefont {Leach},
  \citenamefont {Sasaki}, \citenamefont {Wands},\ and\ \citenamefont
  {Liddle}}]{Leach:2001zf}%
  \BibitemOpen
  \bibfield  {author} {\bibinfo {author} {\bibfnamefont {S.~M.}\ \bibnamefont
  {Leach}}, \bibinfo {author} {\bibfnamefont {M.}~\bibnamefont {Sasaki}},
  \bibinfo {author} {\bibfnamefont {D.}~\bibnamefont {Wands}}, \ and\ \bibinfo
  {author} {\bibfnamefont {A.~R.}\ \bibnamefont {Liddle}},\ }\href {\doibase
  10.1103/PhysRevD.64.023512} {\bibfield  {journal} {\bibinfo  {journal} {Phys.
  Rev.}\ }\textbf {\bibinfo {volume} {D64}},\ \bibinfo {pages} {023512}
  (\bibinfo {year} {2001})},\ \Eprint {http://arxiv.org/abs/astro-ph/0101406}
  {arXiv:astro-ph/0101406 [astro-ph]} \BibitemShut {NoStop}%
%%CITATION = ASTRO-PH/0101406;%%
\bibitem [{\citenamefont {Carr}(1975)}]{Carr:1975qj}%
  \BibitemOpen
  \bibfield  {author} {\bibinfo {author} {\bibfnamefont {B.~J.}\ \bibnamefont
  {Carr}},\ }\href {\doibase 10.1086/153853} {\bibfield  {journal} {\bibinfo
  {journal} {Astrophys. J.}\ }\textbf {\bibinfo {volume} {201}},\ \bibinfo
  {pages} {1} (\bibinfo {year} {1975})}\BibitemShut {NoStop}%
%%CITATION = ASJOA,201,1;%%
\bibitem [{\citenamefont {Green}\ \emph {et~al.}(2004)\citenamefont {Green},
  \citenamefont {Liddle}, \citenamefont {Malik},\ and\ \citenamefont
  {Sasaki}}]{Green:2004wb}%
  \BibitemOpen
  \bibfield  {author} {\bibinfo {author} {\bibfnamefont {A.~M.}\ \bibnamefont
  {Green}}, \bibinfo {author} {\bibfnamefont {A.~R.}\ \bibnamefont {Liddle}},
  \bibinfo {author} {\bibfnamefont {K.~A.}\ \bibnamefont {Malik}}, \ and\
  \bibinfo {author} {\bibfnamefont {M.}~\bibnamefont {Sasaki}},\ }\href
  {\doibase 10.1103/PhysRevD.70.041502} {\bibfield  {journal} {\bibinfo
  {journal} {Phys. Rev.}\ }\textbf {\bibinfo {volume} {D70}},\ \bibinfo {pages}
  {041502} (\bibinfo {year} {2004})},\ \Eprint
  {http://arxiv.org/abs/astro-ph/0403181} {arXiv:astro-ph/0403181 [astro-ph]}
  \BibitemShut {NoStop}%
%%CITATION = ASTRO-PH/0403181;%%
\bibitem [{\citenamefont {Young}\ \emph {et~al.}(2014)\citenamefont {Young},
  \citenamefont {Byrnes},\ and\ \citenamefont {Sasaki}}]{Young:2014ana}%
  \BibitemOpen
  \bibfield  {author} {\bibinfo {author} {\bibfnamefont {S.}~\bibnamefont
  {Young}}, \bibinfo {author} {\bibfnamefont {C.~T.}\ \bibnamefont {Byrnes}}, \
  and\ \bibinfo {author} {\bibfnamefont {M.}~\bibnamefont {Sasaki}},\ }\href
  {\doibase 10.1088/1475-7516/2014/07/045} {\bibfield  {journal} {\bibinfo
  {journal} {JCAP}\ }\textbf {\bibinfo {volume} {1407}},\ \bibinfo {pages}
  {045} (\bibinfo {year} {2014})},\ \Eprint {http://arxiv.org/abs/1405.7023}
  {arXiv:1405.7023 [gr-qc]} \BibitemShut {NoStop}%
%%CITATION = ARXIV:1405.7023;%%
\bibitem [{\citenamefont {Carr}\ \emph {et~al.}(1994)\citenamefont {Carr},
  \citenamefont {Gilbert},\ and\ \citenamefont {Lidsey}}]{Carr:1994ar}%
  \BibitemOpen
  \bibfield  {author} {\bibinfo {author} {\bibfnamefont {B.~J.}\ \bibnamefont
  {Carr}}, \bibinfo {author} {\bibfnamefont {J.~H.}\ \bibnamefont {Gilbert}}, \
  and\ \bibinfo {author} {\bibfnamefont {J.~E.}\ \bibnamefont {Lidsey}},\
  }\href {\doibase 10.1103/PhysRevD.50.4853} {\bibfield  {journal} {\bibinfo
  {journal} {Phys. Rev.}\ }\textbf {\bibinfo {volume} {D50}},\ \bibinfo {pages}
  {4853} (\bibinfo {year} {1994})},\ \Eprint
  {http://arxiv.org/abs/astro-ph/9405027} {arXiv:astro-ph/9405027 [astro-ph]}
  \BibitemShut {NoStop}%
%%CITATION = ASTRO-PH/9405027;%%
\bibitem [{\citenamefont {Green}\ and\ \citenamefont
  {Liddle}(1997)}]{Green:1997sz}%
  \BibitemOpen
  \bibfield  {author} {\bibinfo {author} {\bibfnamefont {A.~M.}\ \bibnamefont
  {Green}}\ and\ \bibinfo {author} {\bibfnamefont {A.~R.}\ \bibnamefont
  {Liddle}},\ }\href {\doibase 10.1103/PhysRevD.56.6166} {\bibfield  {journal}
  {\bibinfo  {journal} {Phys. Rev.}\ }\textbf {\bibinfo {volume} {D56}},\
  \bibinfo {pages} {6166} (\bibinfo {year} {1997})},\ \Eprint
  {http://arxiv.org/abs/astro-ph/9704251} {arXiv:astro-ph/9704251 [astro-ph]}
  \BibitemShut {NoStop}%
%%CITATION = ASTRO-PH/9704251;%%
\bibitem [{\citenamefont {Harada}\ \emph {et~al.}(2013)\citenamefont {Harada},
  \citenamefont {Yoo},\ and\ \citenamefont {Kohri}}]{Harada:2013epa}%
  \BibitemOpen
  \bibfield  {author} {\bibinfo {author} {\bibfnamefont {T.}~\bibnamefont
  {Harada}}, \bibinfo {author} {\bibfnamefont {C.-M.}\ \bibnamefont {Yoo}}, \
  and\ \bibinfo {author} {\bibfnamefont {K.}~\bibnamefont {Kohri}},\ }\href
  {\doibase 10.1103/PhysRevD.88.084051, 10.1103/PhysRevD.89.029903} {\bibfield
  {journal} {\bibinfo  {journal} {Phys. Rev.}\ }\textbf {\bibinfo {volume}
  {D88}},\ \bibinfo {pages} {084051} (\bibinfo {year} {2013})},\ \bibinfo
  {note} {[Erratum: Phys. Rev.D89,no.2,029903(2014)]},\ \Eprint
  {http://arxiv.org/abs/1309.4201} {arXiv:1309.4201 [astro-ph.CO]} \BibitemShut
  {NoStop}%
%%CITATION = ARXIV:1309.4201;%%
\bibitem [{\citenamefont {Germani}\ and\ \citenamefont
  {Musco}(2019)}]{Germani:2018jgr}%
  \BibitemOpen
  \bibfield  {author} {\bibinfo {author} {\bibfnamefont {C.}~\bibnamefont
  {Germani}}\ and\ \bibinfo {author} {\bibfnamefont {I.}~\bibnamefont
  {Musco}},\ }\href {\doibase 10.1103/PhysRevLett.122.141302} {\bibfield
  {journal} {\bibinfo  {journal} {Phys. Rev. Lett.}\ }\textbf {\bibinfo
  {volume} {122}},\ \bibinfo {pages} {141302} (\bibinfo {year} {2019})},\
  \Eprint {http://arxiv.org/abs/1805.04087} {arXiv:1805.04087 [astro-ph.CO]}
  \BibitemShut {NoStop}%
%%CITATION = ARXIV:1805.04087;%%
\bibitem [{\citenamefont {Yoo}\ \emph {et~al.}(2019)\citenamefont {Yoo},
  \citenamefont {Gong},\ and\ \citenamefont {Yokoyama}}]{Yoo:2019pma}%
  \BibitemOpen
  \bibfield  {author} {\bibinfo {author} {\bibfnamefont {C.-M.}\ \bibnamefont
  {Yoo}}, \bibinfo {author} {\bibfnamefont {J.-O.}\ \bibnamefont {Gong}}, \
  and\ \bibinfo {author} {\bibfnamefont {S.}~\bibnamefont {Yokoyama}},\ }\href
  {\doibase 10.1088/1475-7516/2019/09/033} {\bibfield  {journal} {\bibinfo
  {journal} {JCAP}\ }\textbf {\bibinfo {volume} {1909}},\ \bibinfo {pages}
  {033} (\bibinfo {year} {2019})},\ \Eprint {http://arxiv.org/abs/1906.06790}
  {arXiv:1906.06790 [astro-ph.CO]} \BibitemShut {NoStop}%
%%CITATION = ARXIV:1906.06790;%%
\bibitem [{\citenamefont {Musco}(2019)}]{Musco:2018rwt}%
  \BibitemOpen
  \bibfield  {author} {\bibinfo {author} {\bibfnamefont {I.}~\bibnamefont
  {Musco}},\ }\href {\doibase 10.1103/PhysRevD.100.123524} {\bibfield
  {journal} {\bibinfo  {journal} {Phys. Rev.}\ }\textbf {\bibinfo {volume}
  {D100}},\ \bibinfo {pages} {123524} (\bibinfo {year} {2019})},\ \Eprint
  {http://arxiv.org/abs/1809.02127} {arXiv:1809.02127 [gr-qc]} \BibitemShut
  {NoStop}%
%%CITATION = ARXIV:1809.02127;%%
\bibitem [{\citenamefont {Carr}\ \emph {et~al.}(2017)\citenamefont {Carr},
  \citenamefont {Raidal}, \citenamefont {Tenkanen}, \citenamefont {Vaskonen},\
  and\ \citenamefont {Veermäe}}]{Carr:2017jsz}%
  \BibitemOpen
  \bibfield  {author} {\bibinfo {author} {\bibfnamefont {B.}~\bibnamefont
  {Carr}}, \bibinfo {author} {\bibfnamefont {M.}~\bibnamefont {Raidal}},
  \bibinfo {author} {\bibfnamefont {T.}~\bibnamefont {Tenkanen}}, \bibinfo
  {author} {\bibfnamefont {V.}~\bibnamefont {Vaskonen}}, \ and\ \bibinfo
  {author} {\bibfnamefont {H.}~\bibnamefont {Veermäe}},\ }\href {\doibase
  10.1103/PhysRevD.96.023514} {\bibfield  {journal} {\bibinfo  {journal} {Phys.
  Rev.}\ }\textbf {\bibinfo {volume} {D96}},\ \bibinfo {pages} {023514}
  (\bibinfo {year} {2017})},\ \Eprint {http://arxiv.org/abs/1705.05567}
  {arXiv:1705.05567 [astro-ph.CO]} \BibitemShut {NoStop}%
%%CITATION = ARXIV:1705.05567;%%
\bibitem [{\citenamefont {Montero-Camacho}\ \emph {et~al.}(2019)\citenamefont
  {Montero-Camacho}, \citenamefont {Fang}, \citenamefont {Vasquez},
  \citenamefont {Silva},\ and\ \citenamefont
  {Hirata}}]{Montero-Camacho:2019jte}%
  \BibitemOpen
  \bibfield  {author} {\bibinfo {author} {\bibfnamefont {P.}~\bibnamefont
  {Montero-Camacho}}, \bibinfo {author} {\bibfnamefont {X.}~\bibnamefont
  {Fang}}, \bibinfo {author} {\bibfnamefont {G.}~\bibnamefont {Vasquez}},
  \bibinfo {author} {\bibfnamefont {M.}~\bibnamefont {Silva}}, \ and\ \bibinfo
  {author} {\bibfnamefont {C.~M.}\ \bibnamefont {Hirata}},\ }\href {\doibase
  10.1088/1475-7516/2019/08/031} {\bibfield  {journal} {\bibinfo  {journal}
  {JCAP}\ }\textbf {\bibinfo {volume} {1908}},\ \bibinfo {pages} {031}
  (\bibinfo {year} {2019})},\ \Eprint {http://arxiv.org/abs/1906.05950}
  {arXiv:1906.05950 [astro-ph.CO]} \BibitemShut {NoStop}%
%%CITATION = ARXIV:1906.05950;%%
\bibitem [{\citenamefont {Jung}\ and\ \citenamefont
  {Kim}(2019)}]{Jung:2019fcs}%
  \BibitemOpen
  \bibfield  {author} {\bibinfo {author} {\bibfnamefont {S.}~\bibnamefont
  {Jung}}\ and\ \bibinfo {author} {\bibfnamefont {T.}~\bibnamefont {Kim}},\
  }\href@noop {} {\  (\bibinfo {year} {2019})},\ \Eprint
  {http://arxiv.org/abs/1908.00078} {arXiv:1908.00078 [astro-ph.CO]}
  \BibitemShut {NoStop}%
%%CITATION = ARXIV:1908.00078;%%
\bibitem [{\citenamefont {Dasgupta}\ \emph {et~al.}(2019)\citenamefont
  {Dasgupta}, \citenamefont {Laha},\ and\ \citenamefont
  {Ray}}]{Dasgupta:2019cae}%
  \BibitemOpen
  \bibfield  {author} {\bibinfo {author} {\bibfnamefont {B.}~\bibnamefont
  {Dasgupta}}, \bibinfo {author} {\bibfnamefont {R.}~\bibnamefont {Laha}}, \
  and\ \bibinfo {author} {\bibfnamefont {A.}~\bibnamefont {Ray}},\ }\href@noop
  {} {\  (\bibinfo {year} {2019})},\ \Eprint {http://arxiv.org/abs/1912.01014}
  {arXiv:1912.01014 [hep-ph]} \BibitemShut {NoStop}%
%%CITATION = ARXIV:1912.01014;%%
\bibitem [{\citenamefont {Rasanen}\ and\ \citenamefont
  {Tomberg}(2019)}]{Rasanen:2018fom}%
  \BibitemOpen
  \bibfield  {author} {\bibinfo {author} {\bibfnamefont {S.}~\bibnamefont
  {Rasanen}}\ and\ \bibinfo {author} {\bibfnamefont {E.}~\bibnamefont
  {Tomberg}},\ }\href {\doibase 10.1088/1475-7516/2019/01/038} {\bibfield
  {journal} {\bibinfo  {journal} {JCAP}\ }\textbf {\bibinfo {volume} {1901}},\
  \bibinfo {pages} {038} (\bibinfo {year} {2019})},\ \Eprint
  {http://arxiv.org/abs/1810.12608} {arXiv:1810.12608 [astro-ph.CO]}
  \BibitemShut {NoStop}%
%%CITATION = ARXIV:1810.12608;%%
\end{thebibliography}%


%merlin.mbs apsrev4-1.bst 2010-07-25 4.21a (PWD, AO, DPC) hacked
%Control: key (0)
%Control: author (8) initials jnrlst
%Control: editor formatted (1) identically to author
%Control: production of article title (-1) disabled
%Control: page (0) single
%Control: year (1) truncated
%Control: production of eprint (0) enabled
\providecommand{\noopsort}[1]{}\providecommand{\singleletter}[1]{#1}%
\begin{thebibliography}{69}%
\makeatletter
\providecommand \@ifxundefined [1]{%
 \@ifx{#1\undefined}
}%
\providecommand \@ifnum [1]{%
 \ifnum #1\expandafter \@firstoftwo
 \else \expandafter \@secondoftwo
 \fi
}%
\providecommand \@ifx [1]{%
 \ifx #1\expandafter \@firstoftwo
 \else \expandafter \@secondoftwo
 \fi
}%
\providecommand \natexlab [1]{#1}%
\providecommand \enquote  [1]{``#1''}%
\providecommand \bibnamefont  [1]{#1}%
\providecommand \bibfnamefont [1]{#1}%
\providecommand \citenamefont [1]{#1}%
\providecommand \href@noop [0]{\@secondoftwo}%
\providecommand \href [0]{\begingroup \@sanitize@url \@href}%
\providecommand \@href[1]{\@@startlink{#1}\@@href}%
\providecommand \@@href[1]{\endgroup#1\@@endlink}%
\providecommand \@sanitize@url [0]{\catcode `\\12\catcode `\$12\catcode
  `\&12\catcode `\#12\catcode `\^12\catcode `\_12\catcode `\%12\relax}%
\providecommand \@@startlink[1]{}%
\providecommand \@@endlink[0]{}%
\providecommand \url  [0]{\begingroup\@sanitize@url \@url }%
\providecommand \@url [1]{\endgroup\@href {#1}{\urlprefix }}%
\providecommand \urlprefix  [0]{URL }%
\providecommand \Eprint [0]{\href }%
\providecommand \doibase [0]{http://dx.doi.org/}%
\providecommand \selectlanguage [0]{\@gobble}%
\providecommand \bibinfo  [0]{\@secondoftwo}%
\providecommand \bibfield  [0]{\@secondoftwo}%
\providecommand \translation [1]{[#1]}%
\providecommand \BibitemOpen [0]{}%
\providecommand \bibitemStop [0]{}%
\providecommand \bibitemNoStop [0]{.\EOS\space}%
\providecommand \EOS [0]{\spacefactor3000\relax}%
\providecommand \BibitemShut  [1]{\csname bibitem#1\endcsname}%
\let\auto@bib@innerbib\@empty
%</preamble>
\bibitem [{\citenamefont {Bezrukov}\ and\ \citenamefont
  {Shaposhnikov}(2008)}]{Bezrukov:2007ep}%
  \BibitemOpen
  \bibfield  {author} {\bibinfo {author} {\bibfnamefont {F.~L.}\ \bibnamefont
  {Bezrukov}}\ and\ \bibinfo {author} {\bibfnamefont {M.}~\bibnamefont
  {Shaposhnikov}},\ }\href {\doibase 10.1016/j.physletb.2007.11.072} {\bibfield
   {journal} {\bibinfo  {journal} {Phys. Lett.}\ }\textbf {\bibinfo {volume}
  {B659}},\ \bibinfo {pages} {703} (\bibinfo {year} {2008})},\ \Eprint
  {http://arxiv.org/abs/0710.3755} {arXiv:0710.3755 [hep-th]} \BibitemShut
  {NoStop}%
%%CITATION = ARXIV:0710.3755;%%
\bibitem [{\citenamefont {Starobinsky}(1980)}]{Starobinsky:1980te}%
  \BibitemOpen
  \bibfield  {author} {\bibinfo {author} {\bibfnamefont {A.~A.}\ \bibnamefont
  {Starobinsky}},\ }\href {\doibase 10.1016/0370-2693(80)90670-X} {\bibfield
  {journal} {\bibinfo  {journal} {Phys. Lett.}\ }\textbf {\bibinfo {volume}
  {91B}},\ \bibinfo {pages} {99} (\bibinfo {year} {1980})},\ \bibinfo {note}
  {[,771(1980)]}\BibitemShut {NoStop}%
%%CITATION = PHLTA,91B,99;%%
\bibitem [{\citenamefont {Aghanim}\ \emph {et~al.}(2018)\citenamefont {Aghanim}
  \emph {et~al.}}]{Aghanim:2018eyx}%
  \BibitemOpen
  \bibfield  {author} {\bibinfo {author} {\bibfnamefont {N.}~\bibnamefont
  {Aghanim}} \emph {et~al.} (\bibinfo {collaboration} {Planck}),\ }\href@noop
  {} {\  (\bibinfo {year} {2018})},\ \Eprint {http://arxiv.org/abs/1807.06209}
  {arXiv:1807.06209 [astro-ph.CO]} \BibitemShut {NoStop}%
%%CITATION = ARXIV:1807.06209;%%
\bibitem [{\citenamefont {Akrami}\ \emph {et~al.}(2018)\citenamefont {Akrami}
  \emph {et~al.}}]{Akrami:2018odb}%
  \BibitemOpen
  \bibfield  {author} {\bibinfo {author} {\bibfnamefont {Y.}~\bibnamefont
  {Akrami}} \emph {et~al.} (\bibinfo {collaboration} {Planck}),\ }\href@noop {}
  {\  (\bibinfo {year} {2018})},\ \Eprint {http://arxiv.org/abs/1807.06211}
  {arXiv:1807.06211 [astro-ph.CO]} \BibitemShut {NoStop}%
%%CITATION = ARXIV:1807.06211;%%
\bibitem [{\citenamefont {Park}\ and\ \citenamefont
  {Yamaguchi}(2008)}]{Park:2008hz}%
  \BibitemOpen
  \bibfield  {author} {\bibinfo {author} {\bibfnamefont {S.~C.}\ \bibnamefont
  {Park}}\ and\ \bibinfo {author} {\bibfnamefont {S.}~\bibnamefont
  {Yamaguchi}},\ }\href {\doibase 10.1088/1475-7516/2008/08/009} {\bibfield
  {journal} {\bibinfo  {journal} {JCAP}\ }\textbf {\bibinfo {volume} {0808}},\
  \bibinfo {pages} {009} (\bibinfo {year} {2008})},\ \Eprint
  {http://arxiv.org/abs/0801.1722} {arXiv:0801.1722 [hep-ph]} \BibitemShut
  {NoStop}%
%%CITATION = ARXIV:0801.1722;%%
\bibitem [{\citenamefont {Burgess}\ \emph {et~al.}(2009)\citenamefont
  {Burgess}, \citenamefont {Lee},\ and\ \citenamefont
  {Trott}}]{Burgess:2009ea}%
  \BibitemOpen
  \bibfield  {author} {\bibinfo {author} {\bibfnamefont {C.~P.}\ \bibnamefont
  {Burgess}}, \bibinfo {author} {\bibfnamefont {H.~M.}\ \bibnamefont {Lee}}, \
  and\ \bibinfo {author} {\bibfnamefont {M.}~\bibnamefont {Trott}},\ }\href
  {\doibase 10.1088/1126-6708/2009/09/103} {\bibfield  {journal} {\bibinfo
  {journal} {JHEP}\ }\textbf {\bibinfo {volume} {09}},\ \bibinfo {pages} {103}
  (\bibinfo {year} {2009})},\ \Eprint {http://arxiv.org/abs/0902.4465}
  {arXiv:0902.4465 [hep-ph]} \BibitemShut {NoStop}%
%%CITATION = ARXIV:0902.4465;%%
\bibitem [{\citenamefont {Barbon}\ and\ \citenamefont
  {Espinosa}(2009)}]{Barbon:2009ya}%
  \BibitemOpen
  \bibfield  {author} {\bibinfo {author} {\bibfnamefont {J.~L.~F.}\
  \bibnamefont {Barbon}}\ and\ \bibinfo {author} {\bibfnamefont {J.~R.}\
  \bibnamefont {Espinosa}},\ }\href {\doibase 10.1103/PhysRevD.79.081302}
  {\bibfield  {journal} {\bibinfo  {journal} {Phys. Rev.}\ }\textbf {\bibinfo
  {volume} {D79}},\ \bibinfo {pages} {081302} (\bibinfo {year} {2009})},\
  \Eprint {http://arxiv.org/abs/0903.0355} {arXiv:0903.0355 [hep-ph]}
  \BibitemShut {NoStop}%
%%CITATION = ARXIV:0903.0355;%%
\bibitem [{\citenamefont {Burgess}\ \emph {et~al.}(2010)\citenamefont
  {Burgess}, \citenamefont {Lee},\ and\ \citenamefont
  {Trott}}]{Burgess:2010zq}%
  \BibitemOpen
  \bibfield  {author} {\bibinfo {author} {\bibfnamefont {C.~P.}\ \bibnamefont
  {Burgess}}, \bibinfo {author} {\bibfnamefont {H.~M.}\ \bibnamefont {Lee}}, \
  and\ \bibinfo {author} {\bibfnamefont {M.}~\bibnamefont {Trott}},\ }\href
  {\doibase 10.1007/JHEP07(2010)007} {\bibfield  {journal} {\bibinfo  {journal}
  {JHEP}\ }\textbf {\bibinfo {volume} {07}},\ \bibinfo {pages} {007} (\bibinfo
  {year} {2010})},\ \Eprint {http://arxiv.org/abs/1002.2730} {arXiv:1002.2730
  [hep-ph]} \BibitemShut {NoStop}%
%%CITATION = ARXIV:1002.2730;%%
\bibitem [{\citenamefont {Lerner}\ and\ \citenamefont
  {McDonald}(2010)}]{Lerner:2009na}%
  \BibitemOpen
  \bibfield  {author} {\bibinfo {author} {\bibfnamefont {R.~N.}\ \bibnamefont
  {Lerner}}\ and\ \bibinfo {author} {\bibfnamefont {J.}~\bibnamefont
  {McDonald}},\ }\href {\doibase 10.1088/1475-7516/2010/04/015} {\bibfield
  {journal} {\bibinfo  {journal} {JCAP}\ }\textbf {\bibinfo {volume} {1004}},\
  \bibinfo {pages} {015} (\bibinfo {year} {2010})},\ \Eprint
  {http://arxiv.org/abs/0912.5463} {arXiv:0912.5463 [hep-ph]} \BibitemShut
  {NoStop}%
%%CITATION = ARXIV:0912.5463;%%
\bibitem [{\citenamefont {Park}\ and\ \citenamefont
  {Shin}(2019)}]{Park:2018kst}%
  \BibitemOpen
  \bibfield  {author} {\bibinfo {author} {\bibfnamefont {S.~C.}\ \bibnamefont
  {Park}}\ and\ \bibinfo {author} {\bibfnamefont {C.~S.}\ \bibnamefont
  {Shin}},\ }\href {\doibase 10.1140/epjc/s10052-019-7037-4} {\bibfield
  {journal} {\bibinfo  {journal} {Eur. Phys. J.}\ }\textbf {\bibinfo {volume}
  {C79}},\ \bibinfo {pages} {529} (\bibinfo {year} {2019})},\ \Eprint
  {http://arxiv.org/abs/1807.09952} {arXiv:1807.09952 [hep-ph]} \BibitemShut
  {NoStop}%
%%CITATION = ARXIV:1807.09952;%%
\bibitem [{\citenamefont {Bezrukov}\ \emph {et~al.}(2011)\citenamefont
  {Bezrukov}, \citenamefont {Magnin}, \citenamefont {Shaposhnikov},\ and\
  \citenamefont {Sibiryakov}}]{Bezrukov:2010jz}%
  \BibitemOpen
  \bibfield  {author} {\bibinfo {author} {\bibfnamefont {F.}~\bibnamefont
  {Bezrukov}}, \bibinfo {author} {\bibfnamefont {A.}~\bibnamefont {Magnin}},
  \bibinfo {author} {\bibfnamefont {M.}~\bibnamefont {Shaposhnikov}}, \ and\
  \bibinfo {author} {\bibfnamefont {S.}~\bibnamefont {Sibiryakov}},\ }\href
  {\doibase 10.1007/JHEP01(2011)016} {\bibfield  {journal} {\bibinfo  {journal}
  {JHEP}\ }\textbf {\bibinfo {volume} {01}},\ \bibinfo {pages} {016} (\bibinfo
  {year} {2011})},\ \Eprint {http://arxiv.org/abs/1008.5157} {arXiv:1008.5157
  [hep-ph]} \BibitemShut {NoStop}%
%%CITATION = ARXIV:1008.5157;%%
\bibitem [{\citenamefont {Hamada}\ \emph {et~al.}(2014)\citenamefont {Hamada},
  \citenamefont {Kawai}, \citenamefont {Oda},\ and\ \citenamefont
  {Park}}]{Hamada:2014iga}%
  \BibitemOpen
  \bibfield  {author} {\bibinfo {author} {\bibfnamefont {Y.}~\bibnamefont
  {Hamada}}, \bibinfo {author} {\bibfnamefont {H.}~\bibnamefont {Kawai}},
  \bibinfo {author} {\bibfnamefont {K.-y.}\ \bibnamefont {Oda}}, \ and\
  \bibinfo {author} {\bibfnamefont {S.~C.}\ \bibnamefont {Park}},\ }\href
  {\doibase 10.1103/PhysRevLett.112.241301} {\bibfield  {journal} {\bibinfo
  {journal} {Phys. Rev. Lett.}\ }\textbf {\bibinfo {volume} {112}},\ \bibinfo
  {pages} {241301} (\bibinfo {year} {2014})},\ \Eprint
  {http://arxiv.org/abs/1403.5043} {arXiv:1403.5043 [hep-ph]} \BibitemShut
  {NoStop}%
%%CITATION = ARXIV:1403.5043;%%
\bibitem [{\citenamefont {Bezrukov}\ and\ \citenamefont
  {Shaposhnikov}(2014)}]{Bezrukov:2014bra}%
  \BibitemOpen
  \bibfield  {author} {\bibinfo {author} {\bibfnamefont {F.}~\bibnamefont
  {Bezrukov}}\ and\ \bibinfo {author} {\bibfnamefont {M.}~\bibnamefont
  {Shaposhnikov}},\ }\href {\doibase 10.1016/j.physletb.2014.05.074} {\bibfield
   {journal} {\bibinfo  {journal} {Phys. Lett.}\ }\textbf {\bibinfo {volume}
  {B734}},\ \bibinfo {pages} {249} (\bibinfo {year} {2014})},\ \Eprint
  {http://arxiv.org/abs/1403.6078} {arXiv:1403.6078 [hep-ph]} \BibitemShut
  {NoStop}%
%%CITATION = ARXIV:1403.6078;%%
\bibitem [{\citenamefont {Hamada}\ \emph {et~al.}(2015)\citenamefont {Hamada},
  \citenamefont {Kawai}, \citenamefont {Oda},\ and\ \citenamefont
  {Park}}]{Hamada:2014wna}%
  \BibitemOpen
  \bibfield  {author} {\bibinfo {author} {\bibfnamefont {Y.}~\bibnamefont
  {Hamada}}, \bibinfo {author} {\bibfnamefont {H.}~\bibnamefont {Kawai}},
  \bibinfo {author} {\bibfnamefont {K.-y.}\ \bibnamefont {Oda}}, \ and\
  \bibinfo {author} {\bibfnamefont {S.~C.}\ \bibnamefont {Park}},\ }\href
  {\doibase 10.1103/PhysRevD.91.053008} {\bibfield  {journal} {\bibinfo
  {journal} {Phys. Rev.}\ }\textbf {\bibinfo {volume} {D91}},\ \bibinfo {pages}
  {053008} (\bibinfo {year} {2015})},\ \Eprint {http://arxiv.org/abs/1408.4864}
  {arXiv:1408.4864 [hep-ph]} \BibitemShut {NoStop}%
%%CITATION = ARXIV:1408.4864;%%
\bibitem [{\citenamefont {Giudice}\ and\ \citenamefont
  {Lee}(2011)}]{Giudice:2010ka}%
  \BibitemOpen
  \bibfield  {author} {\bibinfo {author} {\bibfnamefont {G.~F.}\ \bibnamefont
  {Giudice}}\ and\ \bibinfo {author} {\bibfnamefont {H.~M.}\ \bibnamefont
  {Lee}},\ }\href {\doibase 10.1016/j.physletb.2010.10.035} {\bibfield
  {journal} {\bibinfo  {journal} {Phys. Lett.}\ }\textbf {\bibinfo {volume}
  {B694}},\ \bibinfo {pages} {294} (\bibinfo {year} {2011})},\ \Eprint
  {http://arxiv.org/abs/1010.1417} {arXiv:1010.1417 [hep-ph]} \BibitemShut
  {NoStop}%
%%CITATION = ARXIV:1010.1417;%%
\bibitem [{\citenamefont {Barbon}\ \emph {et~al.}(2015)\citenamefont {Barbon},
  \citenamefont {Casas}, \citenamefont {Elias-Miro},\ and\ \citenamefont
  {Espinosa}}]{Barbon:2015fla}%
  \BibitemOpen
  \bibfield  {author} {\bibinfo {author} {\bibfnamefont {J.~L.~F.}\
  \bibnamefont {Barbon}}, \bibinfo {author} {\bibfnamefont {J.~A.}\
  \bibnamefont {Casas}}, \bibinfo {author} {\bibfnamefont {J.}~\bibnamefont
  {Elias-Miro}}, \ and\ \bibinfo {author} {\bibfnamefont {J.~R.}\ \bibnamefont
  {Espinosa}},\ }\href {\doibase 10.1007/JHEP09(2015)027} {\bibfield  {journal}
  {\bibinfo  {journal} {JHEP}\ }\textbf {\bibinfo {volume} {09}},\ \bibinfo
  {pages} {027} (\bibinfo {year} {2015})},\ \Eprint
  {http://arxiv.org/abs/1501.02231} {arXiv:1501.02231 [hep-ph]} \BibitemShut
  {NoStop}%
%%CITATION = ARXIV:1501.02231;%%
\bibitem [{\citenamefont {Giudice}\ and\ \citenamefont
  {Lee}(2014)}]{Giudice:2014toa}%
  \BibitemOpen
  \bibfield  {author} {\bibinfo {author} {\bibfnamefont {G.~F.}\ \bibnamefont
  {Giudice}}\ and\ \bibinfo {author} {\bibfnamefont {H.~M.}\ \bibnamefont
  {Lee}},\ }\href {\doibase 10.1016/j.physletb.2014.04.020} {\bibfield
  {journal} {\bibinfo  {journal} {Phys. Lett.}\ }\textbf {\bibinfo {volume}
  {B733}},\ \bibinfo {pages} {58} (\bibinfo {year} {2014})},\ \Eprint
  {http://arxiv.org/abs/1402.2129} {arXiv:1402.2129 [hep-ph]} \BibitemShut
  {NoStop}%
%%CITATION = ARXIV:1402.2129;%%
\bibitem [{\citenamefont {Ema}(2017)}]{Ema:2017rqn}%
  \BibitemOpen
  \bibfield  {author} {\bibinfo {author} {\bibfnamefont {Y.}~\bibnamefont
  {Ema}},\ }\href {\doibase 10.1016/j.physletb.2017.04.060} {\bibfield
  {journal} {\bibinfo  {journal} {Phys. Lett.}\ }\textbf {\bibinfo {volume}
  {B770}},\ \bibinfo {pages} {403} (\bibinfo {year} {2017})},\ \Eprint
  {http://arxiv.org/abs/1701.07665} {arXiv:1701.07665 [hep-ph]} \BibitemShut
  {NoStop}%
%%CITATION = ARXIV:1701.07665;%%
\bibitem [{\citenamefont {Gorbunov}\ and\ \citenamefont
  {Tokareva}(2019)}]{Gorbunov:2018llf}%
  \BibitemOpen
  \bibfield  {author} {\bibinfo {author} {\bibfnamefont {D.}~\bibnamefont
  {Gorbunov}}\ and\ \bibinfo {author} {\bibfnamefont {A.}~\bibnamefont
  {Tokareva}},\ }\href {\doibase 10.1016/j.physletb.2018.11.015} {\bibfield
  {journal} {\bibinfo  {journal} {Phys. Lett.}\ }\textbf {\bibinfo {volume}
  {B788}},\ \bibinfo {pages} {37} (\bibinfo {year} {2019})},\ \Eprint
  {http://arxiv.org/abs/1807.02392} {arXiv:1807.02392 [hep-ph]} \BibitemShut
  {NoStop}%
%%CITATION = ARXIV:1807.02392;%%
\bibitem [{\citenamefont {Salvio}\ and\ \citenamefont
  {Mazumdar}(2015)}]{Salvio:2015kka}%
  \BibitemOpen
  \bibfield  {author} {\bibinfo {author} {\bibfnamefont {A.}~\bibnamefont
  {Salvio}}\ and\ \bibinfo {author} {\bibfnamefont {A.}~\bibnamefont
  {Mazumdar}},\ }\href {\doibase 10.1016/j.physletb.2015.09.020} {\bibfield
  {journal} {\bibinfo  {journal} {Phys. Lett.}\ }\textbf {\bibinfo {volume}
  {B750}},\ \bibinfo {pages} {194} (\bibinfo {year} {2015})},\ \Eprint
  {http://arxiv.org/abs/1506.07520} {arXiv:1506.07520 [hep-ph]} \BibitemShut
  {NoStop}%
%%CITATION = ARXIV:1506.07520;%%
\bibitem [{\citenamefont {Calmet}\ and\ \citenamefont
  {Kuntz}(2016)}]{Calmet:2016fsr}%
  \BibitemOpen
  \bibfield  {author} {\bibinfo {author} {\bibfnamefont {X.}~\bibnamefont
  {Calmet}}\ and\ \bibinfo {author} {\bibfnamefont {I.}~\bibnamefont {Kuntz}},\
  }\href {\doibase 10.1140/epjc/s10052-016-4136-3} {\bibfield  {journal}
  {\bibinfo  {journal} {Eur. Phys. J.}\ }\textbf {\bibinfo {volume} {C76}},\
  \bibinfo {pages} {289} (\bibinfo {year} {2016})},\ \Eprint
  {http://arxiv.org/abs/1605.02236} {arXiv:1605.02236 [hep-th]} \BibitemShut
  {NoStop}%
%%CITATION = ARXIV:1605.02236;%%
\bibitem [{\citenamefont {Wang}\ and\ \citenamefont
  {Wang}(2017)}]{Wang:2017fuy}%
  \BibitemOpen
  \bibfield  {author} {\bibinfo {author} {\bibfnamefont {Y.-C.}\ \bibnamefont
  {Wang}}\ and\ \bibinfo {author} {\bibfnamefont {T.}~\bibnamefont {Wang}},\
  }\href {\doibase 10.1103/PhysRevD.96.123506} {\bibfield  {journal} {\bibinfo
  {journal} {Phys. Rev.}\ }\textbf {\bibinfo {volume} {D96}},\ \bibinfo {pages}
  {123506} (\bibinfo {year} {2017})},\ \Eprint
  {http://arxiv.org/abs/1701.06636} {arXiv:1701.06636 [gr-qc]} \BibitemShut
  {NoStop}%
%%CITATION = ARXIV:1701.06636;%%
\bibitem [{\citenamefont {Ghilencea}(2018)}]{Ghilencea:2018rqg}%
  \BibitemOpen
  \bibfield  {author} {\bibinfo {author} {\bibfnamefont {D.~M.}\ \bibnamefont
  {Ghilencea}},\ }\href {\doibase 10.1103/PhysRevD.98.103524} {\bibfield
  {journal} {\bibinfo  {journal} {Phys. Rev.}\ }\textbf {\bibinfo {volume}
  {D98}},\ \bibinfo {pages} {103524} (\bibinfo {year} {2018})},\ \Eprint
  {http://arxiv.org/abs/1807.06900} {arXiv:1807.06900 [hep-ph]} \BibitemShut
  {NoStop}%
%%CITATION = ARXIV:1807.06900;%%
\bibitem [{\citenamefont {Ema}(2019)}]{Ema:2019fdd}%
  \BibitemOpen
  \bibfield  {author} {\bibinfo {author} {\bibfnamefont {Y.}~\bibnamefont
  {Ema}},\ }\href {\doibase 10.1088/1475-7516/2019/09/027} {\bibfield
  {journal} {\bibinfo  {journal} {JCAP}\ }\textbf {\bibinfo {volume} {1909}},\
  \bibinfo {pages} {027} (\bibinfo {year} {2019})},\ \Eprint
  {http://arxiv.org/abs/1907.00993} {arXiv:1907.00993 [hep-ph]} \BibitemShut
  {NoStop}%
%%CITATION = ARXIV:1907.00993;%%
\bibitem [{\citenamefont {Canko}\ \emph {et~al.}(2019)\citenamefont {Canko},
  \citenamefont {Gialamas},\ and\ \citenamefont {Kodaxis}}]{Canko:2019mud}%
  \BibitemOpen
  \bibfield  {author} {\bibinfo {author} {\bibfnamefont {D.~D.}\ \bibnamefont
  {Canko}}, \bibinfo {author} {\bibfnamefont {I.~D.}\ \bibnamefont {Gialamas}},
  \ and\ \bibinfo {author} {\bibfnamefont {G.~P.}\ \bibnamefont {Kodaxis}},\
  }\href@noop {} {\  (\bibinfo {year} {2019})},\ \Eprint
  {http://arxiv.org/abs/1901.06296} {arXiv:1901.06296 [hep-th]} \BibitemShut
  {NoStop}%
%%CITATION = ARXIV:1901.06296;%%
\bibitem [{\citenamefont {He}\ \emph {et~al.}(2019)\citenamefont {He},
  \citenamefont {Jinno}, \citenamefont {Kamada}, \citenamefont {Park},
  \citenamefont {Starobinsky},\ and\ \citenamefont {Yokoyama}}]{He:2018mgb}%
  \BibitemOpen
  \bibfield  {author} {\bibinfo {author} {\bibfnamefont {M.}~\bibnamefont
  {He}}, \bibinfo {author} {\bibfnamefont {R.}~\bibnamefont {Jinno}}, \bibinfo
  {author} {\bibfnamefont {K.}~\bibnamefont {Kamada}}, \bibinfo {author}
  {\bibfnamefont {S.~C.}\ \bibnamefont {Park}}, \bibinfo {author}
  {\bibfnamefont {A.~A.}\ \bibnamefont {Starobinsky}}, \ and\ \bibinfo {author}
  {\bibfnamefont {J.}~\bibnamefont {Yokoyama}},\ }\href {\doibase
  10.1016/j.physletb.2019.02.008} {\bibfield  {journal} {\bibinfo  {journal}
  {Phys. Lett.}\ }\textbf {\bibinfo {volume} {B791}},\ \bibinfo {pages} {36}
  (\bibinfo {year} {2019})},\ \Eprint {http://arxiv.org/abs/1812.10099}
  {arXiv:1812.10099 [hep-ph]} \BibitemShut {NoStop}%
%%CITATION = ARXIV:1812.10099;%%
\bibitem [{\citenamefont {He}\ \emph {et~al.}(2018)\citenamefont {He},
  \citenamefont {Starobinsky},\ and\ \citenamefont {Yokoyama}}]{He:2018gyf}%
  \BibitemOpen
  \bibfield  {author} {\bibinfo {author} {\bibfnamefont {M.}~\bibnamefont
  {He}}, \bibinfo {author} {\bibfnamefont {A.~A.}\ \bibnamefont {Starobinsky}},
  \ and\ \bibinfo {author} {\bibfnamefont {J.}~\bibnamefont {Yokoyama}},\
  }\href {\doibase 10.1088/1475-7516/2018/05/064} {\bibfield  {journal}
  {\bibinfo  {journal} {JCAP}\ }\textbf {\bibinfo {volume} {1805}},\ \bibinfo
  {pages} {064} (\bibinfo {year} {2018})},\ \Eprint
  {http://arxiv.org/abs/1804.00409} {arXiv:1804.00409 [astro-ph.CO]}
  \BibitemShut {NoStop}%
%%CITATION = ARXIV:1804.00409;%%
\bibitem [{\citenamefont {Gundhi}\ and\ \citenamefont
  {Steinwachs}(2018)}]{Gundhi:2018wyz}%
  \BibitemOpen
  \bibfield  {author} {\bibinfo {author} {\bibfnamefont {A.}~\bibnamefont
  {Gundhi}}\ and\ \bibinfo {author} {\bibfnamefont {C.~F.}\ \bibnamefont
  {Steinwachs}},\ }\href@noop {} {\  (\bibinfo {year} {2018})},\ \Eprint
  {http://arxiv.org/abs/1810.10546} {arXiv:1810.10546 [hep-th]} \BibitemShut
  {NoStop}%
%%CITATION = ARXIV:1810.10546;%%
\bibitem [{\citenamefont {Ema}\ \emph {et~al.}(2017)\citenamefont {Ema},
  \citenamefont {Jinno}, \citenamefont {Mukaida},\ and\ \citenamefont
  {Nakayama}}]{Ema:2016dny}%
  \BibitemOpen
  \bibfield  {author} {\bibinfo {author} {\bibfnamefont {Y.}~\bibnamefont
  {Ema}}, \bibinfo {author} {\bibfnamefont {R.}~\bibnamefont {Jinno}}, \bibinfo
  {author} {\bibfnamefont {K.}~\bibnamefont {Mukaida}}, \ and\ \bibinfo
  {author} {\bibfnamefont {K.}~\bibnamefont {Nakayama}},\ }\href {\doibase
  10.1088/1475-7516/2017/02/045} {\bibfield  {journal} {\bibinfo  {journal}
  {JCAP}\ }\textbf {\bibinfo {volume} {1702}},\ \bibinfo {pages} {045}
  (\bibinfo {year} {2017})},\ \Eprint {http://arxiv.org/abs/1609.05209}
  {arXiv:1609.05209 [hep-ph]} \BibitemShut {NoStop}%
%%CITATION = ARXIV:1609.05209;%%
\bibitem [{\citenamefont {Jinno}\ \emph {et~al.}(2020)\citenamefont {Jinno},
  \citenamefont {Kubota}, \citenamefont {Oda},\ and\ \citenamefont
  {Park}}]{Jinno:2019und}%
  \BibitemOpen
  \bibfield  {author} {\bibinfo {author} {\bibfnamefont {R.}~\bibnamefont
  {Jinno}}, \bibinfo {author} {\bibfnamefont {M.}~\bibnamefont {Kubota}},
  \bibinfo {author} {\bibfnamefont {K.-y.}\ \bibnamefont {Oda}}, \ and\
  \bibinfo {author} {\bibfnamefont {S.~C.}\ \bibnamefont {Park}},\ }\href
  {\doibase 10.1088/1475-7516/2020/03/063} {\bibfield  {journal} {\bibinfo
  {journal} {JCAP}\ }\textbf {\bibinfo {volume} {2003}},\ \bibinfo {pages}
  {063} (\bibinfo {year} {2020})},\ \Eprint {http://arxiv.org/abs/1904.05699}
  {arXiv:1904.05699 [hep-ph]} \BibitemShut {NoStop}%
%%CITATION = ARXIV:1904.05699;%%
\bibitem [{\citenamefont {Cheong}\ \emph
  {et~al.}(2019{\natexlab{a}})\citenamefont {Cheong}, \citenamefont {Lee},\
  and\ \citenamefont {Park}}]{Cheong:2018udx}%
  \BibitemOpen
  \bibfield  {author} {\bibinfo {author} {\bibfnamefont {D.~Y.}\ \bibnamefont
  {Cheong}}, \bibinfo {author} {\bibfnamefont {S.~M.}\ \bibnamefont {Lee}}, \
  and\ \bibinfo {author} {\bibfnamefont {S.~C.}\ \bibnamefont {Park}},\ }\href
  {\doibase 10.1016/j.physletb.2018.12.046} {\bibfield  {journal} {\bibinfo
  {journal} {Phys. Lett.}\ }\textbf {\bibinfo {volume} {B789}},\ \bibinfo
  {pages} {336} (\bibinfo {year} {2019}{\natexlab{a}})},\ \Eprint
  {http://arxiv.org/abs/1811.03622} {arXiv:1811.03622 [hep-ph]} \BibitemShut
  {NoStop}%
%%CITATION = ARXIV:1811.03622;%%
\bibitem [{\citenamefont {Park}(2019)}]{Park:2018fuj}%
  \BibitemOpen
  \bibfield  {author} {\bibinfo {author} {\bibfnamefont {S.~C.}\ \bibnamefont
  {Park}},\ }\href {\doibase 10.1088/1475-7516/2019/01/053} {\bibfield
  {journal} {\bibinfo  {journal} {JCAP}\ }\textbf {\bibinfo {volume} {1901}},\
  \bibinfo {pages} {053} (\bibinfo {year} {2019})},\ \Eprint
  {http://arxiv.org/abs/1810.11279} {arXiv:1810.11279 [hep-ph]} \BibitemShut
  {NoStop}%
%%CITATION = ARXIV:1810.11279;%%
\bibitem [{\citenamefont {Degrassi}\ \emph {et~al.}(2012)\citenamefont
  {Degrassi}, \citenamefont {Di~Vita}, \citenamefont {Elias-Miro},
  \citenamefont {Espinosa}, \citenamefont {Giudice}, \citenamefont {Isidori},\
  and\ \citenamefont {Strumia}}]{Degrassi:2012ry}%
  \BibitemOpen
  \bibfield  {author} {\bibinfo {author} {\bibfnamefont {G.}~\bibnamefont
  {Degrassi}}, \bibinfo {author} {\bibfnamefont {S.}~\bibnamefont {Di~Vita}},
  \bibinfo {author} {\bibfnamefont {J.}~\bibnamefont {Elias-Miro}}, \bibinfo
  {author} {\bibfnamefont {J.~R.}\ \bibnamefont {Espinosa}}, \bibinfo {author}
  {\bibfnamefont {G.~F.}\ \bibnamefont {Giudice}}, \bibinfo {author}
  {\bibfnamefont {G.}~\bibnamefont {Isidori}}, \ and\ \bibinfo {author}
  {\bibfnamefont {A.}~\bibnamefont {Strumia}},\ }\href {\doibase
  10.1007/JHEP08(2012)098} {\bibfield  {journal} {\bibinfo  {journal} {JHEP}\
  }\textbf {\bibinfo {volume} {08}},\ \bibinfo {pages} {098} (\bibinfo {year}
  {2012})},\ \Eprint {http://arxiv.org/abs/1205.6497} {arXiv:1205.6497
  [hep-ph]} \BibitemShut {NoStop}%
%%CITATION = ARXIV:1205.6497;%%
\bibitem [{\citenamefont {Bezrukov}\ \emph {et~al.}(2015)\citenamefont
  {Bezrukov}, \citenamefont {Rubio},\ and\ \citenamefont
  {Shaposhnikov}}]{Bezrukov:2014ipa}%
  \BibitemOpen
  \bibfield  {author} {\bibinfo {author} {\bibfnamefont {F.}~\bibnamefont
  {Bezrukov}}, \bibinfo {author} {\bibfnamefont {J.}~\bibnamefont {Rubio}}, \
  and\ \bibinfo {author} {\bibfnamefont {M.}~\bibnamefont {Shaposhnikov}},\
  }\href {\doibase 10.1103/PhysRevD.92.083512} {\bibfield  {journal} {\bibinfo
  {journal} {Phys. Rev. D}\ }\textbf {\bibinfo {volume} {92}},\ \bibinfo
  {pages} {083512} (\bibinfo {year} {2015})},\ \Eprint
  {http://arxiv.org/abs/1412.3811} {arXiv:1412.3811 [hep-ph]} \BibitemShut
  {NoStop}%
\bibitem [{\citenamefont {Markkanen}\ \emph
  {et~al.}(2018{\natexlab{a}})\citenamefont {Markkanen}, \citenamefont
  {Rajantie},\ and\ \citenamefont {Stopyra}}]{Markkanen:2018pdo}%
  \BibitemOpen
  \bibfield  {author} {\bibinfo {author} {\bibfnamefont {T.}~\bibnamefont
  {Markkanen}}, \bibinfo {author} {\bibfnamefont {A.}~\bibnamefont {Rajantie}},
  \ and\ \bibinfo {author} {\bibfnamefont {S.}~\bibnamefont {Stopyra}},\ }\href
  {\doibase 10.3389/fspas.2018.00040} {\bibfield  {journal} {\bibinfo
  {journal} {Front. Astron. Space Sci.}\ }\textbf {\bibinfo {volume} {5}},\
  \bibinfo {pages} {40} (\bibinfo {year} {2018}{\natexlab{a}})},\ \Eprint
  {http://arxiv.org/abs/1809.06923} {arXiv:1809.06923 [astro-ph.CO]}
  \BibitemShut {NoStop}%
\bibitem [{\citenamefont {Corcella}(2019)}]{Corcella:2019tgt}%
  \BibitemOpen
  \bibfield  {author} {\bibinfo {author} {\bibfnamefont {G.}~\bibnamefont
  {Corcella}},\ }\href {\doibase 10.3389/fphy.2019.00054} {\bibfield  {journal}
  {\bibinfo  {journal} {Front. in Phys.}\ }\textbf {\bibinfo {volume} {7}},\
  \bibinfo {pages} {54} (\bibinfo {year} {2019})},\ \Eprint
  {http://arxiv.org/abs/1903.06574} {arXiv:1903.06574 [hep-ph]} \BibitemShut
  {NoStop}%
\bibitem [{\citenamefont {Zyla}\ \emph {et~al.}(2020)\citenamefont {Zyla} \emph
  {et~al.}}]{Zyla:2020zbs}%
  \BibitemOpen
  \bibfield  {author} {\bibinfo {author} {\bibfnamefont {P.}~\bibnamefont
  {Zyla}} \emph {et~al.} (\bibinfo {collaboration} {Particle Data Group}),\
  }\href {\doibase 10.1093/ptep/ptaa104} {\bibfield  {journal} {\bibinfo
  {journal} {PTEP}\ }\textbf {\bibinfo {volume} {2020}},\ \bibinfo {pages}
  {083C01} (\bibinfo {year} {2020})}\BibitemShut {NoStop}%
\bibitem [{\citenamefont {Buttazzo}\ \emph {et~al.}(2013)\citenamefont
  {Buttazzo}, \citenamefont {Degrassi}, \citenamefont {Giardino}, \citenamefont
  {Giudice}, \citenamefont {Sala}, \citenamefont {Salvio},\ and\ \citenamefont
  {Strumia}}]{Buttazzo:2013uya}%
  \BibitemOpen
  \bibfield  {author} {\bibinfo {author} {\bibfnamefont {D.}~\bibnamefont
  {Buttazzo}}, \bibinfo {author} {\bibfnamefont {G.}~\bibnamefont {Degrassi}},
  \bibinfo {author} {\bibfnamefont {P.~P.}\ \bibnamefont {Giardino}}, \bibinfo
  {author} {\bibfnamefont {G.~F.}\ \bibnamefont {Giudice}}, \bibinfo {author}
  {\bibfnamefont {F.}~\bibnamefont {Sala}}, \bibinfo {author} {\bibfnamefont
  {A.}~\bibnamefont {Salvio}}, \ and\ \bibinfo {author} {\bibfnamefont
  {A.}~\bibnamefont {Strumia}},\ }\href {\doibase 10.1007/JHEP12(2013)089}
  {\bibfield  {journal} {\bibinfo  {journal} {JHEP}\ }\textbf {\bibinfo
  {volume} {12}},\ \bibinfo {pages} {089} (\bibinfo {year} {2013})},\ \Eprint
  {http://arxiv.org/abs/1307.3536} {arXiv:1307.3536 [hep-ph]} \BibitemShut
  {NoStop}%
%%CITATION = ARXIV:1307.3536;%%
\bibitem [{\citenamefont {Hamada}\ \emph {et~al.}(2017)\citenamefont {Hamada},
  \citenamefont {Kawai}, \citenamefont {Nakanishi},\ and\ \citenamefont
  {Oda}}]{Hamada:2016onh}%
  \BibitemOpen
  \bibfield  {author} {\bibinfo {author} {\bibfnamefont {Y.}~\bibnamefont
  {Hamada}}, \bibinfo {author} {\bibfnamefont {H.}~\bibnamefont {Kawai}},
  \bibinfo {author} {\bibfnamefont {Y.}~\bibnamefont {Nakanishi}}, \ and\
  \bibinfo {author} {\bibfnamefont {K.-y.}\ \bibnamefont {Oda}},\ }\href
  {\doibase 10.1103/PhysRevD.95.103524} {\bibfield  {journal} {\bibinfo
  {journal} {Phys. Rev. D}\ }\textbf {\bibinfo {volume} {95}},\ \bibinfo
  {pages} {103524} (\bibinfo {year} {2017})},\ \Eprint
  {http://arxiv.org/abs/1610.05885} {arXiv:1610.05885 [hep-th]} \BibitemShut
  {NoStop}%
\bibitem [{\citenamefont {Hamada}\ \emph {et~al.}(2020)\citenamefont {Hamada},
  \citenamefont {Kawana},\ and\ \citenamefont {Scherlis}}]{Hamada:2020kuy}%
  \BibitemOpen
  \bibfield  {author} {\bibinfo {author} {\bibfnamefont {Y.}~\bibnamefont
  {Hamada}}, \bibinfo {author} {\bibfnamefont {K.}~\bibnamefont {Kawana}}, \
  and\ \bibinfo {author} {\bibfnamefont {A.}~\bibnamefont {Scherlis}},\
  }\href@noop {} {\  (\bibinfo {year} {2020})},\ \Eprint
  {http://arxiv.org/abs/2007.04701} {arXiv:2007.04701 [hep-ph]} \BibitemShut
  {NoStop}%
\bibitem [{\citenamefont {Ema}\ \emph {et~al.}(2020)\citenamefont {Ema},
  \citenamefont {Mukaida},\ and\ \citenamefont {van~de Vis}}]{Ema:2020evi}%
  \BibitemOpen
  \bibfield  {author} {\bibinfo {author} {\bibfnamefont {Y.}~\bibnamefont
  {Ema}}, \bibinfo {author} {\bibfnamefont {K.}~\bibnamefont {Mukaida}}, \ and\
  \bibinfo {author} {\bibfnamefont {J.}~\bibnamefont {van~de Vis}},\
  }\href@noop {} {\  (\bibinfo {year} {2020})},\ \Eprint
  {http://arxiv.org/abs/2008.01096} {arXiv:2008.01096 [hep-ph]} \BibitemShut
  {NoStop}%
\bibitem [{\citenamefont {Bezrukov}\ \emph {et~al.}(2019)\citenamefont
  {Bezrukov}, \citenamefont {Gorbunov}, \citenamefont {Shepherd},\ and\
  \citenamefont {Tokareva}}]{Bezrukov:2019ylq}%
  \BibitemOpen
  \bibfield  {author} {\bibinfo {author} {\bibfnamefont {F.}~\bibnamefont
  {Bezrukov}}, \bibinfo {author} {\bibfnamefont {D.}~\bibnamefont {Gorbunov}},
  \bibinfo {author} {\bibfnamefont {C.}~\bibnamefont {Shepherd}}, \ and\
  \bibinfo {author} {\bibfnamefont {A.}~\bibnamefont {Tokareva}},\ }\href
  {\doibase 10.1016/j.physletb.2019.06.064} {\bibfield  {journal} {\bibinfo
  {journal} {Phys. Lett. B}\ }\textbf {\bibinfo {volume} {795}},\ \bibinfo
  {pages} {657} (\bibinfo {year} {2019})},\ \Eprint
  {http://arxiv.org/abs/1904.04737} {arXiv:1904.04737 [hep-ph]} \BibitemShut
  {NoStop}%
\bibitem [{\citenamefont {He}\ \emph {et~al.}(2020)\citenamefont {He},
  \citenamefont {Jinno}, \citenamefont {Kamada}, \citenamefont {Starobinsky},\
  and\ \citenamefont {Yokoyama}}]{He:2020ivk}%
  \BibitemOpen
  \bibfield  {author} {\bibinfo {author} {\bibfnamefont {M.}~\bibnamefont
  {He}}, \bibinfo {author} {\bibfnamefont {R.}~\bibnamefont {Jinno}}, \bibinfo
  {author} {\bibfnamefont {K.}~\bibnamefont {Kamada}}, \bibinfo {author}
  {\bibfnamefont {A.~A.}\ \bibnamefont {Starobinsky}}, \ and\ \bibinfo {author}
  {\bibfnamefont {J.}~\bibnamefont {Yokoyama}},\ }\href@noop {} {\  (\bibinfo
  {year} {2020})},\ \Eprint {http://arxiv.org/abs/2007.10369} {arXiv:2007.10369
  [hep-ph]} \BibitemShut {NoStop}%
\bibitem [{\citenamefont {Bezrukov}\ and\ \citenamefont
  {Shepherd}(2020)}]{Bezrukov:2020txg}%
  \BibitemOpen
  \bibfield  {author} {\bibinfo {author} {\bibfnamefont {F.}~\bibnamefont
  {Bezrukov}}\ and\ \bibinfo {author} {\bibfnamefont {C.}~\bibnamefont
  {Shepherd}},\ }\href@noop {} {\  (\bibinfo {year} {2020})},\ \Eprint
  {http://arxiv.org/abs/2007.10978} {arXiv:2007.10978 [hep-ph]} \BibitemShut
  {NoStop}%
\bibitem [{\citenamefont {Katz}\ \emph {et~al.}(2018)\citenamefont {Katz},
  \citenamefont {Kopp}, \citenamefont {Sibiryakov},\ and\ \citenamefont
  {Xue}}]{Katz:2018zrn}%
  \BibitemOpen
  \bibfield  {author} {\bibinfo {author} {\bibfnamefont {A.}~\bibnamefont
  {Katz}}, \bibinfo {author} {\bibfnamefont {J.}~\bibnamefont {Kopp}}, \bibinfo
  {author} {\bibfnamefont {S.}~\bibnamefont {Sibiryakov}}, \ and\ \bibinfo
  {author} {\bibfnamefont {W.}~\bibnamefont {Xue}},\ }\href {\doibase
  10.1088/1475-7516/2018/12/005} {\bibfield  {journal} {\bibinfo  {journal}
  {JCAP}\ }\textbf {\bibinfo {volume} {1812}},\ \bibinfo {pages} {005}
  (\bibinfo {year} {2018})},\ \Eprint {http://arxiv.org/abs/1807.11495}
  {arXiv:1807.11495 [astro-ph.CO]} \BibitemShut {NoStop}%
%%CITATION = ARXIV:1807.11495;%%
\bibitem [{\citenamefont {Jung}\ and\ \citenamefont
  {Kim}(2019)}]{Jung:2019fcs}%
  \BibitemOpen
  \bibfield  {author} {\bibinfo {author} {\bibfnamefont {S.}~\bibnamefont
  {Jung}}\ and\ \bibinfo {author} {\bibfnamefont {T.}~\bibnamefont {Kim}},\
  }\href@noop {} {\  (\bibinfo {year} {2019})},\ \Eprint
  {http://arxiv.org/abs/1908.00078} {arXiv:1908.00078 [astro-ph.CO]}
  \BibitemShut {NoStop}%
%%CITATION = ARXIV:1908.00078;%%
\bibitem [{\citenamefont {Dasgupta}\ \emph {et~al.}(2019)\citenamefont
  {Dasgupta}, \citenamefont {Laha},\ and\ \citenamefont
  {Ray}}]{Dasgupta:2019cae}%
  \BibitemOpen
  \bibfield  {author} {\bibinfo {author} {\bibfnamefont {B.}~\bibnamefont
  {Dasgupta}}, \bibinfo {author} {\bibfnamefont {R.}~\bibnamefont {Laha}}, \
  and\ \bibinfo {author} {\bibfnamefont {A.}~\bibnamefont {Ray}},\ }\href@noop
  {} {\  (\bibinfo {year} {2019})},\ \Eprint {http://arxiv.org/abs/1912.01014}
  {arXiv:1912.01014 [hep-ph]} \BibitemShut {NoStop}%
%%CITATION = ARXIV:1912.01014;%%
\bibitem [{\citenamefont {Ezquiaga}\ \emph {et~al.}(2018)\citenamefont
  {Ezquiaga}, \citenamefont {Garcia-Bellido},\ and\ \citenamefont
  {Ruiz~Morales}}]{Ezquiaga:2017fvi}%
  \BibitemOpen
  \bibfield  {author} {\bibinfo {author} {\bibfnamefont {J.~M.}\ \bibnamefont
  {Ezquiaga}}, \bibinfo {author} {\bibfnamefont {J.}~\bibnamefont
  {Garcia-Bellido}}, \ and\ \bibinfo {author} {\bibfnamefont {E.}~\bibnamefont
  {Ruiz~Morales}},\ }\href {\doibase 10.1016/j.physletb.2017.11.039} {\bibfield
   {journal} {\bibinfo  {journal} {Phys. Lett.}\ }\textbf {\bibinfo {volume}
  {B776}},\ \bibinfo {pages} {345} (\bibinfo {year} {2018})},\ \Eprint
  {http://arxiv.org/abs/1705.04861} {arXiv:1705.04861 [astro-ph.CO]}
  \BibitemShut {NoStop}%
%%CITATION = ARXIV:1705.04861;%%
\bibitem [{\citenamefont {Kannike}\ \emph {et~al.}(2017)\citenamefont
  {Kannike}, \citenamefont {Marzola}, \citenamefont {Raidal},\ and\
  \citenamefont {Veermäe}}]{Kannike:2017bxn}%
  \BibitemOpen
  \bibfield  {author} {\bibinfo {author} {\bibfnamefont {K.}~\bibnamefont
  {Kannike}}, \bibinfo {author} {\bibfnamefont {L.}~\bibnamefont {Marzola}},
  \bibinfo {author} {\bibfnamefont {M.}~\bibnamefont {Raidal}}, \ and\ \bibinfo
  {author} {\bibfnamefont {H.}~\bibnamefont {Veermäe}},\ }\href {\doibase
  10.1088/1475-7516/2017/09/020} {\bibfield  {journal} {\bibinfo  {journal}
  {JCAP}\ }\textbf {\bibinfo {volume} {1709}},\ \bibinfo {pages} {020}
  (\bibinfo {year} {2017})},\ \Eprint {http://arxiv.org/abs/1705.06225}
  {arXiv:1705.06225 [astro-ph.CO]} \BibitemShut {NoStop}%
%%CITATION = ARXIV:1705.06225;%%
\bibitem [{\citenamefont {Germani}\ and\ \citenamefont
  {Prokopec}(2017)}]{Germani:2017bcs}%
  \BibitemOpen
  \bibfield  {author} {\bibinfo {author} {\bibfnamefont {C.}~\bibnamefont
  {Germani}}\ and\ \bibinfo {author} {\bibfnamefont {T.}~\bibnamefont
  {Prokopec}},\ }\href {\doibase 10.1016/j.dark.2017.09.001} {\bibfield
  {journal} {\bibinfo  {journal} {Phys. Dark Univ.}\ }\textbf {\bibinfo
  {volume} {18}},\ \bibinfo {pages} {6} (\bibinfo {year} {2017})},\ \Eprint
  {http://arxiv.org/abs/1706.04226} {arXiv:1706.04226 [astro-ph.CO]}
  \BibitemShut {NoStop}%
%%CITATION = ARXIV:1706.04226;%%
\bibitem [{\citenamefont {Bezrukov}\ \emph {et~al.}(2018)\citenamefont
  {Bezrukov}, \citenamefont {Pauly},\ and\ \citenamefont
  {Rubio}}]{Bezrukov:2017dyv}%
  \BibitemOpen
  \bibfield  {author} {\bibinfo {author} {\bibfnamefont {F.}~\bibnamefont
  {Bezrukov}}, \bibinfo {author} {\bibfnamefont {M.}~\bibnamefont {Pauly}}, \
  and\ \bibinfo {author} {\bibfnamefont {J.}~\bibnamefont {Rubio}},\ }\href
  {\doibase 10.1088/1475-7516/2018/02/040} {\bibfield  {journal} {\bibinfo
  {journal} {JCAP}\ }\textbf {\bibinfo {volume} {1802}},\ \bibinfo {pages}
  {040} (\bibinfo {year} {2018})},\ \Eprint {http://arxiv.org/abs/1706.05007}
  {arXiv:1706.05007 [hep-ph]} \BibitemShut {NoStop}%
%%CITATION = ARXIV:1706.05007;%%
\bibitem [{\citenamefont {Motohashi}\ and\ \citenamefont
  {Hu}(2017)}]{Motohashi:2017kbs}%
  \BibitemOpen
  \bibfield  {author} {\bibinfo {author} {\bibfnamefont {H.}~\bibnamefont
  {Motohashi}}\ and\ \bibinfo {author} {\bibfnamefont {W.}~\bibnamefont {Hu}},\
  }\href {\doibase 10.1103/PhysRevD.96.063503} {\bibfield  {journal} {\bibinfo
  {journal} {Phys. Rev.}\ }\textbf {\bibinfo {volume} {D96}},\ \bibinfo {pages}
  {063503} (\bibinfo {year} {2017})},\ \Eprint
  {http://arxiv.org/abs/1706.06784} {arXiv:1706.06784 [astro-ph.CO]}
  \BibitemShut {NoStop}%
%%CITATION = ARXIV:1706.06784;%%
\bibitem [{\citenamefont {Masina}(2018)}]{Masina:2018ejw}%
  \BibitemOpen
  \bibfield  {author} {\bibinfo {author} {\bibfnamefont {I.}~\bibnamefont
  {Masina}},\ }\href {\doibase 10.1103/PhysRevD.98.043536} {\bibfield
  {journal} {\bibinfo  {journal} {Phys. Rev.}\ }\textbf {\bibinfo {volume}
  {D98}},\ \bibinfo {pages} {043536} (\bibinfo {year} {2018})},\ \Eprint
  {http://arxiv.org/abs/1805.02160} {arXiv:1805.02160 [hep-ph]} \BibitemShut
  {NoStop}%
%%CITATION = ARXIV:1805.02160;%%
\bibitem [{\citenamefont {Drees}\ and\ \citenamefont
  {Xu}(2019)}]{Drees:2019xpp}%
  \BibitemOpen
  \bibfield  {author} {\bibinfo {author} {\bibfnamefont {M.}~\bibnamefont
  {Drees}}\ and\ \bibinfo {author} {\bibfnamefont {Y.}~\bibnamefont {Xu}},\
  }\href@noop {} {\  (\bibinfo {year} {2019})},\ \Eprint
  {http://arxiv.org/abs/1905.13581} {arXiv:1905.13581 [hep-ph]} \BibitemShut
  {NoStop}%
%%CITATION = ARXIV:1905.13581;%%
\bibitem [{\citenamefont {Cheong}\ \emph
  {et~al.}(2019{\natexlab{b}})\citenamefont {Cheong}, \citenamefont {Lee},\
  and\ \citenamefont {Park}}]{Cheong:2019vzl}%
  \BibitemOpen
  \bibfield  {author} {\bibinfo {author} {\bibfnamefont {D.~Y.}\ \bibnamefont
  {Cheong}}, \bibinfo {author} {\bibfnamefont {S.~M.}\ \bibnamefont {Lee}}, \
  and\ \bibinfo {author} {\bibfnamefont {S.~C.}\ \bibnamefont {Park}},\
  }\href@noop {} {\  (\bibinfo {year} {2019}{\natexlab{b}})},\ \Eprint
  {http://arxiv.org/abs/1912.12032} {arXiv:1912.12032 [hep-ph]} \BibitemShut
  {NoStop}%
\bibitem [{\citenamefont {Green}\ \emph {et~al.}(2004)\citenamefont {Green},
  \citenamefont {Liddle}, \citenamefont {Malik},\ and\ \citenamefont
  {Sasaki}}]{Green:2004wb}%
  \BibitemOpen
  \bibfield  {author} {\bibinfo {author} {\bibfnamefont {A.~M.}\ \bibnamefont
  {Green}}, \bibinfo {author} {\bibfnamefont {A.~R.}\ \bibnamefont {Liddle}},
  \bibinfo {author} {\bibfnamefont {K.~A.}\ \bibnamefont {Malik}}, \ and\
  \bibinfo {author} {\bibfnamefont {M.}~\bibnamefont {Sasaki}},\ }\href
  {\doibase 10.1103/PhysRevD.70.041502} {\bibfield  {journal} {\bibinfo
  {journal} {Phys. Rev.}\ }\textbf {\bibinfo {volume} {D70}},\ \bibinfo {pages}
  {041502} (\bibinfo {year} {2004})},\ \Eprint
  {http://arxiv.org/abs/astro-ph/0403181} {arXiv:astro-ph/0403181 [astro-ph]}
  \BibitemShut {NoStop}%
%%CITATION = ASTRO-PH/0403181;%%
\bibitem [{\citenamefont {Harada}\ \emph {et~al.}(2013)\citenamefont {Harada},
  \citenamefont {Yoo},\ and\ \citenamefont {Kohri}}]{Harada:2013epa}%
  \BibitemOpen
  \bibfield  {author} {\bibinfo {author} {\bibfnamefont {T.}~\bibnamefont
  {Harada}}, \bibinfo {author} {\bibfnamefont {C.-M.}\ \bibnamefont {Yoo}}, \
  and\ \bibinfo {author} {\bibfnamefont {K.}~\bibnamefont {Kohri}},\ }\href
  {\doibase 10.1103/PhysRevD.88.084051, 10.1103/PhysRevD.89.029903} {\bibfield
  {journal} {\bibinfo  {journal} {Phys. Rev.}\ }\textbf {\bibinfo {volume}
  {D88}},\ \bibinfo {pages} {084051} (\bibinfo {year} {2013})},\ \bibinfo
  {note} {[Erratum: Phys. Rev.D89,no.2,029903(2014)]},\ \Eprint
  {http://arxiv.org/abs/1309.4201} {arXiv:1309.4201 [astro-ph.CO]} \BibitemShut
  {NoStop}%
%%CITATION = ARXIV:1309.4201;%%
\bibitem [{\citenamefont {Musco}(2019)}]{Musco:2018rwt}%
  \BibitemOpen
  \bibfield  {author} {\bibinfo {author} {\bibfnamefont {I.}~\bibnamefont
  {Musco}},\ }\href {\doibase 10.1103/PhysRevD.100.123524} {\bibfield
  {journal} {\bibinfo  {journal} {Phys. Rev.}\ }\textbf {\bibinfo {volume}
  {D100}},\ \bibinfo {pages} {123524} (\bibinfo {year} {2019})},\ \Eprint
  {http://arxiv.org/abs/1809.02127} {arXiv:1809.02127 [gr-qc]} \BibitemShut
  {NoStop}%
%%CITATION = ARXIV:1809.02127;%%
\bibitem [{\citenamefont {Carr}(1975)}]{Carr:1975qj}%
  \BibitemOpen
  \bibfield  {author} {\bibinfo {author} {\bibfnamefont {B.~J.}\ \bibnamefont
  {Carr}},\ }\href {\doibase 10.1086/153853} {\bibfield  {journal} {\bibinfo
  {journal} {Astrophys. J.}\ }\textbf {\bibinfo {volume} {201}},\ \bibinfo
  {pages} {1} (\bibinfo {year} {1975})}\BibitemShut {NoStop}%
%%CITATION = ASJOA,201,1;%%
\bibitem [{\citenamefont {Carr}\ \emph {et~al.}(2010)\citenamefont {Carr},
  \citenamefont {Kohri}, \citenamefont {Sendouda},\ and\ \citenamefont
  {Yokoyama}}]{Carr:2009jm}%
  \BibitemOpen
  \bibfield  {author} {\bibinfo {author} {\bibfnamefont {B.~J.}\ \bibnamefont
  {Carr}}, \bibinfo {author} {\bibfnamefont {K.}~\bibnamefont {Kohri}},
  \bibinfo {author} {\bibfnamefont {Y.}~\bibnamefont {Sendouda}}, \ and\
  \bibinfo {author} {\bibfnamefont {J.}~\bibnamefont {Yokoyama}},\ }\href
  {\doibase 10.1103/PhysRevD.81.104019} {\bibfield  {journal} {\bibinfo
  {journal} {Phys. Rev.}\ }\textbf {\bibinfo {volume} {D81}},\ \bibinfo {pages}
  {104019} (\bibinfo {year} {2010})},\ \Eprint {http://arxiv.org/abs/0912.5297}
  {arXiv:0912.5297 [astro-ph.CO]} \BibitemShut {NoStop}%
%%CITATION = ARXIV:0912.5297;%%
\bibitem [{\citenamefont {Young}\ \emph {et~al.}(2014)\citenamefont {Young},
  \citenamefont {Byrnes},\ and\ \citenamefont {Sasaki}}]{Young:2014ana}%
  \BibitemOpen
  \bibfield  {author} {\bibinfo {author} {\bibfnamefont {S.}~\bibnamefont
  {Young}}, \bibinfo {author} {\bibfnamefont {C.~T.}\ \bibnamefont {Byrnes}}, \
  and\ \bibinfo {author} {\bibfnamefont {M.}~\bibnamefont {Sasaki}},\ }\href
  {\doibase 10.1088/1475-7516/2014/07/045} {\bibfield  {journal} {\bibinfo
  {journal} {JCAP}\ }\textbf {\bibinfo {volume} {1407}},\ \bibinfo {pages}
  {045} (\bibinfo {year} {2014})},\ \Eprint {http://arxiv.org/abs/1405.7023}
  {arXiv:1405.7023 [gr-qc]} \BibitemShut {NoStop}%
%%CITATION = ARXIV:1405.7023;%%
\bibitem [{\citenamefont {Codello}\ and\ \citenamefont
  {Jain}(2016)}]{Codello:2015mba}%
  \BibitemOpen
  \bibfield  {author} {\bibinfo {author} {\bibfnamefont {A.}~\bibnamefont
  {Codello}}\ and\ \bibinfo {author} {\bibfnamefont {R.~K.}\ \bibnamefont
  {Jain}},\ }\href {\doibase 10.1088/0264-9381/33/22/225006} {\bibfield
  {journal} {\bibinfo  {journal} {Class. Quant. Grav.}\ }\textbf {\bibinfo
  {volume} {33}},\ \bibinfo {pages} {225006} (\bibinfo {year} {2016})},\
  \Eprint {http://arxiv.org/abs/1507.06308} {arXiv:1507.06308 [gr-qc]}
  \BibitemShut {NoStop}%
%%CITATION = ARXIV:1507.06308;%%
\bibitem [{\citenamefont {Markkanen}\ \emph
  {et~al.}(2018{\natexlab{b}})\citenamefont {Markkanen}, \citenamefont {Nurmi},
  \citenamefont {Rajantie},\ and\ \citenamefont {Stopyra}}]{Markkanen:2018bfx}%
  \BibitemOpen
  \bibfield  {author} {\bibinfo {author} {\bibfnamefont {T.}~\bibnamefont
  {Markkanen}}, \bibinfo {author} {\bibfnamefont {S.}~\bibnamefont {Nurmi}},
  \bibinfo {author} {\bibfnamefont {A.}~\bibnamefont {Rajantie}}, \ and\
  \bibinfo {author} {\bibfnamefont {S.}~\bibnamefont {Stopyra}},\ }\href
  {\doibase 10.1007/JHEP06(2018)040} {\bibfield  {journal} {\bibinfo  {journal}
  {JHEP}\ }\textbf {\bibinfo {volume} {06}},\ \bibinfo {pages} {040} (\bibinfo
  {year} {2018}{\natexlab{b}})},\ \Eprint {http://arxiv.org/abs/1804.02020}
  {arXiv:1804.02020 [hep-ph]} \BibitemShut {NoStop}%
%%CITATION = ARXIV:1804.02020;%%
\bibitem [{\citenamefont {De~Simone}\ \emph {et~al.}(2009)\citenamefont
  {De~Simone}, \citenamefont {Hertzberg},\ and\ \citenamefont
  {Wilczek}}]{DeSimone:2008ei}%
  \BibitemOpen
  \bibfield  {author} {\bibinfo {author} {\bibfnamefont {A.}~\bibnamefont
  {De~Simone}}, \bibinfo {author} {\bibfnamefont {M.~P.}\ \bibnamefont
  {Hertzberg}}, \ and\ \bibinfo {author} {\bibfnamefont {F.}~\bibnamefont
  {Wilczek}},\ }\href {\doibase 10.1016/j.physletb.2009.05.054} {\bibfield
  {journal} {\bibinfo  {journal} {Phys. Lett.}\ }\textbf {\bibinfo {volume}
  {B678}},\ \bibinfo {pages} {1} (\bibinfo {year} {2009})},\ \Eprint
  {http://arxiv.org/abs/0812.4946} {arXiv:0812.4946 [hep-ph]} \BibitemShut
  {NoStop}%
%%CITATION = ARXIV:0812.4946;%%
\bibitem [{\citenamefont {Huang}(2014)}]{Huang:2013hsb}%
  \BibitemOpen
  \bibfield  {author} {\bibinfo {author} {\bibfnamefont {Q.-G.}\ \bibnamefont
  {Huang}},\ }\href {\doibase 10.1088/1475-7516/2014/02/035} {\bibfield
  {journal} {\bibinfo  {journal} {JCAP}\ }\textbf {\bibinfo {volume} {1402}},\
  \bibinfo {pages} {035} (\bibinfo {year} {2014})},\ \Eprint
  {http://arxiv.org/abs/1309.3514} {arXiv:1309.3514 [hep-th]} \BibitemShut
  {NoStop}%
%%CITATION = ARXIV:1309.3514;%%
\bibitem [{\citenamefont {Sebastiani}\ \emph {et~al.}(2014)\citenamefont
  {Sebastiani}, \citenamefont {Cognola}, \citenamefont {Myrzakulov},
  \citenamefont {Odintsov},\ and\ \citenamefont
  {Zerbini}}]{Sebastiani:2013eqa}%
  \BibitemOpen
  \bibfield  {author} {\bibinfo {author} {\bibfnamefont {L.}~\bibnamefont
  {Sebastiani}}, \bibinfo {author} {\bibfnamefont {G.}~\bibnamefont {Cognola}},
  \bibinfo {author} {\bibfnamefont {R.}~\bibnamefont {Myrzakulov}}, \bibinfo
  {author} {\bibfnamefont {S.~D.}\ \bibnamefont {Odintsov}}, \ and\ \bibinfo
  {author} {\bibfnamefont {S.}~\bibnamefont {Zerbini}},\ }\href {\doibase
  10.1103/PhysRevD.89.023518} {\bibfield  {journal} {\bibinfo  {journal} {Phys.
  Rev.}\ }\textbf {\bibinfo {volume} {D89}},\ \bibinfo {pages} {023518}
  (\bibinfo {year} {2014})},\ \Eprint {http://arxiv.org/abs/1311.0744}
  {arXiv:1311.0744 [gr-qc]} \BibitemShut {NoStop}%
%%CITATION = ARXIV:1311.0744;%%
\bibitem [{\citenamefont {Kamada}\ and\ \citenamefont
  {Yokoyama}(2014)}]{Kamada:2014gma}%
  \BibitemOpen
  \bibfield  {author} {\bibinfo {author} {\bibfnamefont {K.}~\bibnamefont
  {Kamada}}\ and\ \bibinfo {author} {\bibfnamefont {J.}~\bibnamefont
  {Yokoyama}},\ }\href {\doibase 10.1103/PhysRevD.90.103520} {\bibfield
  {journal} {\bibinfo  {journal} {Phys. Rev.}\ }\textbf {\bibinfo {volume}
  {D90}},\ \bibinfo {pages} {103520} (\bibinfo {year} {2014})},\ \Eprint
  {http://arxiv.org/abs/1405.6732} {arXiv:1405.6732 [hep-th]} \BibitemShut
  {NoStop}%
%%CITATION = ARXIV:1405.6732;%%
\bibitem [{\citenamefont {Artymowski}\ \emph {et~al.}(2015)\citenamefont
  {Artymowski}, \citenamefont {Lalak},\ and\ \citenamefont
  {Lewicki}}]{Artymowski:2014nva}%
  \BibitemOpen
  \bibfield  {author} {\bibinfo {author} {\bibfnamefont {M.}~\bibnamefont
  {Artymowski}}, \bibinfo {author} {\bibfnamefont {Z.}~\bibnamefont {Lalak}}, \
  and\ \bibinfo {author} {\bibfnamefont {M.}~\bibnamefont {Lewicki}},\ }\href
  {\doibase 10.1088/1475-7516/2015/06/031} {\bibfield  {journal} {\bibinfo
  {journal} {JCAP}\ }\textbf {\bibinfo {volume} {1506}},\ \bibinfo {pages}
  {031} (\bibinfo {year} {2015})},\ \Eprint {http://arxiv.org/abs/1412.8075}
  {arXiv:1412.8075 [hep-th]} \BibitemShut {NoStop}%
%%CITATION = ARXIV:1412.8075;%%
\bibitem [{\citenamefont {Cheong}\ \emph {et~al.}(2020)\citenamefont {Cheong},
  \citenamefont {Lee},\ and\ \citenamefont {Park}}]{Cheong:2020rao}%
  \BibitemOpen
  \bibfield  {author} {\bibinfo {author} {\bibfnamefont {D.~Y.}\ \bibnamefont
  {Cheong}}, \bibinfo {author} {\bibfnamefont {H.~M.}\ \bibnamefont {Lee}}, \
  and\ \bibinfo {author} {\bibfnamefont {S.~C.}\ \bibnamefont {Park}},\ }\href
  {\doibase 10.1016/j.physletb.2020.135453} {\bibfield  {journal} {\bibinfo
  {journal} {Phys. Lett. B}\ }\textbf {\bibinfo {volume} {805}},\ \bibinfo
  {pages} {135453} (\bibinfo {year} {2020})},\ \Eprint
  {http://arxiv.org/abs/2002.07981} {arXiv:2002.07981 [hep-ph]} \BibitemShut
  {NoStop}%
\end{thebibliography}%

\end{document}